%% file: main.tex
\documentclass[reprint, amsmath, amssymb, showpacs, floatfix, longbibliography, aps, twocolumn, superscriptaddress, nofootinbib, nobalancelastpage]{revtex4-1}
\usepackage{CJKutf8}
\usepackage{amsmath,amssymb,bm,mathtools}

\usepackage{amsthm}
\usepackage{graphicx,booktabs}
\usepackage{array}[=2016-10-06]
\usepackage{tabularx}
\usepackage{placeins}
\usepackage[dvipsnames]{xcolor}
\usepackage{tikz}
\usetikzlibrary{arrows.meta,calc}
\usepackage[linktocpage]{hyperref}
\hypersetup{colorlinks=true,citecolor=blue,linkcolor=blue,urlcolor=blue,breaklinks=true,pdftitle={Higher-Group Statistics on Lattices},pdfauthor={Chen Yang and Yu-An Chen}}
\newcommand{\Z}{\mathbb Z}
\newcommand{\R}{\mathbb R}

\newcommand{\supp}{\operatorname{supp}}

\newcommand{\Gg}{\mathbb G}

\newcommand{\Sq}{\operatorname{Sq}}

\newcommand{\vac}{\mathrm{vac}}

\newcommand{\cfg}{\mathfrak c}
\newcommand{\ket}[1]{\lvert #1\rangle}
\newcommand{\bra}[1]{\langle #1\rvert}

\theoremstyle{definition}
\newtheorem{theorem}{Theorem}[section]
\newtheorem{proposition}[theorem]{Proposition}
\newtheorem{lemma}[theorem]{Lemma}

\newtheorem{definition}[theorem]{Definition}

\newtheorem{remark}{Remark}[section]
\graphicspath{{figures/}}

\begin{document}
\title{Higher-Group Statistics on Lattices}
\author{Chen Yang~\begin{CJK*}{UTF8}{gbsn}(杨晨)\end{CJK*}}
\affiliation{International Center for Quantum Materials, School of Physics, Peking University, Beijing 100871, China}
\author{Yu-An Chen~\begin{CJK*}{UTF8}{bsmi}(陳昱安)\end{CJK*}}
\email[E-mail: ]{yuanchen@pku.edu.cn}
\affiliation{International Center for Quantum Materials, School of Physics, Peking University, Beijing 100871, China}
\date{\today}

\begin{abstract}
Crossing a domain wall can permute anyons, and intersecting strings can create a particle.
These couplings between defects of various dimensions are specified by higher groups.
We develop a lattice framework defining higher-group statistics as Berry phases of closed sequences of hopping operators, invariant under local modifications of the operators.
Our central example is a closed Majorana chain, a 1-dimensional topological superconductor, embedded in 3 dimensions. We construct a loop-flip process with phase $+i$, refining the usual $\pm 1$ loop-flip invariant through intermediate transfer of fermion parity between the chain and physical fermions.
In two dimensions, charge-wall statistics distinguish two $\mathbb Z_4$ toric codes enriched by $\mathbb Z_2$ charge-conjugation symmetry, with identical anyon statistics but distinct gauged theories: the quantum doubles of the dihedral group $D_4$ and the quaternion group $Q_8$.
We provide two complementary constructions of statistical processes.
The first extends the Smith normal form algorithm in \href{https://doi.org/10.1103/6k88-w52n}{[Phys.\ Rev.\ X \textbf{16}, 011010 (2026)]} to higher groups, generating processes from lattice geometry and prescribed local moves.
For a finite higher group $\mathbb G$, we prove that statistical processes on a $d$-sphere are classified by $H_{d+2}(B\mathbb G,\mathbb Z)$.
The second constructs processes directly from homology classes without examining exponentially many defect configurations, enabling us to study statistics for complex higher-group structures where the previous method is computationally prohibitive.
These processes detect boundary anomalies of higher-group symmetry-protected topological phases. For example, we distinguish 2-group anomalies in 1, 2, and 3 dimensions. Higher-group statistics provide new invariants for characterizing coupled defects in quantum many-body systems.
\end{abstract}
\maketitle

\hypersetup{bookmarksdepth=2}
\begingroup
\makeatletter
\def\appendix@toc{%
  \let\l@subsection\@gobbletwo
  \let\l@subsubsection\@gobbletwo
}
\tableofcontents
\makeatother
\endgroup

\section{Introduction}
\label{sec:introduction}

\begin{table*}[t]
 \centering\small
 \setlength{\tabcolsep}{5pt}
 \newcommand{\measurementcell}[1]{%
   \parbox[c][2.8\baselineskip][c]{\linewidth}{\raggedright #1}}
 \begin{tabularx}{\textwidth}{@{}cllcX@{}}
 \toprule
 \textbf{Spatial $d$} & \textbf{Statistical process} &
 \textbf{Anomaly cocycle} & \textbf{Phase} & \textbf{What it measures}\\
 \midrule
 $2$ & Charge-wall $\mathcal W_{10}$ \eqref{ex:inversion-word-main}
 & $\Omega^{(\epsilon)}$ \eqref{ex:inversion-anomaly}
 & $(-1)^\epsilon$ & \measurementcell{Distinguishes symmetry fractionalization despite identical intrinsic anyon statistics.}\\
 $3$ & Kitaev-chain flip $\mathcal K_{24}$ \eqref{eq:ftc-word}
 & $\Omega_{\rm K}$ \eqref{eq:ftc-restricted-anomaly}
 & $+i$ & \measurementcell{Loop flip with parity transfer between the Kitaev chain and physical fermions.}\\
 $1$ & Wall transport $\mathcal D_6$ \eqref{spt:d1-six}
 & $\alpha_3$ \eqref{spt:alpha-three}
 & $-1$ & \measurementcell{Compares a junction hop in two wall backgrounds on a circle.}\\
 $2$ & Particle exchange $\mathcal T_6$ \eqref{spt:d2-short-word}
 & $\alpha_4$ \eqref{spt:d2-anomaly}
 & $-1$ & \measurementcell{Measures the particle exchange at a T-junction.}\\
 $3$ & Loop flip $\mu_{24}$ \eqref{spt:typeI-phase}
 & $\mu\alpha_5^{\mathrm I}+\nu\alpha_5^{\mathrm{II}}$ \eqref{spt:anomalies}
 & $(-1)^\mu$ & \measurementcell{Isolates the coefficient $\mu$ using loop flipping.}\\
 $3$ & Wall-loop $\mathcal N_{50}$ \eqref{spt:mixed-protocol}
 & $\mu\alpha_5^{\mathrm I}+\nu\alpha_5^{\mathrm{II}}$ \eqref{spt:anomalies}
 & $(-1)^\nu$ & \measurementcell{Isolates the coefficient $\nu$ using wall and loop moves.}\\
 \bottomrule
 \end{tabularx}
 \caption{
Higher-group statistical processes in $d$ spatial dimensions.
Each sequence restores the starting configuration, and its phase
is invariant under local redefinitions of the hopping operators. The
charge-wall process $\mathcal W_{10}$ distinguishes two
$\Z_4$ toric codes with the same $\Z_2$ charge-conjugation
symmetry. Gauging $\Z_2$ symmetry gives the quantum doubles of
$D_4$ for $\epsilon=0$ and $Q_8$ for $\epsilon=1$.
The process $\mathcal K_{24}$ flips a closed Kitaev chain and yields $+i$, with physical fermions appearing during the process.
The remaining rows probe boundaries of symmetry-protected topological (SPT) phases with a nonsplit 2-group coupling a $\Z_2$ 0-form symmetry to a $\Z_2$ 1-form symmetry. In $d=1, 2$, domain-wall transport and particle exchange give $-1$ for the respective nontrivial anomalies. In $d=3$, the loop flip and the wall-loop process separately determine the anomaly coefficients $\mu$ and $\nu$.
}
\label{intro:measurements}
\label{intro:process-lengths}
\end{table*}

A particle crossing a domain wall can change its anyon
type~\cite{Bombin2010Topological, Barkeshli2019Symmetry, Barkeshli2013Twist}.
In three dimensions, a flux loop crossing a twist string can
create a point charge~\cite{Barkeshli2023Codimension}.
Twisting a Kitaev-chain defect can also transfer fermion parity
between the chain and physical
fermions~\cite{Kitaev2001Unpaired, Tarantino2016Discrete, Wang2018Towards, Wang2020Construction}.
These couplings between defects of different dimensions are
encoded by higher-group
structures~\cite{Gaiotto2015Generalized,Kapustin2017Higher,Benini2019TwoGroup,Cordova2019Exploring}.
Such structures occur in finite gauge theories and in the algebra
of logical gates of topological stabilizer
codes~\cite{Barkeshli2024Higher,Yoshida2015Topological,Yoshida2016Topological}.

Statistical phases are Berry phases acquired during closed sequences of defect moves, including anyon braiding, loop flips, and comparisons between fusion paths~\cite{Arovas1984Fractional, Kitaev2003FaultTolerant, Kitaev2006Anyons, Wang2014Braiding, Kawagoe2020Microscopic, Fidkowski2022Gravitational, Carolyn2023entanglers, Kobayashi2026Generalized, Levin2003Fermions, Putrov2017Braiding, Liang2024Extracting, Sun2026Majorana}.
These statistics distinguish phases of matter and constrain possible phase transitions~\cite{Jiang2014Generalized, Jian2014Layer, Lin2015Loop, Wen2019Emergent, McGreevy2023Generalized}. For example, in a two-dimensional bosonic
topological phase, Abelian anyons that condense together must be
bosons and have trivial mutual braiding. Excitations that braid
nontrivially with the condensate are confined, while those that
braid trivially can remain deconfined~\cite{Kesselring2024Anyon, Burnell2018Anyon, Kong2014Anyon, Barkeshli2013Theory, Ellison2022Pauli, Liang2024Operator}.
The statistics of symmetry defects can also diagnose symmetry
anomalies~\cite{Levin2012Braiding, Else2014Classifying, Wang2016Bulk, Liujun2024Symmetries, Kapustin2025Higher, Feng2026Higher, Hsin2026Bockstein,Wang2015Topological,Hsin2019Comments,Tu2026Anomalies}.
A nontrivial anomaly detected in this way rules out a
symmetry-preserving short-range-entangled ground
state~\cite{Bravyi2025HowMuch, Li2025HowMuch, Shirley2026AnomalyFree}.

The statistics of individual defect species need not determine
those of their coupled processes. We develop a lattice framework
that extracts this additional information from closed sequences
of microscopic hopping operators. These sequences are chosen
so that their phases are invariant under local modifications
of the operators~\cite{Kawagoe2020Microscopic, Kobayashi2026Generalized}.
For our lattice construction on a spatial $d$-sphere, we prove
that statistical processes for a finite higher group $\mathbb G$
are classified by the homology group
$H_{d+2}(B\mathbb G,\mathbb Z)$.
Assignments of statistical phases are classified by the dual cohomology group
$H^{d+2}(B\mathbb G,U(1))$.\footnote{Here, $B\mathbb G$
is the classifying space of $\mathbb G$~\cite{Lan2019Fermion}.
There is a canonical isomorphism
$H^{d+2}(B\mathbb G,U(1))\cong
\operatorname{Hom}(H_{d+2}(B\mathbb G,\mathbb Z),U(1))$.
Appendix~\ref{app:conv:sec} gives mathematical preliminaries.}
Physically, these statistical phases detect boundary anomalies
of symmetry-protected topological (SPT) phases with
higher-group symmetry $\mathbb G$ in $d+1$ spatial
dimensions~\cite{Chen2013Symmetry,Kapustin2017Higher}.

Sec.~\ref{ex:section} presents two microscopic examples.
Our central example is a closed Kitaev chain, a
one-dimensional topological superconductor embedded in a
three-dimensional fermionic
system~\cite{Kitaev2001Unpaired,Barkeshli2023Codimension}.
We construct a process that reverses the loop orientation and
has phase $+i$, refining the usual $\Z_2$ loop-flip
invariant~\cite{Fidkowski2022Gravitational,Kobayashi2026Generalized,Chen2023Exactly}.
During the process, membrane operators transfer fermion parity
between the chain and physical fermions, so the intermediate
configurations involve both defect species.
A second example is a two-dimensional charge-wall process
that gives opposite signs in two $\Z_4$ toric codes enriched
by $\Z_2$ charge-conjugation symmetry. The theories have
identical intrinsic anyon statistics and the same permutation
of anyon types across a wall, but the 2-group statistics distinguishes their symmetry enrichment before gauging. Gauging the symmetry gives the quantum doubles of the dihedral group $D_4$ and the quaternion group $Q_8$~\cite{Tarantino2016Symmetry}.

\input{figures/inversion_framed_history_figure}

Sec.~\ref{loc:sec} extends the Smith normal form algorithm of
Ref.~\cite{Kobayashi2026Generalized} to coupled defects with higher-group structures.
Lattice geometry fixes the operator supports, and the hopping
rules track changes in all affected defect species. From these
data and the locality relations of the operators, we identify
the independent statistical processes.

Sec.~\ref{dc:sec} gives an explicit one-to-one correspondence
between homology classes and statistical process classes
for the specified spherical lattice and full set of moves.
Given a cycle, we can write down the corresponding hopping
sequence in closed form, without enumerating all allowed
defect configurations or relying on computer assistance.
This establishes the correspondence conjectured in
Ref.~\cite{Kobayashi2026Generalized} and realizes the statistical
phases predicted by the hopping-operator
algebra~\cite{Wen2026Holographic}.

Sec.~\ref{sec:spt-examples} applies the construction to a 2-group
that couples a $\Z_2$ 0-form symmetry to a $\Z_2$ 1-form symmetry.
In one, two, and three dimensions, the resulting statistical
processes distinguish the boundary anomalies of SPT phases in one
higher dimension.
Table~\ref{intro:measurements} summarizes the main results and examples.

\section{Microscopic examples}
\label{ex:section}

We illustrate higher-group statistics in two microscopic settings.
In two dimensions, a closed sequence of charge and wall moves gives
opposite phases in two $\Z_4$ toric codes with the same charge-conjugation
symmetry. In three dimensions, membrane operators move a closed Kitaev chain, while creating or annihilating physical fermions. The loop flip gives phase $+i$. In both examples, we follow the defects through the sequence and check that the local phase shifts cancel.

\subsection{Charge--wall statistics in \texorpdfstring{$\Z_2$-enriched $\Z_4$}{Z2-enriched Z4} toric codes}
\label{sec:inversion-comparison}

\input{sections/microscopic_inversion}

\subsection{Kitaev-chain flip statistics}
\label{sec:fermionic-toric}

\input{sections/microscopic_fermionic_toric}

\section{Lattice framework for higher-group statistics}
\label{loc:sec}

We describe higher-group defect configurations and the local operators
that move them. Their configuration updates and supports let us test
whether a sequence restores the full configuration and whether its
phase is invariant under allowed local redefinitions of the hopping
operators. We then use locality identities to identify equivalent
processes and compute the statistical group by Smith normal form,
extending Refs.~\cite{Kobayashi2026Generalized, Wen2026Holographic}.
Appendix~\ref{app:conv:sec} summarizes our mathematical conventions and the description of higher groups and their classifying spaces.

\subsection{Configurations and local moves}
\label{pattern}
\label{newpattern}
\label{realization}
\label{dc:microscopic}
\label{loc:coupled-configurations}

A configuration $\cfg$ records the positions, labels, and junctions
of all defects, together with any global sector labels needed to
specify the state. In the SET example of
Sec.~\ref{sec:inversion-comparison}, it specifies the closed walls
and framed charges. In the Kitaev-chain example of
Sec.~\ref{sec:fermionic-toric}, it specifies $(n_2,n_3)$ subject to
Eq.~\eqref{eq:ftc-parent-constraint}. The framework applies to
general spatial lattices once the configurations, local moves, and
their supports are specified. For simplicity, we introduce the
notation on the boundary of the auxiliary simplex
$T=[0,\ldots,d+1]$. A degree-$r$ gauge field on an auxiliary
$r$-face labels a defect of spatial dimension $d-r$ on the
complementary physical face.

A local hopping unitary sweeps a defect across a physical cell.
Acting on the vacuum, it creates the defect on that cell's boundary.
A string creates a pair of point defects at its endpoints, while a membrane
creates a loop along its boundary~\cite{Levin2003Fermions,Levin2006Quantum,Feng2026Pauli}. In the presence of defects,
these operators can move or deform them. Higher-group couplings
can also change lower-dimensional defects within the support.
For example, moving a charge-conjugation wall reverses an enclosed charge,
and a Kitaev membrane can change point occupations according to
the incoming chain configuration.

We specify an elementary move $s$ by its invertible configuration
update $\tau_s$ and prescribed support $\supp(s)$. On this simplex
lattice, label a move of a codimension-$r$ defect by $s=(r,f,u)$ and
write $U(s)=U_{r,f}(u)$, where $G_r$ is its fusion group and $u\in G_r$
is the move parameter on an auxiliary $(r-1)$-face $f$ of $T$.
The prescribed support $\supp(s)=f^\vee$ is the complementary
physical simplex spanned by the vertices of $T$ outside $f$,
including its boundary. It has dimension $d-r+1$, one higher than
that of the defect it moves. In two dimensions, for example, a wall
move has a vertex parameter and sweeps the opposite triangle.
These physical simplices form the mesoscopic lattice; a microscopic
hopping circuit acts within a narrow neighborhood of its prescribed
simplex. The coupled updates
are higher-group gauge transformations, with the parameter
supported on the auxiliary face $f$.
Appendix~\ref{app:transport} specifies the updates for this
simplex construction.

Following Ref.~\cite{Kobayashi2026Generalized}, we call the
configurations reachable from the vacuum $\cfg_0$, together with
the elementary moves, their updates, and their supports, an
\emph{excitation model} $\mathfrak m$. Its statistical group also
uses the local phase freedoms and locality identities specified
below.

\begin{definition}[Microscopic realization]
A realization assigns an orthogonal normalized state $\ket\cfg$ to
each configuration and local unitary hopping operators satisfying
\begin{equation}
 U(s)\ket\cfg=e^{i\theta_s(\cfg)}\ket{\tau_s\cfg}.
 \label{loc:hop-action}
\end{equation}
The operators obey the specified locality identities on the
subspace spanned by these states.
\end{definition}
The excitation model fixes the updates $\tau_s$. A realization supplies
the hopping phases $\theta_s(\cfg)$. Each complete configuration
specifies one state ray, which can be a superposition of microscopic
basis states. Transport on a multidimensional fusion space would
instead require matrix-valued amplitudes.

The strict inverse of the same operation acts as
\begin{equation}
 U(s)^{-1}\ket{\cfg'}=e^{-i\theta_s(\cfg)}\ket\cfg,
 \qquad \cfg'=\tau_s\cfg.
 \label{loc:inverse-action}
\end{equation}
An inverse occurrence contributes minus the forward angle evaluated
at its outgoing configuration. Its definition fixes both the reverse
update and the inverse phase.

\subsection{Invariance under local operator modifications}
\label{loc:phase-invariance}

A local redefinition can assign different phase shifts to the
same hop in different defect backgrounds. We therefore track
the configuration at which each hop occurs. A word
$W=s_N^{\epsilon_N}\cdots s_1^{\epsilon_1}$ specifies
$U(W)=U(s_N)^{\epsilon_N}\cdots U(s_1)^{\epsilon_1}$ with
$\epsilon_j=\pm1$. The rightmost operator acts first. The
arrow-count group $E=\bigoplus_{s,\cfg}\Z(s,\cfg)$ is free on
the forward arrows $(s,\cfg)$. Starting from $\cfg_{\rm in}$,
the word has count expression $\langle W\rangle_{\cfg_{\rm in}}$.
In any expression $e=\sum_{s,\cfg}n_{s,\cfg}(s,\cfg)$,
$n_{s,\cfg}$ is the signed count of its forward arrow.
A forward occurrence from $\cfg$ adds one to $n_{s,\cfg}$.
An inverse occurrence from $\cfg$ subtracts one from
$n_{s,\tau_s^{-1}\cfg}$, since $\tau_s^{-1}\cfg$ is the
source of the corresponding forward arrow. The phase angle is
\begin{equation}
 \Theta(e)=\sum_{s,\cfg}n_{s,\cfg}\theta_s(\cfg)\pmod{2\pi}.
 \label{loc:phase-counts}
\end{equation}
For a word with fixed starting configuration, $\Theta(W)$ is the
angle of its count expression. We take the operators to act in
succession without free evolution.

A statistical process must return the system to its initial state up
to a phase. For ordinary group-valued excitations in the construction
of Ref.~\cite{Kobayashi2026Generalized}, an invariant sequence gives
the same phase in every initial configuration. For a higher group, a
move can induce lower-dimensional defects according to the defects
already present. The same word can thus follow different paths
through configuration space, and its closure and phase can depend
on the initial configuration.
We use the vacuum as the common starting state,
so closed processes can be concatenated there. The closure condition is
\begin{equation}
 U(W)\ket0=e^{i\Theta(W)}\ket0.
 \label{loc:history-phase}
\end{equation}
A process that closes on a prepared state can be based at the
vacuum by preparing that state, applying the process, and undoing
the preparation. In Sec.~\ref{sec:fermionic-toric}, for example,
two membrane moves prepare the closed Kitaev chain. After
$\mathcal K_{24}$ restores its configuration, their strict
inverses return it to the vacuum. The preparation and its
inverse cancel in the signed counts.
Closure balances incoming and outgoing arrows at every configuration:
\begin{equation}
 \sum_s\bigl[n_{s,\tau_s^{-1}\cfg}-n_{s,\cfg}\bigr]=0
 \quad\text{for every }\cfg.
 \label{loc:flow-closure}
\end{equation}
Under a change of state phase $\ket\cfg\mapsto e^{i\alpha(\cfg)}\ket\cfg$,
the forward angle changes by $\alpha(\cfg)-\alpha(\tau_s\cfg)$.
These contributions cancel by flow closure. Concatenating
vacuum-based closed processes adds their signed counts, and reversing
a process negates its counts. These operations give an Abelian
subgroup of $E$ even when the microscopic moves do not commute.

Allowed local modifications of $U(s)$ preserve the update $\tau_s$.
For example, a charge hop can be multiplied by local operators at its
endpoints that preserve the prescribed transport. On the simplex
lattice, define $d_v\cfg$ by restricting the fields and their
coefficient frames to the auxiliary facet opposite $v$. In physical
terms, this keeps the labels on the cells incident on vertex $v$.
The allowed phase shifts are sums of functions $\varphi_{s,v}(d_v\cfg)$
over vertices $v$ in the physical support of $s$. For an inverse
occurrence, the relevant forward input is its outgoing configuration,
as in Eq.~\eqref{loc:inverse-action}. Cancellation for every
allowed function requires
\begin{equation}
 \sum_{\cfg:\,d_v\cfg=\xi}n_{s,\cfg}=0
 \label{loc:cancellation-main}
\end{equation}
for each move $s$, support vertex $v$, and restricted input $\xi$.
Forward and inverse occurrences therefore balance separately for
each local input. Their pairing can differ between support vertices.
In the separated SET geometry, we apply the same condition using the
physical neighborhoods of the wall vertices, charge sites, and
path crossings.

The invariant subgroup $E_{\rm inv}\subset E$ consists of the
expressions satisfying Eqs.~\eqref{loc:flow-closure} and
\eqref{loc:cancellation-main}. Appendix~\ref{app:phase-conditions}
derives the two conditions from local cancellation, and Lemma~\ref{cell:invariant:word}
expresses them in terms of restrictions to faces.

\subsection{Locality identities and quantization}
\label{loc:identities}

Locality identities relate statistical processes without changing
their phases. Operators with disjoint supports commute. More
generally, let $U_j$ be a hopping operator $U(s_j)$ or its strict
inverse, with prescribed mesoscopic support $\supp(s_j)$.
These operators obey the locality identity
\begin{equation}
 [U_k,[\cdots,[U_2,U_1]]]=1
 \quad\text{if}\quad
 \bigcap_{j=1}^k\supp(s_j)=\varnothing,
 \label{loc:locality-identity}
\end{equation}
where $[X,Y]=X^{-1}Y^{-1}XY$.
Finite-depth circuits of bounded-range microscopic gates have
strictly bounded lightcones: conjugation spreads local operators
only a bounded distance~\cite{Bravyi2026Identity}.
For each finite sequence of such hopping operators, we choose the
mesoscopic separations larger than its total propagation range.
In a commutator, gates outside the lightcone of the
overlap cancel with their inverses. Applying this cancellation
successively leaves the two factors in the final commutator with
disjoint microscopic supports when the mesoscopic simplices have
no common intersection. The factors commute exactly, giving
Eq.~\eqref{loc:locality-identity}.

We record the signed occurrences of the elementary moves in each
locality identity, at every allowed starting configuration. Their
integer linear combinations form the subgroup $E_{\rm id}$.
At each physical vertex, at least one operator is absent, and the
remaining local sequence cancels.
Appendix~\ref{app:hopping-and-locality} constructs the local moves,
derives the invariance conditions, and characterizes the operator
words generated by locality identities (Lemma~\ref{cell:lem:word}).
Appendix~\ref{app:complete:sec} proves $E_{\rm id}\subset E_{\rm inv}$
(Proposition~\ref{cell:prop:firstthree}) and establishes the
correspondence between the statistical group defined below and homology.
We therefore define the statistical group by identifying invariant
processes that differ by locality identities.
\begin{definition}[Statistical group]
The statistical group of the excitation model $\mathfrak m$ is
\begin{equation}
 T(\mathfrak m)=E_{\rm inv}/E_{\rm id}.
 \label{statistics}
\end{equation}
\end{definition}

A realization assigns a phase character $[e]\mapsto e^{i\Theta(e)}$
to $T(\mathfrak m)$. The two SET realizations give $+1$ and $-1$ for the
charge-wall process $\mathcal W_{10}$.
The configuration updates determine the signed counts, and the
microscopic hopping phases determine their evaluation.

If $N$ times a count expression $e$ is a sum of locality identities,
then
\begin{equation}
 Ne\in E_{\rm id}\quad\Longrightarrow\quad
 e^{iN\Theta(e)}=1.
 \label{loc:quantization}
\end{equation}
The condition $Ne\in E_{\rm id}$ means that the order of $[e]$
divides $N$. This bound is saturated if a realization gives a
primitive $N$th-root phase.
The charge-wall process $\mathcal W_{10}$ in
Sec.~\ref{sec:inversion-comparison} detects the mixed $\Z_2$
anomaly of Eq.~\eqref{ex:inversion-anomaly} and gives $-1$ in the
quaternion realization. A computer calculation verifies that two
copies of $\mathcal W_{10}$ are a sum of locality identities, so its
statistical class has order~$2$. For the Kitaev-chain flip $\mathcal K_{24}$
in Sec.~\ref{sec:fermionic-toric}, a computer calculation verifies
that four copies are a sum of locality identities.
Its phase $+i$ therefore shows that its statistical class has
order~$4$.

The Smith normal form algorithm starts directly from locality
identities~\cite{Kobayashi2026Generalized}. Record the signed counts
of a generating set of locality identities, including their allowed
starting configurations. Each count refers to a forward arrow
$(s,\cfg)$, with negative counts representing inverse hops. These
vectors form the rows of an integer relation matrix,
with one column for each arrow. We seek integer combinations of
the rows whose counts have a common factor $N>1$. Dividing all
counts by $N$ can give a nontrivial statistical process, even though
the undivided combination is a sum of locality identities.

Smith normal form systematically identifies independent such
processes. Integer row operations recombine locality identities,
while invertible integer column operations change the basis of
arrow counts. Each diagonal entry $N>1$ gives an integer count
expression $e$ such that $Ne$ is a sum of locality identities and
no smaller positive multiple of $e$ is. The divided expression
$e$ automatically satisfies closure and local cancellation:
Eqs.~\eqref{loc:flow-closure} and \eqref{loc:cancellation-main}
are homogeneous integer equations, so their vanishing for $Ne$
implies their vanishing for $e$. Thus, $[e]$ has order $N$ in
$T(\mathfrak m)$, and its phase is an $N$th root of unity.
The entries greater than one give all nontrivial finite cyclic
factors of the statistical group. An entry $1$ gives a process
that is already a sum of locality identities.

Keeping track of the basis changes gives the divided arrow counts
in the original basis. To obtain an operator sequence, arrange these
signed arrows into closed routes and connect each route to the
vacuum by a preparation path and its strict inverse. The added
paths cancel in the counts (Appendix~\ref{app:phase-conditions}).
The construction uses only the lattice geometry and the prescribed
configuration updates and supports of the local moves.

Forming this presentation requires enumerating the configurations
and locality relations. The configuration space can grow exponentially
with the number of independently labeled cells. This growth can make
direct enumeration computationally prohibitive in higher dimensions
or for more complicated higher groups. The next section constructs
a process for a chosen cycle without this enumeration.

\section{Statistical processes and homology}
\label{dc:sec}

The phase of a statistical process is invariant under local modifications of the hopping operators. For defects of a fixed dimension, Ref.~\cite{Kobayashi2026Generalized} conjectured that integral homology
classifies statistical processes modulo locality identities and verified
this correspondence in examples. In this section, we construct the correspondence
explicitly for coupled defects in our higher-group lattice model on the
boundary of a simplex. Appendix~\ref{app:complete:sec} gives the
detailed proof.
Therefore, we can associate each such process with an integral cycle in $B\mathbb G$, so its phase in a cocycle realization can be
computed by pairing the cycle with the anomaly.
Conversely, a
prescribed cycle can guide the construction of a local process
whose phase gives the corresponding response. We introduce the conversion between cycles and processes. This conversion is then used to establish a classification theorem for higher-group statistics.

\subsection{From processes to cycles}
\label{loc:anomaly-response}

\begin{figure*}[t]
 \centering
 \input{figures/cyl_j_correspondence.tex}
 \caption{The cylinder of an elementary move and the correspondence
 between statistical processes and detecting cycles. The cylinder
 map sums the oriented simplices of the cylinders. The closure
 condition cancels the ends, and local cancellation pairs the side faces.
 The lower arrows show the inverse maps on classes established in
 Sec.~\ref{loc:universality}.}
 \label{fig:cyl-j-correspondence}
\end{figure*}

We first associate a cycle with a statistical process, following
Ref.~\cite{Feng2026Anyonic}. To motivate this construction,
consider evaluating the process using hopping operators on the
boundary of an SPT phase specified by a cocycle. The process
thus assigns a $U(1)$ phase to each SPT cocycle. We construct
a cycle such that evaluating any such cocycle on it reproduces
the corresponding statistical phase.

The spatial lattice is selected as $\partial T$, where $T$ is an auxiliary
simplex of dimension $m=d+1$. For each elementary move $(s,\cfg)$ of the process, construct a cylinder $T\times I$. Put the complete configurations $\cfg$ and $\tau_s\cfg$ on the bottom and top surfaces of the cylinder. Then conduct the field interpolation in
Appendix~\ref{app:transport} to extend these labels into the interior of the cylinder with the higher-group constraints
satisfied throughout. A process, a series of moves, can then be related to a long cylinder obtained by gluing the cylinders of individual moves end to end. 

Each elementary cylinder is given a standard triangulation.
The sum of its oriented labeled simplices defines an integral
chain $\operatorname{Cyl}(s,\cfg)\in C_{d+2}(B\Gg;\Z)$,
as illustrated in Fig.~\ref{fig:cyl-j-correspondence}.
For signed hopping counts
$e=\sum_{s,\cfg}n_{s,\cfg}(s,\cfg)$, we define
$\operatorname{Cyl}(e)$ by summing the chains
$\operatorname{Cyl}(s,\cfg)$ with multiplicities $n_{s,\cfg}$.
An inverse hop contributes the negative of the cylinder
evaluated at the corresponding forward source.
For an ordered process, this agrees with the chain obtained
by gluing its elementary cylinders end to end.
For $e\in E_{\mathrm{inv}}$, the closure condition cancels
the end faces, while local phase cancellation pairs the side
faces with identical restricted labels and opposite signs,
as proved in Proposition~\ref{cell:prop:firstthree}.
Thus $\partial\operatorname{Cyl}(e)=0$, and
$\operatorname{Cyl}(e)\in Z_{d+2}(B\Gg;\Z)$ is an integral cycle.

In a cocycle realization, the hopping angle of each move is
$2\pi$ times the pairing of its cylinder with the anomaly cocycle
$\Omega$~\cite{Feng2026Anyonic}. The phase of the process is therefore
\begin{equation}
 e^{2\pi i\langle e,\Omega\rangle},\qquad
 \langle e,\Omega\rangle
 :=\langle\operatorname{Cyl}(e),\Omega\rangle.
 \label{probe}
\end{equation}
A locality identity is proven to map to an integral boundary, so the cylinder
map descends to equivalence classes,
\begin{equation}
 \overline{\operatorname{Cyl}}:
 T(\mathfrak m)\longrightarrow H_{d+2}(B\mathbb G;\Z).
 \label{loc:cylinder-map}
\end{equation}

\subsection{From cycles to processes}
\label{sec:J-construction}
\label{dc:construction}
\label{dc:evaluation}
\label{dc:cocycles}

Now, we construct a mapping from cycles to processes, dubbed the $J$-map. Start with a $(d+2)$-dimensional cycle. Write $D=d+2$. The cycle can be represented by a formal sum of cells in $B\Gg$, each corresponding to a field distribution on the simplex $\Delta^{d+2}$ subject to the higher-group constraints.
\begin{equation}
 z=\sum_a n_a\sigma_a,\qquad \partial z=0,\qquad n_a\in\Z.
 \label{dc:cycle}
\end{equation}
The target phase in a cocycle realization is the exponential of
the pairing $\langle z,\Omega\rangle=\sum_a n_a\Omega(\sigma_a)$
in $\R/\Z$. We build a process by assigning local operations to the
cells and combining them with the multiplicities $n_a$.

Take a cell $\sigma=[0,\ldots, D]$ labeled by higher-group gauge fields.
Deleting the last vertex gives a source configuration $\cfg^-=d_D\sigma $ on $[0,\ldots, D-1]$.
Deleting the penultimate vertex gives the target configuration $\cfg^+=d_{D-1}\sigma$ on $[0,\ldots,D-2,D]$.
Both faces can be regarded as configurations on the spatial lattice $\partial T$ through an identification. Specifically, in the target, vertex $D$ supplies the labels at the last slot of $T$.
Both configurations retain all their wall and higher-field labels. The $(d+2)$-cell thus specifies an interpolation from $\cfg^-$ to $\cfg^+$. By distorting the simplex into a cylinder with $\cfg^\pm$ on two ends, we obtain a prototype for the process corresponding to $\sigma$.
To obtain a process, we decompose this cylinder into a cylinder of basic moves, which we call the sweep map $S(\sigma)$. Next, we specify how to prepare both configurations $\cfg^\pm$ at the ends using a given process $P(\cfg^\pm)$. Then we introduce the sweep $S(\sigma)$.
Finally, we use techniques such as an alternating sum to obtain the process $\mathcal{J}(\sigma)$, which becomes locally invariant after being assembled into $\mathcal{J}(z)$.

\begin{figure*}[t]
 \centering
 \input{figures/anomaly_to_process.tex}
 \caption{Constructing a statistical process from a detecting cycle.
 A preparation, sweep, and inverse preparation give $L(\sigma)$.
 Reference comparisons cancel local changes of the sweep phases.
 Combining the resulting words $\mathcal J(\sigma)$ with the signed multiplicities
 of the cycle cancels the preparation phases and gives $W(z)$.
 The labeled cells are schematic. Face matching retains every
 restricted field.}
 \label{fig:anomaly-to-process}
\end{figure*}

\paragraph*{Preparing the endpoints.}
A preparation $P(\cfg)$ takes the vacuum configuration $\cfg_0$ to a specified
configuration $\cfg$. The preparation adjusts the higher-group gauge fields from lower order to higher order. At each order, we compare the difference between the current configuration and the target configuration, and then eliminate that difference with basic moves. Appendix~\ref{cell:sec:time}
gives the general recipe.

For a 2-group example, take binary wall and point fields $a,B$
with $\delta B=a\cup a\cup a$ on $\sigma=[0123\,3']$.
The auxiliary vertex $3'$ supplies the final slot of the target
configuration. Set $a_{01}=a_{03'}=1$, $a_{02}=a_{03}=0$, and
$B_{023'}=B_{03\,3'}=1$, with the other $B_{0ij}$ zero. Flatness gives
$a_{ij}=a_{0i}+a_{0j}$, and Eq.~\eqref{app:conv:reconstruction}
determines the remaining point labels. For example, $B_{123}=0$
and $B_{123'}=1+a_{01}a_{12}a_{23'}=0$ modulo two. On $\partial T$ with $T=[0123]$, the endpoint configurations are
\begin{equation}
 \begin{array}{c|c|c}
 & (a_{01},a_{02},a_{03}) & (B_{012},B_{013},B_{023},B_{123}) \\
 \hline
 \cfg^- &(1,0,0)&(0,0,0,0)\\
 \cfg^+  &(1,0,1)&(0,0,1,0)
 \end{array}\notag
 \label{dc:worked-two-group-endpoints}
\end{equation}
We label the physical vertex dual to the triangle omitting $i$
by $v_i$. The nonzero $B_{023}$ in $\cfg^+$ is thus a point at
$v_1$. Its value comes from $B_{023'}$ of $\sigma$, since vertex
$3'$ replaces slot $3$ in $\cfg^+$.

Let $V_i$ be the wall move with unit parameter at $i$, and let
$E_{ij}$ be the point hop with unit parameter on $ij$. In
Eq.~\eqref{app:conv:gauge}, $V_i$ toggles the wall labels on
edges incident to $i$ and includes the induced point change.
The hop $E_{ij}$ toggles the two triangle labels that contain
$ij$. Read the wall-move parameters from the nonzero $a_{0i}$
in the table, placing them on the auxiliary vertices
$d_0[0i]=[i]$. After setting the wall field, place point-hop
parameters on the auxiliary edges $d_0[0ij]=[ij]$ for which
$B_{0ij}$ still differs from its target value. Acting on the vacuum,
$V_1$ sets the wall field of $\cfg^-$ with all point labels zero. To reach $\cfg^+$,
apply $V_3$ and then $V_1$. At the latter step the term
$a'_{12}a'_{23}=1$ turns on $B_{123}$. The hop $E_{23}$ then
moves this point from $v_0$ to $v_1$.
Thus the preparations are
\begin{equation}
 P(\cfg^-)=V_1,\qquad P(\cfg^+)=E_{23}V_1V_3.
 \label{dc:worked-two-group-preparation}
\end{equation}

\paragraph*{Sweeping between the endpoints.}

The cell $\sigma$ specifies the interpolation between the two
prepared configurations. Its sweep $S(\sigma)$ carries out that
interpolation as elementary moves from $\cfg^-$ to $\cfg^+$. The sweep uses labels involving both endpoint vertices. The key idea is the same as in the preparation $P(\cfg)$, except we now start from $\cfg^-$ instead of the vacuum. See Appendix~\ref{cell:sec:sweep} for details.

For the example above with $\sigma=[0123\,3']$, start at $\cfg^-$. To track the sweep on
the auxiliary cell, first look at the mixed wall label $a_{3\,3'}=1$ of $\sigma$. It prescribes
$V_{3'}$ on the cell, which acts as $V_3$ on the physical lattice.
The incoming value of $a_{3\,3'}$ is still zero, so the term
$a_{i3}a_{3\,3'}$ leaves every mixed point label unchanged.
On the physical lattice, $V_3$ sets the target wall field and
toggles $B_{013}$ through the term $a_{01}a_{13}=1$. This
creates a point at $v_2$, dual to $[013]$.

Next, look at the configuration of $B_{i33'}$, which determines the sweep move for the field $B$.
In the example, the mixed point label $B_{03\,3'}=1$ of $\sigma$ still differs from its current value and prescribes a
point hop on the auxiliary edge $[03']$, which sets this label
and leaves the other mixed point labels zero. Since $3'$
supplies physical slot $3$, the next physical move is $E_{03}$. It
moves the point from $v_2$ to $v_1$, dual to $[023]$,
and leaves the wall field unchanged, giving
$\cfg^+$ with the required odd point parity. The sweep is
\begin{equation}
 S(\sigma)=E_{03}V_3.
 \label{dc:worked-two-group-sweep}
\end{equation}

\paragraph*{Closing the comparison.}
To obtain a process based at the vacuum, first prepare $\cfg^-$, then
sweep, and finally undo the direct preparation of $\cfg^+$:
\begin{equation}
 L(\sigma)=P(\cfg^+)^{-1}S(\sigma)P(\cfg^-).
 \label{cell:eq:44}
\end{equation}
The sequence returns to the vacuum $\cfg_0$. This process can be alternatively viewed as a phase comparison between two routes to $\cfg^+$, namely $S(\sigma)P(\cfg^-)$ and $P(\cfg^+)$. For the binary cell, the word is
$(E_{23}V_1V_3)^{-1}(E_{03}V_3)V_1$. 

\paragraph*{Alternating sum for local cancellation.}
Directly substituting $\sigma$ in a cycle $z$ with the word $L(\sigma)$, however, does not give a locally invariant process. To satisfy local cancellations, we combine comparisons
$L(\tau_\epsilon)$ with alternating signs. For each bit string
$\epsilon=(\epsilon_0,\ldots,\epsilon_{D-2})\in\{0,1\}^{D-1}$,
define $\tau_\epsilon$ by reading the labels of $\sigma$ at
vertex $j+\epsilon_j$ in each slot $j\leq D-2$ and keeping the
last two vertices fixed. Thus $\tau_0=\sigma$, while a nonzero
$\epsilon$ repeats a vertex and gives a degenerate cell supported
on a proper face of $\sigma$. Define the sign as
$c_\epsilon=(-1)^{|\epsilon|}$, with
$|\epsilon|=\sum_j\epsilon_j$ counting replacements
[Eq.~\eqref{cell:eq:45}]. Two choices that differ only in slot
$v$ have the same face obtained by deleting $v$. Their sweeps
therefore have the same moves and forward-source labels visible
at physical vertex $v$. Since their signs are opposite, the local
phase changes there cancel. Therefore, by applying the alternating sum to $L(\sigma)$, we obtain a process that becomes locally invariant after being substituted into a cycle, dubbed $\mathcal{J}(\sigma)$.

For the binary cell, the replaceable slots are $0,1,2$. The
choices $(0,0,0)$ and $(1,0,0)$ give $[0123\,3']$ and $[1123\,3']$
with signs $+1$ and $-1$, for example. For $[1123\,3']$, the mixed triangles
read $B_{13\,3'}$, $B_{13\,3'}$, and $B_{23\,3'}$ of the original cell.
Both labels vanish by the point constraint, and the last two slots
retain $a_{3\,3'}=1$, so its reference sweep consists of $V_3$ alone.
Deleting slot $0$ gives the same restricted source and target
for both sweeps. In the first sweep, $E_{03}$ follows $V_3$ and
acts on the complementary edge $12$, away from the physical vertex
$v_0$, so it leaves the local configuration there unchanged. The $V_3$
phase shift at $v_0$ therefore enters the two comparisons with opposite
signs. The degenerate cell supplies a necessary comparison even
though it vanishes in normalized chains.

\paragraph*{Assembling the cycle.}
Let $\mathcal J(\sigma)$ be the product of
$L(\tau_\epsilon)^{c_\epsilon}$. A negative sign means taking
the strict inverse of the entire comparison, which reverses
each signed count at its forward source. Because every comparison
starts and ends at $\cfg_0$, the product has the same signed
counts and the same phase on $\ket0$ in any order. We choose
lexicographic order to specify the word. The count map is
\begin{align}
 J(\sigma)&=\langle\mathcal J(\sigma)\rangle_{\cfg_0}
   =\sum_\epsilon c_\epsilon\langle L(\tau_\epsilon)\rangle_{\cfg_0},
 \nonumber\\
 J(z)&=\sum_a n_a J(\sigma_a).
 \label{dc:count-map}
\end{align}
Here $\sigma$ is a chain generator, and degenerate chain terms
are zero. A degenerate reference cell $\tau_\epsilon$ still
contributes its comparison word to $\mathcal J(\sigma)$.
For cells ordered as
$\sigma_1,\ldots,\sigma_s$, the product
\begin{equation}
 W(z)=\mathcal J(\sigma_s)^{n_s}\cdots
       \mathcal J(\sigma_1)^{n_1}
 \label{dc:cycle-word}
\end{equation}
executes them and has signed counts $J(z)=\langle W(z)\rangle_{\cfg_0}$.
An integer multiplicity prescribes repetitions of the block or its
inverse. Here again the blocks close at $\cfg_0$, so their order
does not affect the counts or the phase on $\ket0$. Each cell
supplies $2^{d+1}$ reference comparisons, comparable to the
total number of faces of all dimensions on the simplex boundary, while the computational complexity of the traditional Smith normal form approach is exponential in this number. Each preparation and sweep uses at most a fixed number
of local moves per face.

The closed word $W(z)$ is invariant under local phase changes. In a cocycle realization, the phase is
\begin{equation}
 W(z)\ket0=\exp \bigl(2\pi i\langle z,\Omega\rangle\bigr)\ket0,
 \label{dc:response}
\end{equation}
by Eq.~\eqref{dc:detector-identity} below.

\subsection{Classification and anomaly detection}
\label{loc:universality}
\label{loc:completeness}

It can be shown that the cylinder map sends invariant counts to cycles, and if we change a process by a locality identity, its cylinder changes
by a boundary. Conversely, $J$ sends cycles to invariant counts,
and changing a cycle by a boundary changes those counts by
locality identities. Eq.~\eqref{cell:eq:47} makes these two
statements precise, and a proof is given there. Thus $\mathrm{Cyl},J$ can be reduced to maps $\overline{\mathrm{Cyl}}, \bar{J}$ on $T(\mathfrak m)$ and $H_D(B\mathbb G;\Z)$, which identify different representatives of the same process or cycle.

Furthermore, these reduced maps $\overline{\rm Cyl}$ and $\bar{J}$ are mutually inverse.
Comparing the sweep with the labeled interpolation shows that
the cylinder of the constructed process and the input cycle
differ by an integral boundary. In the other direction, start
with invariant counts $e$, form their cylinder, and reconstruct
the counts using $J$. The reconstructed and original counts
differ by a sum of locality identities. These statements give
\begin{align}
 \operatorname{Cyl}(J(z))-z&=\partial H(z) &\implies \overline{\rm Cyl}\circ \bar{J}=1,
 \label{dc:detector-identity}\\
 J(\operatorname{Cyl}(e))-e&\in E_{\mathrm{id}} &\implies \bar{J}\circ \overline{\rm Cyl}=1,
 \label{JCyl}
\end{align}
for $z$ an integral cycle and $e\in E_{\mathrm{inv}}$.
The cocycle pairing vanishes on the boundary in the first
identity, giving
Eq.~\eqref{dc:response}. The second identity recovers the
original statistical class and hence its phase in every
allowed microscopic realization. Appendix~\ref{app:complete:sec} proves these statements. These statements further prove the following classification theorem:

\begin{theorem}[Cycle--process correspondence]
\label{dc:correspondence-theorem}
Let $\mathbb G$ be a finite $n$-group with $n\le d$. On the
boundary $\partial\Delta^{d+1}$, use the full set of local moves
specified in Appendix~\ref{app:transport} and their strict inverses, forming the excitation model $\mathfrak m(\partial\Delta^{d+1},\mathbb G)$.
For each layer $r$, this includes $U_{r,f}(u)$ for every auxiliary
$(r-1)$-face $f$ and every nonidentity $u\in G_r=\pi_r(B\mathbb G)$,
supported on the complementary physical face. Then
the cylinder map induces an isomorphism
\begin{equation}
 T\bigl( \mathfrak m(\partial\Delta^{d+1},\mathbb G)\bigr)
 \simeq H_{d+2}(B\mathbb G;\Z).
 \label{classification}
\end{equation}
The induced map $\bar J$ is its inverse.
\end{theorem}

The bound $n\le d$ puts each field label on a face contained
in at least one spatial facet. However, the one-dimensional example
of Sec.~\ref{spt:one-dimensional} has a locally invisible
sector and is handled as an exception. 

For the full
set of moves in the theorem, the statistical group is finite
and each class has a representative in a spatial ball
[Appendices~\ref{app:finiteorder} and \ref{app:disk:localization}].

A microscopic realization assigns a phase character to $T(\mathfrak m)$.
Eq.~\eqref{classification} identifies this character with a
character on the integral homology. With $D=d+2$, the universal
coefficient theorem gives
\begin{equation}
 H^D(B\mathbb G;\R/\Z)\cong
 \operatorname{Hom}(H_D(B\mathbb G;\Z),\R/\Z).
 \label{cell:eq:112}
\end{equation}
The $\operatorname{Ext}$ term vanishes because $\R/\Z$ is
injective~\cite{Hatcher2002Algebraic}. Under this identification, a cocycle
class acts on a cycle by the pairing $\langle z,\Omega\rangle$.
For a cocycle realization, this is exactly the accumulated angle
in units of $2\pi$ by Eq.~\eqref{probe}. In a microscopic
realization the same character can be read from the hopping
phases of the processes. The isomorphism therefore allows the
anomaly class to be read from a complete set of statistical
phases.

Choose cyclic generators $\omega_i$ of the anomaly group with
orders $N_i$. The dual homology generators $[z_i]$ satisfy
$\langle z_i,\omega_j\rangle=\delta_{ij}/N_i$ in $\R/\Z$. This can be viewed as using $z_i$ to detect a specific anomaly generator, while the choice of detector cycle is not unique.
The process $W(z_i)$ has counts $J(z_i)$ and gives the phase
\begin{equation}
 e^{2\pi i\langle J(z_i),\omega_j\rangle}
 =e^{2\pi i\delta_{ij}/N_i}
\end{equation}
for the generator $\omega_j$. For an anomaly
$[\Omega]=\sum_j k_j[\omega_j]$, the process $W(z_i)$
gives the phase $e^{2\pi i k_i/N_i}$. Each phase determines
one coefficient modulo its order, and the collection
determines the full class.

\input{sections/spt_examples}
\input{sections/spt_d1}
\input{sections/spt_d2}
\input{sections/spt_d3_typeI}
\input{sections/spt_d3_typeII}

\section{Discussion}
\label{sec:discussion}

We have developed a lattice framework for the statistics of coupled
defects with higher-group structure. A statistical process restores
the complete defect configuration and cancels the phase changes
induced by allowed local modifications of its hopping operators.
The locality construction generates such processes from the lattice
geometry and specified local moves. The cycle construction gives a
complementary route: once a detecting cycle is specified, it gives
an explicit operator sequence without enumerating the full
configuration space.
For the spherical simplex lattice and full set of moves in
Theorem~\ref{dc:correspondence-theorem}, these constructions identify
the statistical group with $H_{d+2}(B\mathbb G;\Z)$.

This extends the locality-based approach of
Ref.~\cite{Kobayashi2026Generalized} to defects whose conservation
laws are coupled. The essential additional requirement is to track
every defect species changed by a local move. Kapustin and Xu
construct symmetry anomalies from localized transformations and
commutator pairings on intersections of spatial
regions~\cite{Kapustin2026Anomalies}, giving a complementary
perspective on the locality identities used here. The holographic
approach of Ref.~\cite{Wen2026Holographic} classifies the
corresponding hopping-operator algebras by higher-group cohomology.
Our construction supplies explicit lattice processes that measure
these statistical phases, together with the cancellations that make
them independent of local operator choices. In particular,
the Majorana-chain flip realizes the fourth-root statistics predicted
for intertwined fermion--string conservation laws in that work.

The examples show what can be learned by following the coupled
configurations. For a closed Kitaev chain, the loop-flip phase $+i$
requires keeping track of the physical fermions that exchange parity
with the chain during its deformation. The two symmetry-enriched
$\Z_4$ toric codes share their intrinsic anyon statistics and
charge-conjugation action, but their charge--wall processes give
opposite signs. At the three-dimensional SPT boundaries with
nonsplit 2-group symmetry, the loop and wall--loop processes
separately measure the two anomaly coefficients. Thus mixed
statistical processes can distinguish
realizations that the statistics of individual defect species leave
unresolved.

Locality identities provide a criterion for deciding which
apparently different processes carry independent statistics. An
example is the cluster-state twist string in the $(3+1)$D
$\Z_2\times\Z_2$ toric
code~\cite{Barkeshli2023Codimension}. Two magnetic flux loops linked
with this defect can have nontrivial 3-loop braiding. Preparing
the linking between the twist string and an $m_1$ loop creates an
$e_2$ charge, which can be detected by braiding the $m_2$ loop.
The commutator of the twist and $m_1$ membrane operators is precisely the Wilson
string that transports this induced charge. Its commutator with the
$m_2$ membrane gives the ordinary particle--loop braiding sign.
Thus, for this cluster-string realization, the 3-loop phase is fixed
by the higher-group charge-production rule and the usual mutual
statistics; it introduces no additional independent phase parameter.

Several questions concern the dependence on geometry and available
operations. The classification theorem uses the full set of moves on
the simplex boundary. Although every class has a representative in a
spatial ball, extending the correspondence to refined lattices and
other spatial topologies requires further work. Restricting the moves
or allowing defects with open ends can also change the statistical
group. A related question is which coupled classes remain distinct
after quotienting by the processes available in the wall-free sector.

The relation to phase stability deserves a separate analysis.
Simultaneously conjugating the Hamiltonian, configuration states,
and hopping operators by a finite-depth local unitary $V$ gives
$H'=VHV^\dagger$, $\ket{\cfg}'=V\ket{\cfg}$, and
$U'(s)=VU(s)V^\dagger$, with
\begin{equation}
 U'(W)\ket0'=V U(W)\ket0
             =e^{i\Theta(W)}\ket0'.
 \label{discussion:transported-phase}
\end{equation}
Transporting the defect labels, support assignments, and allowed
phase freedoms along with these operators preserves their locality
relations; microscopic supports broaden by a bounded amount.
A symmetry-preserving $V$ therefore preserves the response in this
transported realization. Establishing stability under a general
gapped Hamiltonian deformation with independently chosen hopping
operators remains an open problem.

There are also practical questions about constructing short
processes. A given detecting cycle avoids the enumeration of all
configurations, but its preparations, sweeps, and reference
comparisons can still yield a long sequence. Finding simple
detecting cycles and reducing their processes by locality identities
are complementary optimization problems. Optimizing the signed counts
alone need not minimize the number of hops: connecting separate
closed components can require a path and its inverse whose net
counts vanish.

Finally, the construction relates lattice statistics to topological
responses on closed manifolds. A background $f:M\to B\mathbb G$ on
a closed oriented $D$-manifold, with $D=d+2$, together with a
triangulation of $M$, supplies a cycle $z_M$ representing $f_*[M]$.
The resulting process measures the
response~\cite{Dijkgraaf1990Topological}
\begin{equation}
 Z_\Omega(M,f)=\exp\!\left(2\pi i\int_M f^*\Omega\right).
 \label{dc:discussion-manifold-response}
\end{equation}
This response depends on $(M,f)$ only through the homology class
$f_*[M]$. An open direction is to ask whether finer information
about the manifold and its background can define additional lattice
statistical observables.

A separate direction is to extend the construction to noninvertible
defects, whose fusion spaces can carry internal states and whose
local moves act by matrices~\cite{Kitaev2006Anyons,Walker2021Universal}.
The challenge is to extend the cycle construction to these fusion
states and identify the locality relations obeyed by the resulting
matrix-valued processes.

\begingroup
\makeatletter
\let\addcontentsline\@gobblethree
\makeatother
\section*{Acknowledgments}

We thank Yitao Feng, Zheng-Cheng Gu, Po-Shen Hsin, Anton Kapustin, Ryohei Kobayashi, Ran Luo, Xingyu Ren, Shu-Heng Shao, Wilbur Shirley, Meng Sun, Nathanan Tantivasadakarn, Qing-Rui Wang, Yi-Nan Wang, Rui Wen, Xiao-Gang Wen, and Yi Zhang for helpful discussions.

Y.-A. C. is supported by the National Natural Science Foundation of China
(Grant No.~12474491) and the Fundamental Research Funds for the Central Universities, Peking University.
\par
\endgroup

\appendix

\section{Mathematical preliminaries and conventions}
\label{app:conv:sec}
\label{app:op:section}

We fix the chain, cochain, and orientation conventions, then
introduce the simplicial description of higher-group classifying
spaces.

\subsection{Conventions}
Operator products act from right to left. Lists and tables give
chronological order; reversing a process reverses the list and
replaces each operator by its strict inverse.

\paragraph*{Chains and orientations.}
In the simplicial model of $B\mathbb G$ described below, each
nondegenerate labeled $\ell$-simplex gives an $\ell$-cell.
On a labeled simplex, define the face map $d_i$ by deleting the
vertex in slot $i$ and restricting the fields to the remaining face.
Define the degeneracy map $s_i$ by repeating that vertex and pulling
back the fields along the order-preserving map that identifies the two
consecutive copies.
A collapsed edge carries the identity of $G_1$. A higher cochain
vanishes on a face whose vertices have repeated images.

Integral $\ell$-chains are sums of the corresponding generators
$[\cfg]$ with integer coefficients.
The boundary is
    \begin{equation}
    \partial[\cfg]=\sum_{i=0}^{\ell}(-1)^i[d_i\cfg].
    \label{hg:eq:boundary}
    \end{equation}
    All simplices are ordered. Their orientation convention is
    \begin{equation}
     \partial[0\cdots n]=\sum_{i=0}^n(-1)^i[0\cdots\widehat i\cdots n].
     \label{app:conv:boundary-sign}
    \end{equation}
Physical defect chains use the coefficient groups specified by
their labels. For integral chains $x$ and $y$ of degrees
$p$ and $q$, the shuffle cross product satisfies
    \begin{equation}
     \partial(x\star y)=(\partial x)\star y
                        +(-1)^p x\star\partial y.
     \label{app:op:boundary}
    \end{equation}
When $x$ and $y$ are cycles, the class of $x\star y$ depends only
on their integral homology classes. Products of ordered simplices
in Appendix~\ref{app:complete:sec} use the same shuffle triangulation.
\paragraph*{Cochains and products.}
We write $\langle z,\Omega\rangle$ for the evaluation of a cochain
$\Omega$ on an integral chain $z$.

The coboundary is dual to the chain boundary:
    \begin{equation}
        (\delta\lambda)(\sigma)=\lambda(\partial\sigma).
    \end{equation}
For untwisted coefficients, the Alexander--Whitney cup convention is
    \begin{equation}
     (x\cup y)(0\cdots p+q)=x(0\cdots p)y(p\cdots p+q)
     \label{app:conv:cup}
    \end{equation}
for degrees $p$ and $q$. With local coefficients, the second
factor is first transported to the frame at vertex $0$, then paired
with the first factor by the equivariant coefficient pairing.
Products retain their displayed order; commuting cochains requires
the appropriate coboundary correction.

For integer cochains $x,y$ of degrees $p,q\ge1$, the first higher cup
product has degree $p+q-1$. We use the signed convention of
Ref.~\cite{Tsui2020Lattice}:
\begin{equation*}
 \begin{aligned}
 &(x\cup_1y)(0\cdots p+q-1)\\
 &\quad=\sum_{j=0}^{p-1}(-1)^{(p-j)(q+1)}\\
 &\qquad\quad\times x(0\cdots j,j+q\cdots p+q-1)y(j\cdots j+q).
 \end{aligned}
\end{equation*}
It corrects the failure of graded commutativity at the cochain
level~\cite{Steenrod1947Products,Chen2019Bosonization,Chen2023Higher}:
\begin{equation*}
 \begin{aligned}
 \delta(x\cup_1y)={}&\delta x\cup_1y+(-1)^p x\cup_1\delta y\\
 &+(-1)^{p+q-1}x\cup y
 +(-1)^{pq+p+q}y\cup x.
 \end{aligned}
\end{equation*}
For example, when $p=q=2$,
\begin{equation*}
 (x\cup_1y)_{0123}=x_{023}y_{012}-x_{013}y_{123}.
\end{equation*}
For $\Z_2$ cochains, all signs reduce to plus signs. The binary
higher cup products used below follow this reduction.

For $A\in Z^r(-;\Z_N)$,
choose the canonical lift $\widetilde A$ and define
    \begin{equation}
        \beta_N A=\frac{\delta\widetilde A}{N}\in Z^{r+1}(-;\Z).
        \label{app:conv:bockstein}
    \end{equation}
Changing the lift changes this cocycle by an integral coboundary.
The definition requires $A$ to be closed modulo $N$; a higher-group
field with nonzero Postnikov coboundary requires a corresponding
correction.

\paragraph*{Field notation.}

A field assignment on the faces of an ordered simplex specifies a
simplex of the model for $B\Gg$ when it satisfies the higher-group
constraints. Nondegenerate assignments label the cells of $B\Gg$.
Appendix~\ref{app:higher-group} gives the constraints and the construction.

Field symbols are specific to each model. In the charge-conjugation example,
$C$ denotes the wall pattern, $q$ the framed charges, and $(A,E)$
the microscopic edge labels. In the Kitaev-chain example, $(n_2,n_3)$
record the chain and physical fermions. The general
construction uses $a$ for the ordinary group field and $B_r$ for
higher fields. A complete configuration $\cfg$ includes every field
and its coefficient frame. A word $W$ with source $\cfg$ has signed
count expression $\langle W\rangle_{\cfg}\in E$. We use $e$ for
general count expressions. The preparation $P(\cfg)$, sweep
$S(\sigma)$, comparison $L(\sigma)$, and reference
$\mathcal J(\sigma)$ are operator words. The map $J$ extends the
counts of the reference words linearly.

\paragraph*{Dimensions and normalization.}
We use $d$ for the spatial dimension, $d+1$ for physical spacetime,
and $D=d+2$ for the anomaly degree. A degree-$r$ background field
is associated with an $(r-1)$-form symmetry layer. In the complementary simplex
realization, its defects have dimension $d-r$ and its elementary
operators sweep regions of dimension $d-r+1$.

Cocycles act on normalized integral chains and evaluate to zero on
degenerate simplices. We perform field pullbacks before discarding
degenerate cells. The values on faces with distinct projected vertices
can remain nontrivial even when the simplex is degenerate.

\subsection{Higher groups and their classifying spaces}
\label{app:higher-group}
\label{app:conv:independent}
\label{app:conv:two-group}
\label{app:conv:boundary}

Higher-group symmetry couples the defect labels: changing one field
can require a junction in another. We express these rules as cochain
constraints and use them to determine the allowed updates. We consider
background degrees $1\le r\le d$; the 2-group formulas require
$d\ge2$. These are the updates used in the correspondence proof in
Appendix~\ref{app:complete:sec}.

\paragraph*{Definition and physical implications.}

An ordinary group describes invertible transformations and their
composition. A higher group also includes invertible transformations
between these transformations, and further levels of the same
structure. Multiplication acts on every level, with associativity
and unit laws implemented by compatible transformations at the next
level. An $n$-group has elements at level zero and transformations
through level $n-1$. A $1$-group is an ordinary group; a $2$-group
has elements and invertible arrows, with an associator satisfying
the pentagon identity~\cite{Baez2004Higher,Benini2019TwoGroup}.

More precisely, an $n$-group $\mathbb{G}$ is a group-like monoidal
$(n-1)$-groupoid. A groupoid has invertible arrows, and a higher
groupoid has invertible arrows at every level. ``Monoidal'' specifies
multiplication and its coherence data; ``group-like'' means that
multiplication makes the equivalence classes of elements a group.
Equivalently, $\mathbb{G}$ is a homotopy $(n-1)$-type with a coherently
associative multiplication and inverses up to homotopy. A homotopy
$k$-type is a space whose homotopy groups vanish above degree $k$.

We consider finite $n$-groups. Their symmetry groups are
\begin{equation}
\begin{aligned}
 G_1&=\pi_0(\mathbb{G})\cong\pi_1(B\mathbb{G},e),\\
 G_r&=\pi_{r-1}(\mathbb{G},e)\cong\pi_{r}(B\mathbb{G},e),\qquad 2\le r\le n.
\end{aligned}
\label{hg:eq:groups}
\end{equation}
Here $\pi_0$ denotes connected components, $e$ is the identity, and
$\pi_j$ denotes the $j$th homotopy group. All groups in
Eq.~\eqref{hg:eq:groups} are finite; $G_1$ can be non-Abelian, while
the others are Abelian. Therefore we write $G_1$ multiplicatively and $G_r$ additively for $r>1$. In symmetry terminology, $G_r$ is the $(r-1)$-form symmetry group.
Conjugation by an ordinary symmetry acts on each higher-form symmetry group through
\begin{equation}
 \rho_r:G_1\longrightarrow\operatorname{Aut}(G_r).
\label{hg:eq:action}
\end{equation}
Thus, transporting a higher symmetry label can change it by $\rho_r$. The groups, these actions, and the Postnikov classes described below specify how the symmetry levels fit together.

\paragraph*{Classifying space and cell structure.}

The classifying space $B\mathbb{G}$ is a pointed, connected homotopy $n$-type whose based loop space recovers $\mathbb{G}$:
\begin{equation}
\begin{aligned}
 \Omega B\mathbb{G}&\simeq\mathbb{G},\\
 \pi_r(B\mathbb{G})&=G_r\quad(2\le r\le n).
\end{aligned}
\label{hg:eq:classifying}
\end{equation}
Here $\Omega B\mathbb G$ denotes the based loop space, whereas
$\Omega$ elsewhere denotes an anomaly cocycle; $\simeq$ denotes
homotopy equivalence. The Abelian property of $G_r$ follows from the Abelian
property of homotopy groups in degrees $r\ge2$.
On a space $Y$ with a CW decomposition, homotopy classes of maps
$Y\to B\mathbb{G}$ classify principal $\mathbb{G}$-bundles.
In the finite setting, these maps describe flat higher-group background
fields, and homotopies describe their gauge equivalences.

The Postnikov construction builds this space one homotopy group at a
time. Write $X^{(r)}$ for the stage retaining the groups through degree
$r$. Start with $X^{(1)}=BG_1$, the ordinary classifying space. For
$2\le r\le n$, the next stage is a fibration
\begin{equation}
 K(G_r,r)\longrightarrow X^{(r)}
 \longrightarrow X^{(r-1)}.
\label{hg:eq:tower}
\end{equation}
The Eilenberg--Mac Lane space $K(M,r)$ has homotopy group $M$ in
degree $r$ and zero homotopy groups in all other positive degrees.
After the last stage, $X^{(n)}\simeq B\mathbb{G}$.

The fibration at stage $r$ has the prescribed action $\rho_r$ and
a class
\begin{equation}
 k_{r}\in
 H^{r+1}\bigl(X^{(r-1)};G_r^{\rho_r}\bigr).
\label{hg:eq:postnikov}
\end{equation}
These are the $k$-invariants, also called Postnikov invariants.
The coefficient system $G_r^{\rho_r}$ means that transport along a
loop labeled by $g\in G_1$ acts by $\rho_r(g)$.
For a map $f:Y\to X^{(r-1)}$ specifying the lower fields, the
obstruction to lifting it to $X^{(r)}$ is $f^*k_r$.
When this class vanishes, a lift supplies the next field.
The successive classes, together with the groups and actions, determine
the homotopy type, up to compatible changes of these data~\cite{Hatcher2002Algebraic,Kapustin2017Higher}.
If all actions and all Postnikov classes are trivial, the construction
gives
\begin{equation}
 B\mathbb{G}\simeq \prod_{r=1}^n K(G_r,r).
\label{hg:eq:product}
\end{equation}

For explicit fields, choose a normalized cocycle $\kappa_{r+1}$
representing each $k_r$. A simplicial model $X$ of
$B\mathbb{G}$ then describes an $\ell$-simplex by fields
\begin{equation}
 \cfg=(a,B_2,\ldots,B_n)
\label{hg:eq:fields}
\end{equation}
on the ordered simplex $\Delta^\ell=[0,\ldots,\ell]$.
Each oriented edge carries $a_{ij}\in G_1$, and each ordered $r$-face
carries $B_r(i_0,\ldots,i_r)\in G_r$. An $r$-cochain means an
assignment of such values to $r$-faces. Values are expressed in the
coefficient frame at the first vertex. Fields of degree greater than
$\ell$ have no components on this simplex.

The edge field is flat, and the higher fields obey
\begin{equation}
\begin{aligned}
 a_{ij}a_{jk}&=a_{ik},\\
 \delta_a B_r&=\kappa_{r+1}(a,B_2,\ldots,B_{r-1}).
\end{aligned}
\label{hg:eq:constraints}
\end{equation}
\begingroup\makeatletter
\edef\@currentlabel{\p@equation\theequation}\label{dc:field-constraints}
\makeatother\endgroup
The first equation holds on every triangle. The second holds on every
$(r+1)$-face, with its right-hand side evaluated on the lower fields
of that face. It is the cochain form of the lifting condition in the
Postnikov tower.

For a $G_r$-valued $q$-cochain $c$, the twisted coboundary on
$[0,\ldots,q+1]$ is
\begin{equation}
\begin{aligned}
 (\delta_a c)_{0\cdots q+1}
=\rho_r(a_{01})c_{1\cdots q+1}
+\sum_{j=1}^{q+1}(-1)^j
 c_{0\cdots\widehat{j}\cdots q+1}.
\end{aligned}
\label{hg:eq:twisted}
\end{equation}
\begingroup\makeatletter
\edef\@currentlabel{\p@equation\theequation}\label{dc:twisted-coboundary}
\makeatother\endgroup
Subscripts denote evaluation on the listed vertices, and a hat denotes
omission. The first term transports its coefficient from vertex $1$
to vertex $0$. Flatness of $a$ gives $\delta_a^2=0$.
The cocycle condition on the Postnikov representative is
\begin{equation}
 \delta_a\kappa_{r+1}(a,B_2,\ldots,B_{r-1})=0
\label{hg:eq:closed}
\end{equation}
whenever the lower fields satisfy their own constraints.
It ensures consistency of Eq.~\eqref{hg:eq:constraints} on
$(r+2)$-faces. On a simplex, the flat coefficient system is
trivializable and positive-degree cohomology vanishes, so this closed
right-hand side admits a solution $B_r$.
Replacing a representative by
$\kappa_{r+1}+\delta_a\lambda_r$ is absorbed by
$B_r\mapsto B_r+\lambda_r$, where $\lambda_r$ is an $r$-cochain
on the lower-stage model, evaluated on the lower fields.

For example, a $2$-group is specified by $G_1$, $G_2$, $\rho_2$,
and $k_2\in H^3(BG_1;G_2^{\rho_2})$. Choose its normalized
cocycle representative $\kappa_3$ and write $B_{ijk}=B_2(i,j,k)$.
On the tetrahedron $[0,1,2,3]$, the field equation is
\begin{equation}
 \kappa_3(a_{01},a_{12},a_{23})=\rho_2(a_{01})B_{123}-B_{023}+B_{013}-B_{012}.
\label{hg:eq:two-group}
\end{equation}
In a presentation with one representative for each component $g\in G_1$,
$\kappa_3(g,h,k)$ specifies the associator. Its twisted cocycle
condition expresses the pentagon identity.

Let $X_\ell$ be the set of these field configurations on $\Delta^\ell$.
The normalized Postnikov cocycles vanish on degenerate simplices.
The face and degeneracy maps defined above preserve the field constraints,
so the sets $X_\ell$ form a simplicial set. Its geometric realization
has the homotopy type of $B\mathbb{G}$.

\section{Microscopic \texorpdfstring{$\Z_2$}{Z2}-enriched \texorpdfstring{$\Z_4$}{Z4} toric codes}
\label{app:tc}
\label{app:tc:inversion}
\label{app:tc:inversion-reduction}
\label{app:mixed:sec}

We derive the configuration states and operator relations used in
Sec.~\ref{ex:inversion}, then evaluate $\mathcal W_{10}$ directly
on the configurations in Fig.~\ref{fig:inversion-framed-history}.
Throughout, $C_{ij}$ labels walls on the black graph and $q_{i'}$
labels the charge at blue site $i'$. On a microscopic refinement,
the same data are represented by wall-crossing bits $A_e$ and
vertex charges $Q_v$, with $Q_{i'}=q_{i'}$.

\input{figures/inversion_square_hopping_figure}

\subsection{Microscopic realization and configuration states}

\paragraph*{Origin of the two realizations.}
The parent groups are
\begin{equation}
 E^{(\epsilon)}=\langle r,s\mid r^4=1,\ srs^{-1}=r^{-1},\
 s^2=r^{2\epsilon}\rangle,
 \label{ex:inversion-extensions}
\end{equation}
giving $D_4$ for $\epsilon=0$ and $Q_8$ for $\epsilon=1$.
Partially gauging $\langle r\rangle=\Z_4$ leaves the quotient
$\Z_2$ as the global charge-conjugation symmetry. Both realizations have
the split wall--electric 2-group
\begin{equation}
 G=\Z_2,\qquad \mathcal A=\Z_4^e,\qquad
 \rho_D(b)=-b,\qquad \beta=0.
 \label{app:mixed:data}
\end{equation}
Their difference is the magnetic fractionalization class
$t_\epsilon(D,D)=m^{2\epsilon}$~\cite{Tarantino2016Symmetry,Benini2019TwoGroup}.
It is a twisted $H^2$ class: its coboundary vanishes because
charge conjugation sends $2\epsilon$ to $-2\epsilon=2\epsilon$ modulo four.
Attaching an anyon $x$ to the wall changes this representative by
$x+\rho_Dx=0$, so the two choices remain distinct. This affects
defect fusion and braiding phases, while the pointlike twist-defect
fusion channels are unchanged: multiplication by $m^2$ permutes
$1,e^2,m^2,e^2m^2$.

\paragraph*{Electric basis and flux.}
Use an ordered triangulation of $S^2$, with edges $ij$ oriented
from $i$ to $j$ for $i<j$. The wall and qudit Pauli conventions
are those of Eq.~\eqref{ex:inversion-edge-paulis-main}. The Fourier
basis is
\begin{equation}
 \begin{gathered}
 \ket{\lambda}_x=\frac12\sum_{E=0}^3i^{\lambda E}\ket E,
 \qquad \lambda\in\Z_4,\\
 \widetilde X\ket{\lambda}_x=i^{-\lambda}\ket{\lambda}_x,
 \qquad \widetilde Z\ket{\lambda}_x=\ket{\lambda+1}_x.
 \end{gathered}
 \label{app:tc:inversion-electric-basis}
\end{equation}
The charge operator in Eq.~\eqref{ex:inversion-charge-main} has
eigenvalue $i^{q_v(A,\lambda)}$, where
\begin{equation}
 q_v(A,\lambda)=(I_A\lambda)_v
 =\sum_{v\to j}\lambda_{vj}
 -\sum_{i\to v}(-1)^{A_{iv}}\lambda_{iv}\pmod4.
 \label{app:tc:inversion-charge}
\end{equation}
We reserve $Q_v$ for its prescribed eigenvalue label.

Write a parent link as $r^{E_{ij}}s^{A_{ij}}$. On a triangle with
even wall parity, group flatness requires
\begin{equation}
 E_{ij}+(-1)^{A_{ij}}E_{jk}-E_{ik}
 =2\epsilon A_{ij}A_{jk}\pmod4.
 \label{app:tc:inversion-groupflat}
\end{equation}
Fourier transformation gives the triangular flux operator
\begin{equation}
 \widehat{\mathcal B}_{ijk}[A]
 =(-1)^{\epsilon A_{ij}A_{jk}}
 \widetilde Z_{ij}\widetilde Z_{jk}^{(-1)^{A_{ij}}}
 \widetilde Z_{ik}^{-1}.
 \label{app:tc:inversion-bareflux}
\end{equation}
Its electric shift has divergence
$(-1)^{A_{ik}}-(-1)^{A_{ij}+A_{jk}}$ at $k$ and zero elsewhere.
It therefore preserves every charge at even wall parity. As in
the main-text square operator, we set it to the identity at odd
parity:
\begin{equation}
 \begin{aligned}
 F_{ijk}^A&=Z_{ij}Z_{jk}Z_{ik},\qquad
 \Pi_{ijk}^A=\tfrac12(1+F_{ijk}^A),\\
 \overline{\mathcal B}_{ijk}
 &=(1-\Pi_{ijk}^A)+\Pi_{ijk}^A
       \widehat{\mathcal B}_{ijk}[\widehat A].
 \end{aligned}
 \label{app:tc:inversion-flux}
\end{equation}
\begingroup\makeatletter
\edef\@currentlabel{\p@equation\theequation}\label{ex:inversion-flux-main}
\makeatother\endgroup
Here $\widehat A$ means control by the wall qubits. Charge and
flux operators commute on the full Hilbert space; flux operators
read the same wall bits and apply commuting electric shifts.
Thus the triangular version of Eq.~\eqref{ex:inversion-H-main}
is a commuting Hamiltonian. A nonzero charge or flux costs four,
and an occupied wall edge costs two.

\paragraph*{One state for each configuration.}
For closed walls, choose an exterior reference vertex $v_*$ and
paths $\gamma_v$ from $v_*$ to $v$. The crossing parity
$A(\gamma_v)$ is path independent on $S^2$. Neutrality is
\begin{equation}
 \sum_v(-1)^{A(\gamma_v)}Q_v=0\pmod4.
 \label{app:tc:inversion-framed-dictionary}
\end{equation}
This is Eq.~\eqref{ex:inversion-neutral-main} on the separated
lattices of Fig.~\ref{fig:inversion-framed-history}.
The links $s^{A(\gamma_i)}s^{-A(\gamma_j)}$ form a flat parent
connection with
$E^0_{ij}(A)=2\epsilon A_{ij}A(\gamma_j)$ modulo four.
The corresponding normalized vacuum-flux state is
\begin{equation}
 \ket{A,Q}^{(\epsilon)}=\ket A\otimes\frac1{\sqrt{N_L}}
 \sum_{I_A\lambda=Q}i^{-\langle\lambda,E^0(A)\rangle}
                         \ket\lambda_x,
 \label{app:tc:inversion-reduced-state}
\end{equation}
\begingroup\makeatletter
\edef\@currentlabel{\p@equation\theequation}\label{app:tc:inversion-state}
\makeatother\endgroup
where the pairing sums over stored edges and
$N_L=4^{|L_1|-|L_0|+1}$ for vertex and edge sets $L_0,L_1$.
To see this count, rephase rows and columns of $I_A$ by the
crossing signs to obtain the ordinary incidence matrix. Neutrality
is its only constraint. On $S^2$, triangle boundaries connect
all solutions, and the vacuum-flux conditions fix their relative
phases. Each allowed wall and charge configuration therefore
specifies one ray. In Fig.~\ref{fig:inversion-framed-history},
all charges away from the four blue sites are zero, giving the
main-text state $\ket{C,q}$.

\begin{table*}[t]
 \input{figures/inversion_framed_cancellation_table}
\end{table*}

\subsection{Hopping operators and symmetry}
\label{app:tc:inversion-square-details}

\paragraph*{Wall hopping from partial gauging.}
For a region indicator $h_v\in\Z_2$, put $A'=A+\delta h$ and
$(S_h\lambda)_{ij}=(-1)^{h_i}\lambda_{ij}$. The region version
of Eq.~\eqref{ex:inversion-vertex-hop} is
\begin{equation}
 \begin{aligned}
 U_h^{(\epsilon)}\ket{A,\lambda}_x
 ={}&(-1)^{\epsilon\sum_{ij}(D_h(A))_{ij}\lambda_{ij}}\\
 &\times(-1)^{\epsilon\sum_vh_v\mathbf1_{q_v=1}}
 \ket{A+\delta h,S_h\lambda}_x,\\
 (D_h(A))_{ij}&=(h_i+h_j)A_{ij}+h_j(1+h_i)\pmod2.
 \end{aligned}
 \label{ex:inversion-wall-main}
\end{equation}
All phases use the incoming labels. The gate flips the wall bits
on the region boundary and reverses its enclosed charges.

The charge-dependent sign follows directly from the ungauged
parent. Set $f(Q)=\mathbf1_{Q=3}$ and choose the on-site charge
states
\begin{equation}
 \ket{\varphi_Q}=\frac1{\sqrt8}\sum_{k=0}^3\sum_{a=0}^1
 i^{-Qk}(-1)^{\epsilon af(Q)}\ket{r^ks^a}.
 \label{app:tc:inversion-parent}
\end{equation}
Left multiplication by $r^b$ gives $i^{Qb}$, while $s$ sends
$\ket{\varphi_Q}$ to
$(-1)^{\epsilon\mathbf1_{Q=1}}\ket{\varphi_{-Q}}$.
The invariant state $\ket{\varphi_0}$ supplies a trivial
product-state parent before gauging.
After gauging $\Z_4$, the Gauss constraint is $Q=I_A\lambda$.
Fixing the matter variable to $s^a$ leaves a spin
$(\ket0+(-1)^{\epsilon f(Q)}\ket1)/\sqrt2$ at each vertex.
The local circuit
\begin{equation}
 K^{(\epsilon)}\ket{A,a,\lambda}_x
 =(-1)^{\epsilon\sum_va_vf(q_v)}\ket{A,a,\lambda}_x
 \label{app:tc:inversion-disentangler}
\end{equation}
disentangles these spins. Before this circuit, the wall gate has
the same register updates as above, also sends $a$ to $a+h$,
and has edge phase
$(-1)^{\epsilon\langle D_h(A)+\delta(h\cup a),\lambda\rangle}$.
Conjugating by $K^{(\epsilon)}$ cancels the incoming-spin phases
and leaves $h_v f(-q_v)=h_v\mathbf1_{q_v=1}$, giving
Eq.~\eqref{ex:inversion-wall-main}. The transformed spins are
decoupled in $\ket+$. This gives the edge-only Hamiltonian and
states used throughout Sec.~\ref{ex:inversion}.

\paragraph*{Charge, flux, and composition.}
The identity
$I_{A+\delta h}S_h=\operatorname{diag}((-1)^h)I_A$
gives the charge update. In binary cochains,
$D_h(A)=h\cup A+(A+\delta h)\cup h$, so on flat triangles
\begin{equation}
 \delta D_h(A)=(A+\delta h)\cup(A+\delta h)-A\cup A.
 \label{app:tc:inversion-descent}
\end{equation}
This is exactly the change in the plaquette sign. The electric
reversal supplies its orientation factor, while the
charge-dependent phase is unchanged by a flux shift. Hence
\begin{equation}
 \begin{aligned}
 (U_h^{(\epsilon)})^\dagger Z_{ij}U_h^{(\epsilon)}
   &=(-1)^{h_i+h_j}Z_{ij},\\
 (U_h^{(\epsilon)})^\dagger\widehat{\mathcal A}_vU_h^{(\epsilon)}
   &=\widehat{\mathcal A}_v^{\,(-1)^{h_v}},\\
 (U_h^{(\epsilon)})^\dagger\overline{\mathcal B}_pU_h^{(\epsilon)}
   &=\overline{\mathcal B}_p^{\,(-1)^{h_i}}.
 \end{aligned}
 \label{ex:inversion-wall-checks-main}
\end{equation}
Here $i$ is the initial vertex of the ordered plaquette. Odd
wall parity is preserved, so the identity flux operator in that
sector obeys the same relation.

For successive moves, the edge phases satisfy
$D_h(A)+D_k(A+\delta h)=D_{h+k}(A)+\delta(h\cup k)$.
The extra term contributes $\sum_vh_vk_vq_v$ modulo two.
The charge-dependent signs cancel it because
$\mathbf1_{q=1}+\mathbf1_{-q=1}=q$ modulo two. Thus
\begin{equation}
 U_k^{(\epsilon)}U_h^{(\epsilon)}=U_{h+k}^{(\epsilon)},
 \qquad (U_h^{(\epsilon)})^2=1,
 \label{ex:inversion-hop-laws-main}
\end{equation}
with the updated wall bits used in the second hop.

For the square in Fig.~\ref{fig:inversion-square-hamiltonian}(c),
even wall parity allows a diagonal $il$ for which both triangles
$[ijl]$ and $[ikl]$ are flat. Their common diagonal
shifts cancel, and their signs multiply to
$(-1)^{\epsilon(A_{ij}A_{jl}+A_{ik}A_{kl})}$, the two controlled-$Z$
factors in Eq.~\eqref{ex:inversion-square-flux-main}. At odd wall
parity, the two boundary paths transport opposite charge signs,
so their difference would change the charge at $l$ by two.
The identity extension avoids this change and uses only the four
boundary edges. The diagonal is only a device for deriving the
square operator.

Setting $h=\mathbf1$ gives the global symmetry
\begin{equation}
 U^{(\epsilon)}\ket{A,\lambda}_x
 =(-1)^{\epsilon\sum_v\mathbf1_{q_v=1}}\ket{A,-\lambda}_x.
 \label{app:tc:inversion-reduced-global}
\end{equation}
It leaves the wall pattern fixed and conjugates both charges and
fluxes. Its square is one because $\sum_vq_v$ is even, and
inverse charge and flux labels have equal energies. Both
Hamiltonians are therefore invariant under charge conjugation. Omitting the
wall-tension term $-\sum_e Z_e$ makes wall positions degenerate
without changing these operator relations.

\paragraph*{Wall fusion at fixed background.}
The fractionalization comparison holds the background fixed,
whereas a physical wall hop updates it. This distinction is
particularly explicit before disentangling the ancillary spins:
\begin{equation}
 \begin{aligned}
 U_h^{(\epsilon),\mathrm{aux}}[A]^2
   &=\prod_v\widehat{\mathcal A}_v[A]^{2\epsilon h_v},\\
 U_h^{(\epsilon),\mathrm{aux}}[A+\delta h]
 U_h^{(\epsilon),\mathrm{aux}}[A]&=1.
 \end{aligned}
 \label{app:tc:inversion-squares}
\end{equation}
\begingroup\makeatletter
\edef\@currentlabel{\p@equation\theequation}\label{ex:inversion-fixed-versus-updated}
\makeatother\endgroup
The first line transports $m^{2\epsilon}$ around the region and
gives $(-1)^{\epsilon Q}$ for enclosed charge $Q$. The second
line is the composition used in $\mathcal W_{10}$. The change
of basis is $K^{(\epsilon)}[A]$ at the source and
$K^{(\epsilon)}[A+\delta h]$ at the target, so the fixed-background
matrix square is not the square of the reduced edge-only matrix.
The closed magnetic transport commutes with $K^{(\epsilon)}$;
its braiding response is unchanged.

\input{sections/inversion_operator_details}

\input{sections/inversion_framed_details}

On other spatial topologies, the configurations may also require
Wilson-loop labels. Open walls would require microscopic
twist-defect cores; neither is needed for the closed-wall process
on $S^2$ considered here.

\input{sections/fermionic_toric_details}

\section{Details of higher-group statistics}
\label{app:hopping-and-locality}

We give the local moves and invariance conditions used in
Secs.~\ref{loc:sec} and \ref{dc:sec}. We first construct the moves
and their strict inverses by filling a cylinder between configurations,
including the changes induced in higher layers. A 2-group example
makes the updates and cocycle hopping phases explicit. We then
derive the flow balance and local phase cancellation conditions
that define invariant statistical processes.

\subsection{Cylinder construction of the local moves}
\label{app:transport}

We construct each move by filling a cylinder between its incoming and
outgoing configurations. Imposing the field constraints throughout
the cylinder determines both the direct change and the junctions it
induces in higher layers. Let $T=[0,\ldots,m]$, where $m=d+1$ and
$D=m+1=d+2$. Physical space is $\partial T$, with auxiliary filling
$T$. Thus $T\times I$ has dimension $D$, one dimension above the
physical spacetime process.

Write $i^t=(i,t)$ for a vertex on the bottom or top of the cylinder.
We use the ordered triangulation
\begin{equation}
 \begin{aligned}
 C_j&=[0^0,\ldots,j^0,j^1,\ldots,m^1],\\
 \epsilon_j&=(-1)^{m-j},\qquad 0\le j\le m.
 \end{aligned}
 \label{cell:eq:15}
\end{equation}
The coefficient $\epsilon_j$ records the orientation of each simplex.
Let $\pi:T\times I
\to T$ be the projection and $i_0:T\to T\times I$ the bottom inclusion.
For a $q$-cochain $c$ and vertices $v_j=i_j^{t_j}$, the contraction
to the bottom is defined as
\begin{equation}
 (\mathsf hc)(v_0,\ldots,v_{q-1})
 =\sum_{j=0}^{q-1}(-1)^j
 c(i_0^0,\ldots,i_j^0,v_j,\ldots,v_{q-1}).
 \label{cell:eq:16}
\end{equation}
\begingroup\makeatletter
\edef\@currentlabel{\p@equation\theequation}\label{dc:prism-contraction}
\makeatother\endgroup
Repeated vertices are interpreted using normalized cochains. For
twisted coefficients, this formula is evaluated after transporting
the fields to the coefficient system pulled back from the bottom.

An elementary move has a parameter on one auxiliary face $f$.
For a degree-one move, $f$ is one vertex and its parameter is a
group element $h_i$ there; all other $h_i$ are the identity.
For a higher-degree move, all $h_i$ are the identity and the
appropriate higher cochain parameter is supported on $f$.
Set $k_{i,0}=e$ and $k_{i,1}=h_i$. If $a$ is the incoming flat
edge field, its interpolation is
\begin{equation}
 \widehat a_{i^s j^t}=k_{i,s}^{-1}a_{ij}k_{j,t}.
 \label{cell:eq:19}
\end{equation}
\begingroup\makeatletter
\edef\@currentlabel{\p@equation\theequation}\label{dc:elementary-edge}
\makeatother\endgroup
Thus the bottom is $a$ and the top is the transformed edge field.

The higher fields are determined successively. Once the fields below
degree $r$ have been filled, evaluate their Postnikov cocycle
$\kappa_{r+1}$ on the cylinder. Express it in the bottom coefficient
frame, denoted by a bar: on a simplex beginning at $v_0$,
$\overline\kappa_{r+1}=\rho_r(k_{v_0})\kappa_{r+1}$.
Extend the spatial move parameter to an $(r-1)$-cochain
$\widehat\lambda_r$ by

\begin{equation}
 \begin{gathered}
 \widehat \lambda_r((i_0,t_0),\ldots,(i_{r-1},t_{r-1}))\\
 =
 \begin{cases}
 \lambda_r(i_0,\ldots,i_{r-1}),&
 \substack{i_0<\cdots<i_{r-1},\\t_{r-1}=1,}\\
 0,&\text{otherwise},
 \end{cases}
 \end{gathered}
 \label{app:conv:time-extension}
\end{equation}
for an admissible tuple $((i_0,t_0),\ldots,(i_{r-1},t_{r-1}))$ corresponding to a simplex in the cylinder. On a cylinder face,
it equals the parameter on the projected face when the spatial
vertices are distinct and the last vertex lies on the top; it is
zero otherwise. This fixes the mixed faces as well as the bottom
and top values. Then set
\begin{equation}
 \widehat B_r
 =\rho_r(k^{-1}_{v_0})\bigl(\pi^*B_r
 +\mathsf h\overline\kappa_{r+1}
 +\delta_{\pi^*a}\widehat\lambda_r\bigr).
 \label{cell:eq:21}
\end{equation}
\begingroup\makeatletter
\edef\@currentlabel{\p@equation\theequation}\label{dc:elementary-higher}
\makeatother\endgroup
The two added terms have distinct roles. The contraction supplies
the correction required by the lower fields, while the coboundary
implements the chosen layer-$r$ move. Even when $\lambda_r=0$,
a move in a lower layer can induce a correction in this layer.

To verify the constraint, use the contraction identity in the bottom
frame, with $\delta_a$ denoting the pulled-back twisted coboundary:
\begin{equation}
 \delta_a\mathsf h+\mathsf h\delta_a=1-\pi^*i_0^*,
 \qquad \mathsf h\pi^*=0.
 \label{cell:eq:17}
\end{equation}
The incoming field obeys
$\delta_aB_r=i_0^*\overline\kappa_{r+1}$. Since the lower cylinder
fields satisfy their constraints, their Postnikov cochain is closed.
The coboundary of the parameter term vanishes, and
Eq.~\eqref{cell:eq:17} gives
\begin{equation}
\begin{aligned}
 \delta_{\pi^*a}\overline{\widehat B}_r
 &=\pi^*i_0^*\overline\kappa_{r+1}
   +\delta_{\pi^*a}\mathsf h\overline\kappa_{r+1}\\
 &=\pi^*i_0^*\overline\kappa_{r+1}
   +\overline\kappa_{r+1}
   -\pi^*i_0^*\overline\kappa_{r+1}
   =\overline\kappa_{r+1}.
\end{aligned}
\label{cell:eq:22}
\end{equation}
Transforming back to the cylinder frame yields the required twisted
Postnikov equation. Induction therefore constructs an admissible
interpolation at every layer.

On a spatial face that omits the active parameter face, the
interpolation is stationary. The construction also commutes with
restriction to a face, including its coefficient frame, so adjacent
cylinders glue consistently. The top configuration defines the move
$\tau_s$. To invert it, first recover the bottom edge field from
Eq.~\eqref{cell:eq:19}, then recover the higher fields successively
from Eq.~\eqref{cell:eq:21}. This gives the strict inverse update;
it need not be obtained by simply negating a parameter in the
forward formula.

The labeled simplices in Eq.~\eqref{cell:eq:15}, summed with
coefficients $\epsilon_j$, give the integral $D$-chain
$\operatorname{Cyl}(s,\cfg)$ of the forward arrow $(s,\cfg)$.
This chain lies in $B\mathbb G$, with degenerate simplices set to
zero. Extend the assignment linearly to $E$. An inverse starting
at $\cfg$ contributes $-\operatorname{Cyl}(s,\tau_s^{-1}\cfg)$. By
Proposition~\ref{cell:prop:firstthree}, invariant expressions
give cycles and locality identities give boundaries.

\subsubsection*{Example: 2-group updates and phases}
\label{app:transport:two-group}

For a 2-group, write $G=G_1$ multiplicatively and $A=G_2$
additively, with action $\rho=\rho_2$ and normalized Postnikov
cocycle $\beta=\kappa_3$. The constraints are
\begin{equation}
 \begin{split}
 a_{ij}a_{jk}&=a_{ik},\\
 \rho(a_{ij})B_{jkl}-B_{ikl}+B_{ijl}-B_{ijk}
 &=\beta(a_{ij},a_{jk},a_{kl}).
 \label{dc:twogroup-flat}
 \end{split}
\end{equation}
 
Here $\rho(a_{ij})$ transports $B_{jkl}$ to the frame at $i$.
The twisted cocycle condition is
\begin{equation}
 \begin{split}
 0={}&\rho(g)\beta(h,k,l)-\beta(gh,k,l)
 +\beta(g,hk,l)\\
 &-\beta(g,h,kl)+\beta(g,h,k).
 \label{app:conv:postnikov}
 \end{split}
\end{equation}
Solving Eq.~\eqref{dc:twogroup-flat} in terms of the independent
labels $a_{0i}$ and $B_{0ij}$ gives
\begin{align}
 a_{ij}&=a_{0i}^{-1}a_{0j},\nonumber\\
 B_{ijk}&=\rho(a_{0i}^{-1})
 \bigl[B_{0jk}-B_{0ik}+B_{0ij}
 +\beta(a_{0i},a_{ij},a_{jk})\bigr].
 \label{app:conv:reconstruction}
\end{align}
For vertex parameters $h_i\in G$ and edge parameters
$\Lambda_{ij}\in A$, restricting the cylinder fields to the top gives
\begin{equation}
 \begin{split}
 a'_{ij}&=h_i^{-1}a_{ij}h_j,\\
 \Gamma_h(a)_{ijk}
 &=\beta(a_{ij},a_{jk},h_k)-\beta(a_{ij},h_j,a'_{jk})\\
 &\quad+\beta(h_i,a'_{ij},a'_{jk}),\\
 B'_{ijk}&=\rho(h_i^{-1})
 \bigl[B_{ijk}+\rho(a_{ij})\Lambda_{jk}-\Lambda_{ik}\\
 &\qquad\qquad+\Lambda_{ij}+\Gamma_h(a)_{ijk}\bigr].
 \label{app:conv:gauge}
 \end{split}
\end{equation}
Flatness follows directly from the transformation of $a$. The three
terms in $\Gamma_h$ evaluate $\beta$ on the three simplices of a
triangle's cylinder. By Eq.~\eqref{cell:eq:17}, their twisted
coboundary is $\beta(a')$ transported to the incoming frame minus
$\beta(a)$. The $\Lambda$ term has zero twisted coboundary, so
$\delta_{a'}B'=\beta(a')$.

For a wall move $V_v(k)$, set $h_v=k$, with all other vertex
parameters trivial and $\Lambda=0$. For a string move $E_{pq}(t)$,
set $\Lambda_{pq}=t$, with all other edge parameters zero and every
$h_i=e$. Write $s$ for either move and $j$ for the last vertex of
its parameter face. The degeneracy $s_j\cfg$ repeats vertex $j$.
On this larger simplex, $\widetilde s$ uses the same parameter
and replaces $j$ in its parameter face with the second copy.
The labeling $\tau_{\widetilde s}(s_j\cfg)$ is the only
potentially nondegenerate staircase cell of the cylinder. For a
normalized $\mathbb R/\mathbb Z$-valued $D$-cocycle $\Omega$,
\begin{equation}
 U(s)\ket{\cfg}
 = e^{2\pi i(-1)^{m-j}\Omega(\tau_{\widetilde s}(s_j\cfg))}
   \ket{\tau_s\cfg}.
 \label{app:conv:coupled-operators}
\end{equation}
\begingroup\makeatletter
\edef\@currentlabel{\p@equation\theequation}\label{app:conv:coupled-cells}
\makeatother\endgroup
The sign is the staircase orientation of Eq.~\eqref{cell:eq:15}.
The strict inverse negates the angle at its forward source.

\paragraph*{Binary wall-and-point example.}
Take $G=A=\Z_2$, trivial action, and $\beta(g,h,k)=ghk$. On
the boundary of a tetrahedron, the wall labels $a_{ij}$ are flat
and the point labels $B_{ijk}$ obey
\begin{equation}
 B_{123}+B_{023}+B_{013}+B_{012}=a_{01}a_{12}a_{23}
 \pmod2.
 \label{loc:binary-constraint}
\end{equation}
The right-hand side fixes point parity from the walls. Start with
$a_{01}=a_{12}=1$, $a_{23}=0$, and $B=0$. Label each physical
vertex by its dual auxiliary triangle. The move $V_3(1)$ sweeps
the triangle with physical vertices $123,023,013$. Its cylinder
gives
\begin{equation}
 \begin{split}
 a'_{i3}&=a_{i3}+1,\qquad 0\le i<3,\\
 B'_{ij3}&=B_{ij3}+a_{ij}a_{j3},\qquad 0\le i<j<3.
 \end{split}
 \label{loc:binary-wall-move}
\end{equation}
The correction uses incoming labels, with addition modulo two.
Other labels stay fixed. Since $a_{13}=1$ and $a_{23}=0$, only
$B_{013}$ changes among the point labels. The new $a_{23}=1$ makes the right-hand side of
Eq.~\eqref{loc:binary-constraint} odd, and the cylinder supplies the
required point at $013$. The string move $E_{01}(1)$ toggles
$B_{013}$ and $B_{012}$, moving the point to $012$ along their dual
edge while leaving the wall labels and total point parity fixed.

\subsection{Closure and local-phase cancellation conditions}
\label{app:phase-conditions}

Let $\cfg_j$ be the configuration after step $j$ of a word
$W=s_N^{\epsilon_N}\cdots s_1^{\epsilon_1}$ starting at the vacuum
$\cfg_0$. Eq.~\eqref{loc:inverse-action} gives
\begin{equation}
 \begin{aligned}
 \Theta(W)&=\sum_{j=1}^N\epsilon_j\theta_{s_j}(\cfg_j^+),\\
 \cfg_j^+&=
 \begin{cases}
 \cfg_{j-1},&\epsilon_j=+1,\\
 \cfg_j,&\epsilon_j=-1.
 \end{cases}
 \end{aligned}
 \label{app:phase-sources}
\end{equation}
Collect these signs by forward arrow $(s,\cfg)$ to obtain
$e=\langle W\rangle_{\cfg_0}=\sum_{s,\cfg}n_{s,\cfg}(s,\cfg)$.
Rephasing $\ket\cfg$ by $e^{i\alpha(\cfg)}$ changes each forward
angle by $\alpha(\cfg)-\alpha(\tau_s\cfg)$. In the total angle,
the coefficient of $\alpha(\cfg)$ is the outgoing count minus the
incoming count. Thus invariance under arbitrary state rephasings
is equivalent to
\begin{equation}
 \sum_s\bigl[n_{s,\tau_s^{-1}\cfg}-n_{s,\cfg}\bigr]=0
 \quad\text{for every }\cfg.
 \label{loc:flow}
\end{equation}
This is flow balance for a closed word. Conversely, a finite
integer expression with balanced flow decomposes into directed
circuits. Reachability from the vacuum lets us attach each circuit
to a preparation path and its strict inverse, giving a vacuum-based
closed word with the same counts.

For a simplex move with parameter face $f_s$, let $d_v\cfg$
be the restriction to the auxiliary facet opposite $v$, including
all fields and their frames. At each support vertex $v\notin f_s$,
an arbitrary function of that input can shift the hopping phase
according to
\begin{equation}
 \theta_s(\cfg)\longmapsto\theta_s(\cfg)
 +\sum_{v\notin f_s}\varphi_{s,v}(d_v\cfg).
\end{equation}
The contribution of $\varphi_{s,v}(\xi)$ has coefficient
$\sum_{\cfg:\,d_v\cfg=\xi}n_{s,\cfg}$. Since the values of the
$\mathbb R/2\pi\mathbb Z$-valued functions are independent, local
phase cancellation is equivalent to
\begin{equation}
 \sum_{\cfg:\,d_v\cfg=\xi}n_{s,\cfg}=0
 \quad\text{for every }s,\ v\notin f_s,\ \xi.
 \label{loc:cancellation}
\end{equation}

Eqs.~\eqref{loc:flow} and \eqref{loc:cancellation}
define $E_{\mathrm{inv}}$. Proposition~\ref{cell:prop:firstthree}
shows that the counts of the locality identities lie in this group.
$E_{\rm id}$ is generated by the counts of these nested commutators, evaluated at every allowed source and with strict inverses.
The statistical group is $E_{\mathrm{inv}}/E_{\mathrm{id}}$.

\input{sections/AppendixF_lemmas}

\input{sections/AppendixFnew}

\input{sections/Appendix_finite_order_and_localization}

\input{sections/spt_examples_details}

\FloatBarrier
\bibliographystyle{apsrev4-2}
\makeatletter
\immediate\write\@auxout{\string\citation{PRXControl}}
\makeatother
\interlinepenalty=10000
\bibliography{bibliography}
\end{document}

%% file: figures/inversion_framed_history_figure.tex
\begin{figure*}[t]
    \centering
    \input{figures/inversion_framed_history.tex}
    \caption{The charge--wall process $\mathcal W_{10}$ in
    Eq.~\eqref{ex:inversion-word-main}. The final arrow returns to the initial vacuum. Dashed black and blue edges show the wall and charge lattices,
    respectively. Gold edges are $\Z_2$ charge-conjugation walls; violet
    dots and labels mark charges $q\in\Z_4$ at the blue vertices.
    Charges are neutral when transported to a common frame, with
    their signs reversed at each wall crossing. Thus, the two $+1$
    charges in panel $2$ lie on opposite sides of the wall and sum
    to zero in a common frame, satisfying Eq.~\eqref{ex:inversion-neutral-main}.
    The membrane operator $U_{123}$ toggles the wall along triangle $123$ and reverses the charge at $0'$. The hopping operator $V_{0'j'}$ removes one unit at $0'$ and deposits $+1$ at $j'$ without a wall crossing or $-1$ with one.
    At each microscopic vertex, the local configuration before a
    forward hop matches that after a paired inverse hop, so their
    local phase shifts cancel. For example, at $0'$, this cancellation
    pairs $U_{123}$ at step $2$ with $U_{123}^{-1}$ at step $8$.
    Table~\ref{ex:inversion-local-pairings-main} in
    Appendix~\ref{app:tc:inversion} lists all local pairings.
    }
    \label{fig:inversion-framed-history}
\end{figure*}

%% file: figures/inversion_framed_history.tex
\begingroup%
\definecolor{invFrWall}{HTML}{C88A2A}%
\definecolor{invFrCharge}{HTML}{7561A8}%
\definecolor{invFrBlue}{HTML}{287DB8}%
\definecolor{invFrInk}{HTML}{30343B}%
\definecolor{invFrMesh}{HTML}{444B52}%
\definecolor{invFrText}{HTML}{46515B}%
\definecolor{invFrRule}{HTML}{E1E5E9}%
\input{figures/inversion_dual_geometry.tex}%
\begin{tikzpicture}[
 x=1cm,y=1cm,
 font=\fontsize{9}{10.5}\selectfont,
 line cap=round,line join=round,
 mesh/.style={draw=invFrMesh,line width=.62pt,
              dash pattern=on 2.2pt off 1.5pt},
 framing/.style={draw=invFrBlue,draw opacity=.75,line width=.70pt,
                 dash pattern=on 2.2pt off 1.5pt},
 wall/.style={draw=invFrWall,line width=2pt,
              preaction={draw=white,line width=3.1pt}},
 hop/.style={draw=invFrBlue,line width=1.25pt},
 charge/.style={draw=white,fill=invFrCharge,line width=.6pt},
 empty charge/.style={draw=invFrBlue,fill=white,line width=.6pt},
 old label/.style={text=invFrText,fill=white,inner sep=.6pt,
                   font=\fontsize{8}{9.5}\selectfont},
 frame label/.style={text=invFrBlue,fill=white,inner sep=.6pt,
                     font=\fontsize{8}{9.5}\selectfont},
 charge label/.style={text=invFrCharge,fill=white,inner sep=.7pt,
                      font=\fontsize{10}{11.5}\selectfont},
 heading/.style={text=invFrInk,inner sep=0pt},
 transition/.style={draw=invFrInk,line width=.7pt,
                    -{Stealth[length=4pt,width=3.8pt]}},
 operator/.style={text=invFrInk,inner sep=1.7pt,
                  font=\fontsize{10}{11.5}\selectfont},
 legend text/.style={anchor=west,text=invFrInk,inner sep=0pt,
                     font=\fontsize{9}{10.5}\selectfont}
 ]
 \path[use as bounding box] (0,0) rectangle (17.8,6.86);

 \newcommand{\invFrPosition}[2]{%
  \pgfmathsetmacro{\invFrX}{1.37+3.5*(#1)}
  \pgfmathsetmacro{\invFrY}{5.24-3.23*#2}
 }
 \newcommand{\invFrPanel}[9]{%
  \invFrPosition{#1}{#2}
  \begin{scope}[shift={(\invFrX,\invFrY)},x=.55cm,y=.55cm]
   \invDualCoordinates
   \node[heading] at (0,2.73) {\textbf{#3}};
   \invDualBlackPaths{mesh}
   \invDualBluePaths{framing}
   \ifnum#4=1\relax
    \invDualBoundary{wall}
   \fi
   \if\relax\detokenize{#8}\relax
   \else
    \ifnum#8=1\relax
     \coordinate (hopStart) at (q0);
     \coordinate (hopEnd) at (q1);
    \else
     \coordinate (hopStart) at (q0);
     \coordinate (hopEnd) at (q2);
    \fi
    \draw[hop] (hopStart)--(hopEnd);
    \ifnum#9=1\relax
     \draw[draw=invFrBlue,line width=.9pt,
           -{Stealth[length=3.8pt,width=3.5pt]}]
       ($(hopStart)!.38!(hopEnd)$)--($(hopStart)!.66!(hopEnd)$);
    \else
     \draw[draw=invFrBlue,line width=.9pt,
           -{Stealth[length=3.8pt,width=3.5pt]}]
       ($(hopStart)!.66!(hopEnd)$)--($(hopStart)!.38!(hopEnd)$);
    \fi
   \fi
   \foreach \v in {1,2,3}{
    \fill[invFrMesh] (p\v) circle[radius=1.3pt];
   }
   \foreach \v/\q in {0/#5,1/#6,2/#7,3/0}{
    \ifnum\q=0\relax
     \filldraw[empty charge] (q\v) circle[radius=1.7pt];
    \else
     \filldraw[charge] (q\v) circle[radius=3.6pt];
     \ifcase\v
      \node[charge label,font=\fontsize{8}{9.5}\selectfont]
        at (.44,.75) {$\q$};
     \or
      \node[charge label,anchor=north] at ([yshift=-4.25pt]q1) {$\q$};
     \or
      \node[charge label,anchor=north] at ([yshift=-4.25pt]q2) {$\q$};
     \or
      \node[charge label,anchor=west] at (.18,2.17) {$\q$};
     \fi
    \fi
   }
   \ifnum#1=0\relax\ifnum#2=0\relax
    \node[old label,anchor=west] at (.80,.40) {$1$};
    \node[old label,anchor=east] at (-.80,.40) {$2$};
    \node[old label,anchor=west] at (.16,-.72) {$3$};
    \node[frame label,anchor=west] at (.14,.00) {$0'$};
    \node[frame label,anchor=north] at ({-sqrt(3)*43/40},-1.25) {$1'$};
    \node[frame label,anchor=north] at ({sqrt(3)*43/40},-1.25) {$2'$};
    \node[frame label,anchor=west] at (.18,2.17) {$3'$};
   \fi\fi
  \end{scope}
 }

 \newcommand{\invFrOp}[3]{%
  $#1_{\hbox{\fontsize{8}{9}\selectfont$#2$}}%
  \ifnum#3<0\relax^{\hbox{\fontsize{8}{9}\selectfont$-1$}}\fi$}

 \newcommand{\invFrTransition}[5]{%
  \invFrPosition{#1+.5}{#2}
  \begin{scope}[shift={(\invFrX,\invFrY)}]
   \draw[transition] (-.41,0)--(.41,0)
     node[midway,above=3pt,operator] {\invFrOp{#3}{#4}{#5}};
  \end{scope}
 }
 \draw[invFrRule,line width=.35pt] (.16,3.77)--(17.64,3.77);

 \invFrPanel{0}{0}{0 Vacuum}{0}{0}{0}{0}{}{1}
 \invFrPanel{1}{0}{1}{0}{-1}{+1}{0}{1}{1}
 \invFrPanel{2}{0}{2}{1}{+1}{+1}{0}{}{1}
 \invFrPanel{3}{0}{3}{1}{0}{+1}{-1}{2}{1}
 \invFrPanel{4}{0}{4}{0}{0}{+1}{-1}{}{1}

 \invFrPanel{0}{1}{5}{0}{-1}{+1}{0}{2}{1}
 \invFrPanel{1}{1}{6}{0}{0}{0}{0}{1}{-1}
 \invFrPanel{2}{1}{7}{0}{+1}{0}{-1}{2}{-1}
 \invFrPanel{3}{1}{8}{1}{-1}{0}{-1}{}{1}
 \invFrPanel{4}{1}{9}{1}{0}{0}{0}{2}{-1}

 \invFrTransition{0}{0}{V}{0'1'}{1}
 \invFrTransition{1}{0}{U}{123}{1}
 \invFrTransition{2}{0}{V}{0'2'}{1}
 \invFrTransition{3}{0}{U}{123}{1}
 \invFrTransition{4}{0}{V}{0'2'}{1}

 \invFrTransition{0}{1}{V}{0'1'}{-1}
 \invFrTransition{1}{1}{V}{0'2'}{-1}
 \invFrTransition{2}{1}{U}{123}{-1}
 \invFrTransition{3}{1}{V}{0'2'}{-1}
 \invFrTransition{4}{1}{U}{123}{-1}

 \begin{scope}[shift={(.32,.24)}]
  \draw[mesh] (0,0)--(.40,0);
  \node[legend text] at (.55,0) {wall lattice};
  \draw[framing] (2.54,0)--(2.94,0);
  \node[legend text] at (3.09,0) {charge lattice};
  \draw[wall] (5.49,0)--(5.89,0);
  \node[legend text] at (6.04,0) {conjugation wall};
  \filldraw[charge] (8.72,0) circle[radius=3.6pt];
  \node[legend text] at (9.02,0) {charge $q=\pm1$};
  \filldraw[empty charge] (11.55,0) circle[radius=1.7pt];
  \node[legend text] at (11.85,0) {uncharged site};
  \draw[hop,-{Stealth[length=3.8pt,width=3.5pt]}] (14.18,0)--(14.58,0);
  \node[legend text] at (14.73,0) {hopping just applied};
 \end{scope}
\end{tikzpicture}%
\endgroup%

%% file: sections/microscopic_inversion.tex
\label{ex:inversion}

We construct a mixed charge--wall statistical process and evaluate
it in two $\Z_4$ toric codes enriched by $\Z_2$ charge-conjugation
symmetry. The two models have the same anyon statistics and local
configuration updates but give opposite statistical phases.

\input{sections/inversion_edge_formulation}

%% file: sections/inversion_edge_formulation.tex
\label{app:tc:inversion-edge-formulation}

\begin{figure*}[t]
 \centering
 \input{figures/inversion_square_hamiltonian.tex}
 \caption{Local terms of the square-lattice Hamiltonian
 $H^{(\epsilon)}$ in Eq.~\eqref{ex:inversion-H-main}, with
 $\epsilon=0,1$ labeling the two $\Z_2$-enriched $\Z_4$ toric codes.
 Each solid edge is oriented right or up and carries a $\Z_2$
 wall qubit with Pauli operators $X_e,Z_e$ and a $\Z_4$ qudit
 with shift and clock operators $\widetilde X_e,\widetilde Z_e$.
 Dashed lines form the dual lattice.
 (a) A wall segment (gold) crosses edge $e$ when $Z_e=-1$.
 The edge term $-Z_e$ assigns an energy cost to this segment.
 (b) The vertex operator $\widehat{\mathcal A}_v$ is the product
 of the four indicated $\widetilde X$ powers, with wall-dependent
 exponents on incoming edges. Its eigenvalue $i^{q_v}$ specifies
 the charge $q_v\in\Z_4$.
 (c) At even wall parity around the plaquette, the flux operator
 $\overline{\mathcal B}_p$ combines the indicated $\widetilde Z$
 powers with the controlled-$Z$ gates shown to the left, as in
 Eq.~\eqref{ex:inversion-square-flux-main}. At odd wall parity,
 $\overline{\mathcal B}_p=\mathbf1$.
 }
 \label{fig:inversion-square-hamiltonian}
\end{figure*}

The $\Z_4$ toric code has electric charges $e^q$ and magnetic fluxes
$m^k$, with $q,k\in\Z_4$. The global $\Z_2$ symmetry acts by internal
charge conjugation:
\begin{equation}
 D:e^qm^k\longmapsto e^{-q}m^{-k}.
 \label{ex:inversion-action}
\end{equation}
Crossing a symmetry wall conjugates the anyon.

\paragraph*{Configurations and local moves.}

The process uses the black triangle $123$ in
Fig.~\ref{fig:inversion-framed-history}. Walls occupy black
edges. Electric charges sit at blue sites, with $0'$ inside the
triangle and $1',2',3'$ outside. The triangle belongs to a
tetrahedral graph on the sphere. Its fourth vertex $0$ and the
edges $01,02,03$ are omitted from the figure. These edges remain
wall-free during the process.

Let $C_{ij}\in\Z_2$ indicate an occupied wall edge. A closed wall
has zero or two occupied edges at each black vertex. On the
tetrahedral lattice, the blue site $i'$ lies in the face $F_i$
opposite $i$. The path $i'j'$ crosses the common edge $kl$ of
$F_i$ and $F_j$, so its wall-crossing bit is $w_{i'j'}=C_{kl}$.
For example, $w_{0'2'}=C_{13}$.
Write $q_{i'}\in\Z_4$ for the charge at $i'$ in its local frame.
To test neutrality, transport each charge to the exterior site
$2'$, reversing its sign at each occupied wall crossing. The
parity of crossings is path independent because the walls are
closed on a sphere. The neutrality condition is
\begin{equation}
 (-1)^{C_{13}}q_{0'}+(-1)^{C_{03}}q_{1'}+q_{2'}+(-1)^{C_{01}}q_{3'}=0\pmod4.
 \label{ex:inversion-neutral-main}
\end{equation}
Thus a pair of $+1$ charges on opposite sides of a wall is neutral.
The configuration $(C,q)$ consists of the closed wall pattern and
the four charges satisfying Eq.~\eqref{ex:inversion-neutral-main}.
Write $\ket{C,q}$ for its normalized state.

The wall move $U_{123}$ toggles the three edges of triangle $123$
and reverses the charge at $0'$. The electric move $V_{i'j'}$
removes one unit at $i'$ and deposits it at $j'$ with a sign set
by $w_{i'j'}$.
The two moves act as follows.
\begin{equation}
 \begin{aligned}
 U_{123}:\quad&C_{ab}'=C_{ab}+1\pmod2,\quad ab=12,13,23,\\
 &q'_{0'}=-q_{0'},\\
 V_{i'j'}:\quad&q'_{i'}=q_{i'}-1,\quad
 q'_{j'}=q_{j'}+(-1)^{w_{i'j'}}\pmod 4.
 \end{aligned}
 \label{ex:inversion-updates-main}
\end{equation}
Both moves preserve Eq.~\eqref{ex:inversion-neutral-main}.
A microscopic implementation $O$ of an arrow also contributes a hopping phase.
\begin{equation}
 O\ket{C,q}=e^{i\theta_O(C,q)}\ket{C',q'}~.
 \label{ex:inversion-configuration-action}
\end{equation}
A closed process accumulates the product of these hopping phases.
Their local contributions can cancel with different step pairings
in different neighborhoods.

\paragraph*{The charge--wall process $\mathcal W_{10}$.}
The process uses the wall move $U_{123}$ and the charge hops
$V_{0'1'}$ and $V_{0'2'}$. In the configurations shown in
Fig.~\ref{fig:inversion-framed-history}, the wall is either absent
or occupies all three edges of the triangle. The wall move toggles
those edges and reverses $q_{0'}$. Each charge hop removes one unit at $0'$ and
deposits $+1$ at its outer endpoint without the wall or $-1$ with it.

At the configuration level, $U_{123}V_{0'2'}U_{123}$ reverses
the transfer made by $V_{0'2'}$. Applying this charge hop once more
restores the configuration, giving the closed
sequence $\mathcal P=V_{0'2'}U_{123}V_{0'2'}U_{123}$.
The two $V_{0'2'}$ arrows denote the same source-frame transfer.
The phase of $\mathcal P$ can depend on the starting charges and on
the hopping phases. For example, a common phase change $e^{i\chi}$
of the two forward transfers shifts the angle of $\mathcal P$ by $2\chi$.
We prepare a pair with $V_{0'1'}$, placing $q_{0'}=-1$ inside the
triangle and the compensating charge at $1'$. We run $\mathcal P$,
remove the pair, and run $\mathcal P^{-1}$ on the vacuum, which
contributes the opposite shift $-2\chi$.
With the rightmost operator acting first, the resulting process
\begin{equation}
 \mathcal W_{10}
 :=\mathcal P^{-1}V_{0'1'}^{-1}\mathcal P V_{0'1'}
 \label{ex:inversion-word-main}
\end{equation}
measures the phase of $\mathcal P$ with one enclosed unit of charge
divided by its phase with none. Fig.~\ref{fig:inversion-framed-history}
shows all ten moves from the vacuum.
Appendix~\ref{app:tc:inversion} checks local cancellation
for these ten moves; Table~\ref{ex:inversion-local-pairings-main}
lists the local forward--inverse pairings.

\paragraph*{Two microscopic realizations.}
We obtain the toric-code degrees of freedom by gauging the normal
$\Z_4$ subgroup of the dihedral group $D_4$ of order eight or the
quaternion group $Q_8$~\cite{Tarantino2016Symmetry}, labeled by
$\epsilon=0$ and $1$, respectively. The quotient $\Z_2$ supplies the
global symmetry, and edge qubits record its domain walls.
Let $r$ generate the normal $\Z_4$ subgroup and $s$ lift the
nontrivial element of the quotient $\Z_2$. The relation
$s^2=r^{2\epsilon}$ distinguishes the two parent groups.

The two models use the same configuration path and local
cancellation but give
\begin{equation}
 \mathcal W_{10}^{(\epsilon)}\ket{\vac}^{(\epsilon)}
   =(-1)^\epsilon\ket{\vac}^{(\epsilon)}.
 \label{ex:inversion-result}
\end{equation}
Here $\mathcal W_{10}^{(\epsilon)}$ uses the hopping operators
of model $\epsilon$. The phase is $+1$ in the dihedral realization
and $-1$ in the quaternion realization. We specify these operators
below and evaluate the process in Appendix~\ref{app:tc:inversion}.

Following the group-extension construction of Ref.~\cite{Tarantino2016Symmetry},
we write the Hamiltonian in an edge-only form on a square lattice.
Each oriented edge carries a wall qubit $\ket{A_e}$ and a
toric-code qudit $\ket{E_e}$. The qubit operators are $X_e,Z_e$,
and the qudit shift and clock are $\widetilde X_e,\widetilde Z_e$.
\begin{equation}
 \begin{alignedat}{2}
 X_e\ket{A_e}&=\ket{A_e+1\bmod2},&\quad
 Z_e\ket{A_e}&=(-1)^{A_e}\ket{A_e},\\
 \widetilde X_e\ket{E_e}&=\ket{E_e+1\bmod4},&\quad
 \widetilde Z_e\ket{E_e}&=i^{E_e}\ket{E_e}.
 \end{alignedat}
 \label{ex:inversion-edge-paulis-main}
\end{equation}
A wall occupies the dual edge when $Z_e=-1$. The vertex term is
\begin{equation}
 \widehat{\mathcal A}_v=
 \prod_{e_{vj}}\widetilde X_{vj}^{-1}
 \prod_{e_{iv}}\widetilde X_{iv}^{Z_{iv}}.
 \label{ex:inversion-charge-main}
\end{equation}
The products run over outgoing and incoming edges, respectively.
The exponent $Z_{iv}$ reverses an incoming contribution when a wall
crosses $iv$. The eigenvalue $i^{q_v}$ of the vertex term defines
the charge $q_v$.

For a square with the two oriented paths $i\to j\to l$ and
$i\to k\to l$, even wall parity means
$Z_{ij}Z_{jl}Z_{ik}Z_{kl}=1$. In this sector the flux operator is
\begin{equation}
 \begin{aligned}
 \overline{\mathcal B}_{ijlk}
 ={}&\left(\mathrm{CZ}_{ij,jl}\mathrm{CZ}_{ik,kl}\right)^\epsilon
 \widetilde Z_{ij}\widetilde Z_{jl}^{Z_{ij}}
 \widetilde Z_{kl}^{-Z_{ik}}\widetilde Z_{ik}^{-1},
 \end{aligned}
 \label{ex:inversion-square-flux-main}
\end{equation}
where $\mathrm{CZ}_{a,b}=(-1)^{(1-Z_a)(1-Z_b)/4}$ is the controlled-$Z$ gate.
The two Hamiltonians differ through this controlled sign.
When the wall parity is odd, set $\overline{\mathcal B}_{ijlk}=\mathbf1$.
The Hamiltonian is
\begin{equation}
 \begin{aligned}
 H^{(\epsilon)}={}&-\sum_e Z_e-\sum_v\bigl(1+\widehat{\mathcal A}_v
       +\widehat{\mathcal A}_v^2+\widehat{\mathcal A}_v^\dagger\bigr)\\
 &-\sum_p\bigl(1+\overline{\mathcal B}_p
       +\overline{\mathcal B}_p^{\,2}
       +\overline{\mathcal B}_p^\dagger\bigr),
 \end{aligned}
 \label{ex:inversion-H-main}
\end{equation}
shown in Fig.~\ref{fig:inversion-square-hamiltonian}.
These terms commute on the full Hilbert space. On a sphere,
a closed wall pattern and neutral charges in the vacuum flux sector
select a unique joint eigenstate up to phase. These are the states
$\ket{C,q}$ used above, with charges away from the four blue sites
fixed to zero.

\paragraph*{Microscopic hopping operators.}
The electric hop along an oriented edge $i\to j$ is
\begin{equation}
 V_{ij}=\widetilde Z_{ij}^{-1}.
 \label{ex:inversion-single-edge-main}
\end{equation}
This removes one unit at $i$ and adds $(-1)^{A_{ij}}$ at $j$.
The electric hops are diagonal in the $\ket{A,E}$ basis, so they commute with the wall and flux terms in Eq.~\eqref{ex:inversion-H-main}.

To describe the wall hop at a single vertex $v$, use the basis
$\ket{A,\lambda}_x$ in which $\widetilde X_{ij}$ has eigenvalue
$i^{-\lambda_{ij}}$. The hop toggles the wall bit on every edge
incident on $v$ and reverses the electric labels on outgoing edges.
Let $\boldsymbol v$ be the $\Z_2$ vertex field supported at $v$,
so $(\delta\boldsymbol v)_{ij}=\delta_{iv}+\delta_{jv}$.
The unitary gate is
\begin{equation}
 \begin{aligned}
 U_v^{(\epsilon)}\ket{A,\lambda}_x
 &=(-1)^{\epsilon\Phi_v}\ket{A+ \delta \boldsymbol{v},
                  \{(-1)^{\delta_{iv}}\lambda_{ij}\}}_x,\\
 \Phi_v&=\sum_{v\to j}A_{vj}\lambda_{vj}
 +\sum_{i\to v}(1-A_{iv})\lambda_{iv}+\mathbf1_{q_v=1}.
 \end{aligned}
 \label{ex:inversion-vertex-hop}
\end{equation}
The sums run over outgoing and incoming edges, respectively,
and all factors use the labels before the hop. The last term is one
when the incoming charge is $q_v=1$. The edge terms compensate the
change of the wall-dependent plaquette sign, while the charge term
makes $(U_v^{(\epsilon)})^2=1$; Appendix~\ref{app:tc:inversion} derives this choice from the partially gauged parent. The gate is designed to obey
\begin{equation}
 \begin{aligned}
 \bigl(U_v^{(\epsilon)}\bigr)^\dagger Z_{ij}U_v^{(\epsilon)}
   &=(-1)^{\delta_{iv}+\delta_{jv}}Z_{ij},\\
 \bigl(U_v^{(\epsilon)}\bigr)^\dagger\widehat{\mathcal A}_u U_v^{(\epsilon)}
   &=\widehat{\mathcal A}_u^{\,(-1)^{\delta_{uv}}},\\
 \bigl(U_v^{(\epsilon)}\bigr)^\dagger\overline{\mathcal B}_{ijlk}U_v^{(\epsilon)}
   &=\overline{\mathcal B}_{ijlk}^{\,(-1)^{\delta_{iv}}},
 \end{aligned}
 \label{ex:inversion-vertex-hop-checks}
\end{equation}
where $i$ is the common start of
the two oriented plaquette paths. Taking the inverse of a charge or flux operator
leaves its contribution to Eq.~\eqref{ex:inversion-H-main} invariant.

\input{figures/kitaev_chain_flip}

\input{figures/Kitaev}

\paragraph*{Fractionalization interpretation.}
Gauging $\Z_4$ turns the extension label $\epsilon$ into the
fractionalization cocycle $t_\epsilon(D,D)=m^{2\epsilon}$.
The Abelian anyon $t_\epsilon(D,D)$ encodes the fractionalization
of two charge-conjugation operations: its mutual braiding with a test anyon
gives the corresponding composition phase.
The controlled sign in
Eq.~\eqref{ex:inversion-square-flux-main} implements this cocycle.
The two models have the same twist-defect fusion rules at the
level of superselection sectors and different phases associated
with fusion and braiding~\cite{Tarantino2016Symmetry}.

For the closed walls considered here, the wall-composition
comparison at fixed background is represented by transporting
$m^{2\epsilon}$ around the enclosed region. This closed transport
leaves no additional anyon behind, but its braiding with an enclosed
charge $e^q$ gives
\begin{equation}
 M(e^q,m^{2\epsilon})=(-1)^{\epsilon q}.
 \label{ex:inversion-junction}
\end{equation}
In the defect description of Tarantino, Lindner, and
Fidkowski~\cite{Tarantino2016Symmetry}, this relative sign is detected by winding a
charge twice around a charge-conjugation defect, so that its anyon type
returns to the initial one. The process $\mathcal W_{10}$ measures
the same sign using closed walls and cancels the local hopping phases.

In the wall--electric sector, both realizations have the same split
2-group. Charge conjugation acts on $\Z_4$, and the wall associator has
trivial Postnikov class $\beta=0$. In the 2-group description of Benini,
C\'ordova, and Hsin~\cite{Benini2019TwoGroup}, twisted $H^2$ fractionalization data
specify different couplings to the same abstract 2-group. Here,
$\epsilon$ selects the magnetic component of these data, and $\mathcal W_{10}$ distinguishes
its two realizations.

\paragraph*{The mixed anomaly.}
On a spacetime triangulation, $a_{ij}$ records the parity of wall
crossings of edge $ij$, and $b_{ijk}$ records the charge flow modulo
four through triangle $[ijk]$ in the frame at $i$. The split 2-group $\mathbb G$
equations are
\begin{equation}
 \begin{aligned}
 \delta a&=0\pmod2,\\
 (\delta_a b)_{ijkl}
 &=(-1)^{a_{ij}}b_{jkl}-b_{ikl}+b_{ijl}-b_{ijk}
 =0\pmod4 .
 \end{aligned}
 \label{ex:inversion-background-flatness}
\end{equation}
The first equation closes the wall sheets. The second conserves
charge after accounting for its reversal across a wall. Charge
conjugation preserves charge parity. For this 2-group,
$H^4(B\mathbb G,\R/\Z)\cong\Z_8\oplus\Z_2$~\cite{Benini2019TwoGroup}.
The $\Z_8$ factor classifies intrinsic anyon statistics within the
$\Z_4$ charge sector; this component is trivial in both realizations because
the electric charges are bosons. Their distinction lies in the mixed
$\Z_2$ factor, represented by the 4-cocycle
\begin{equation}
 \Omega^{(\epsilon)}=\frac{\epsilon}{2}\,
 a\cup a\cup b\pmod1 .
 \label{ex:inversion-anomaly}
\end{equation}
The factor $1/2$ selects the parity of $b$, coupling the composition
of two walls to electric charge parity. This gives the response
probed by $\mathcal W_{10}$.
Sec.~\ref{dc:sec} explains how an anomaly gives a lattice statistical
phase. Refs.~\cite{Benini2019TwoGroup,Barkeshli2024Higher} discuss the
general higher-group anomaly framework.

%% file: figures/inversion_square_hamiltonian.tex
\begingroup
\definecolor{sqwall}{HTML}{C88A2A}
\definecolor{sqelectric}{HTML}{7561A8}
\definecolor{sqflux}{HTML}{2A8C89}
\definecolor{sqink}{HTML}{30343B}
\definecolor{sqgray}{HTML}{65717C}
\definecolor{sqmesh}{HTML}{BDC6CE}
\begin{tikzpicture}[x=1cm,y=1cm,>=Stealth,
 font=\fontsize{9}{11}\selectfont,line cap=round,line join=round,
 mesh/.style={draw=sqmesh,line width=.5pt},
 dual/.style={draw=sqmesh!75,line width=.45pt,dash pattern=on 2pt off 2.4pt},
 active/.style={line width=1.15pt},
 field/.style={text=sqink,fill=white,inner sep=1.5pt},
 corner/.style={text=sqgray,fill=white,inner sep=1.1pt},
 head/.style={anchor=west,text=sqink},
 edgearrow/.style={-{Stealth[length=3.2pt,width=2.7pt]}}]
 \path[use as bounding box] (0,1.8) rectangle (16.8,7.05);
 \draw[sqmesh!55,line width=.4pt] (5.6,1.8)--(5.6,6.78);
 \draw[sqmesh!55,line width=.4pt] (11.2,1.8)--(11.2,6.78);
 \node[head] at (.14,6.82) {\textbf{(a)}\enspace Edge term: wall};
 \node[head] at (5.74,6.82) {\textbf{(b)}\enspace Vertex term: charge};
 \node[head] at (11.34,6.82) {\textbf{(c)}\enspace Plaquette term: flux};

 \node[text=sqink] at (2.8,5.91) {$-Z_e$};
 \node[text=sqink] at (8.4,5.91)
  {$-\bigl(1+\widehat{\mathcal A}_v+\widehat{\mathcal A}_v^2+\widehat{\mathcal A}_v^\dagger\bigr)$};
 \node[text=sqink] at (15.25,5.91)
  {$-\bigl(1+\overline{\mathcal B}_p+\overline{\mathcal B}_p^2+\overline{\mathcal B}_p^\dagger\bigr)$};

 \begin{scope}[shift={(2.65,3.6)}]
  \foreach \s in {-1.3,0,1.3} {
   \draw[mesh] (-1.65,\s)--(1.65,\s);
   \draw[mesh] (\s,-1.65)--(\s,1.65);
  }
  \foreach \s in {-.65,.65} {
   \draw[dual] (-1.65,\s)--(1.65,\s);
   \draw[dual] (\s,-1.65)--(\s,1.65);
  }
  \foreach \x in {-1.3,0,1.3} {
   \foreach \y in {-1.3,0,1.3} {
    \fill[sqmesh] (\x,\y) circle[radius=1.25pt];
   }
  }
  \foreach \x in {-1.3,0} {
   \foreach \y in {-1.3,0,1.3} {
    \draw[mesh,edgearrow] (\x+.45,\y)--(\x+.82,\y);
   }
  }
  \foreach \x in {-1.3,0,1.3} {
   \foreach \y in {-1.3,0} {
    \draw[mesh,edgearrow] (\x,\y+.45)--(\x,\y+.82);
   }
  }
  \fill[sqwall!8] (-.65,-.65) rectangle (.65,.65);
  \draw[draw=sqink,active,edgearrow] (0,0)--(1.3,0);
  \draw[sqwall,line width=1.25pt,dash pattern=on 4pt off 2pt]
   (-.65,-.65) rectangle (.65,.65);
  \fill[sqink] (0,0) circle[radius=1.8pt];
  \fill[sqink] (1.3,0) circle[radius=1.8pt];
  \node[corner,anchor=north east] at (-.05,-.06) {$i$};
  \node[corner,anchor=north west] at (1.35,-.06) {$j$};
  \node[field,anchor=south] at (.87,.1) {$e$};
  \node[text=sqwall,anchor=south,fill=white,inner sep=1.5pt]
   at (0,.78) {closed wall};
 \end{scope}
 \begin{scope}[shift={(8.4,3.6)}]
  \foreach \s in {-1.3,0,1.3} {
   \draw[mesh] (-1.65,\s)--(1.65,\s);
   \draw[mesh] (\s,-1.65)--(\s,1.65);
  }
  \foreach \s in {-.65,.65} {
   \draw[dual] (-1.65,\s)--(1.65,\s);
   \draw[dual] (\s,-1.65)--(\s,1.65);
  }
  \foreach \x in {-1.3,0,1.3} {
   \foreach \y in {-1.3,0,1.3} {
    \fill[sqmesh] (\x,\y) circle[radius=1.25pt];
   }
  }
  \foreach \x in {-1.3,0} {
   \foreach \y in {-1.3,1.3} {
    \draw[mesh,edgearrow] (\x+.45,\y)--(\x+.82,\y);
   }
  }
  \foreach \x in {-1.3,1.3} {
   \foreach \y in {-1.3,0} {
    \draw[mesh,edgearrow] (\x,\y+.45)--(\x,\y+.82);
   }
  }
  \draw[sqelectric,active,edgearrow] (-1.3,0)--(0,0);
  \draw[sqelectric,active,edgearrow] (0,0)--(1.3,0);
  \draw[sqelectric,active,edgearrow] (0,-1.3)--(0,0);
  \draw[sqelectric,active,edgearrow] (0,0)--(0,1.3);
  \fill[sqelectric] (0,0) circle[radius=2.6pt];
  \node[corner,anchor=north west] at (.08,-.08) {$v$};
  \node[corner,anchor=east] at (-1.37,0) {$i$};
  \node[corner,anchor=west] at (1.37,0) {$j$};
  \node[corner,anchor=north] at (0,-1.37) {$k$};
  \node[corner,anchor=south] at (0,1.37) {$l$};
  \node[field,anchor=east] at (-.20,.26) {$\widetilde X_{iv}^{Z_{iv}}$};
  \node[field,anchor=south] at (.7,.1) {$\widetilde X_{vj}^{-1}$};
  \node[field,anchor=west] at (.1,-.82) {$\widetilde X_{kv}^{Z_{kv}}$};
  \node[field,anchor=west] at (.1,.95) {$\widetilde X_{vl}^{-1}$};
 \end{scope}
 \begin{scope}[shift={(15.38,3.6)}]
  \foreach \s in {-.65,.65} {
   \draw[mesh] (-1.65,\s)--(1.42,\s);
   \draw[mesh] (\s,-1.65)--(\s,1.65);
  }
  \draw[dual] (-.65,0)--(1.42,0);
  \draw[dual] (0,-1.65)--(0,1.65);
  \fill[sqflux!7] (-.65,-.65) rectangle (.65,.65);
  \draw[sqflux,active,edgearrow] (-.65,-.65)--(.65,-.65);
  \draw[sqflux,active,edgearrow] (.65,-.65)--(.65,.65);
  \draw[sqflux,active,edgearrow] (-.65,.65)--(.65,.65);
  \draw[sqflux,active,edgearrow] (-.65,-.65)--(-.65,.65);
  \foreach \x in {-.65,.65} {
   \foreach \y in {-.65,.65} {
    \fill[sqink] (\x,\y) circle[radius=1.8pt];
   }
  }
  \node[corner,anchor=north east] at (-.7,-.7) {$i$};
  \node[corner,anchor=north west] at (.7,-.7) {$j$};
  \node[corner,anchor=south east] at (-.7,.7) {$k$};
  \node[corner,anchor=south west] at (.7,.7) {$l$};
  \node[field,anchor=north] at (0,-.85) {$\widetilde Z_{ij}$};
  \node[field,anchor=south] at (0,.87) {$\widetilde Z_{kl}^{-Z_{ik}}$};
  \node[field,anchor=base east] (leftflux) at (-.74,-.08) {$\widetilde Z_{ik}^{-1}$};
  \node[field,anchor=base west] at (.86,-.08) {$\widetilde Z_{jl}^{Z_{ij}}$};
  \node[text=sqink,anchor=base east,inner sep=0pt] at ([xshift=0pt]leftflux.base west)
   {$\bigl(\mathrm{CZ}_{ij,jl}\mathrm{CZ}_{ik,kl}\bigr)^\epsilon\times$};
 \end{scope}
\end{tikzpicture}
\endgroup

%% file: figures/kitaev_chain_flip.tex
\begin{figure*}[t]
\centering
\begingroup
\definecolor{ftcInner}{HTML}{258BD2}
\definecolor{ftcPoint}{HTML}{7561A8}
\definecolor{ftcMesh}{HTML}{A9AFB6}
\definecolor{ftcMove}{HTML}{B5243F}
\begin{tikzpicture}[
  x=1cm,y=1cm,line cap=round,line join=round,
  font=\fontsize{9}{10.5}\selectfont,
  ftc mesh/.style={draw=ftcMesh,line width=.4pt,
                  dash pattern=on 1.5pt off 1.2pt},
  ftc outer/.style={draw=black,line width=1.2pt},
  ftc inner/.style={draw=ftcInner,line width=1.25pt},
  ftc point/.style={draw=white,fill=ftcPoint,line width=.45pt},
  ftc label/.style={fill=white,inner sep=.5pt,font=\fontsize{8}{9}\selectfont},
  ftc key/.style={anchor=west,inner sep=0pt},
  ftc transition/.style={->,draw=black!65,line width=.55pt},
  ftc operator/.style={text=ftcMove,inner sep=1pt},
]
  \makeatletter
  \newcommand{\ftcIfOccupied}[3]{%
    \edef\ftcTestEdge{\noexpand\in@{,#2,}{,#1,}}%
    \ftcTestEdge
    \ifin@ #3\fi
  }
  \makeatother
  \newcommand{\ftcBridge}[3]{%
    \draw[white,line width=2.6pt]
      ($(#1)!1.2pt!(#2)$)--($(#1)!1.2pt!(#3)$);
  }
  \newcommand{\ftcOrientation}[1]{%
    \foreach \u/\v/\t in {#1}{%
      \def\ftcLoopStyle{ftc outer}%
      \ifnum\u=4\relax\def\ftcLoopStyle{ftc inner}\fi
      \ifnum\v=4\relax\def\ftcLoopStyle{ftc inner}\fi
      \coordinate (loopArrow) at ($(p\u)!\t!(p\v)$);
      \draw[\ftcLoopStyle,-{Stealth[length=4.2pt,width=4.2pt]}]
        ($(loopArrow)!2.8pt!(p\u)$)--($(loopArrow)!2.8pt!(p\v)$);
    }
  }
  \newcommand{\ftcPanel}[7]{%
    \begin{scope}[shift={(#1,#2)},x=.94cm,y=.94cm]
      \coordinate (p0) at (.37,-.59);
      \coordinate (p1) at (.80,-.04);
      \coordinate (p2) at (-.78,-.04);
      \coordinate (p3) at (-.04,.77);
      \coordinate (p4) at (-.09,.17);
      \coordinate (cross0312) at (intersection of p0--p3 and p1--p2);
      \coordinate (cross0314) at (intersection of p0--p3 and p1--p4);
      \coordinate (cross0412) at (intersection of p0--p4 and p1--p2);
      \foreach \u/\v in {0/1,0/2,0/3,1/2,1/3,2/3,0/4,1/4,2/4,3/4}{%
        \draw[ftc mesh] (p\u)--(p\v);
      }
      \foreach \u/\v in {1/2,0/1,0/2,1/3,2/3,1/4,2/4,3/4,0/4,0/3}{%
        \ftcIfOccupied{#5}{\u/\v}{%
          \ifnum\u=0\relax
            \ifnum\v=3\relax
              \ftcBridge{cross0312}{p0}{p3}
              \ftcBridge{cross0314}{p0}{p3}
            \fi
            \ifnum\v=4\relax
              \ftcBridge{cross0412}{p0}{p4}
            \fi
          \fi
          \ifnum\v=4\relax
            \draw[ftc inner] (p\u)--(p\v);
          \else
            \draw[ftc outer] (p\u)--(p\v);
          \fi
        }
      }
      \foreach \v in {0,...,4}{\fill[ftcMesh] (p\v) circle[radius=.8pt];}
      \foreach \v in {#6}{\filldraw[ftc point] (p\v) circle[radius=3.2pt];}
      \ifnum#4=1\relax
        \node[ftc label,anchor=north] at ([yshift=-2pt]p0) {$0$};
        \node[ftc label,anchor=west] at ([xshift=2pt]p1) {$1$};
        \node[ftc label,anchor=east] at ([xshift=-2pt]p2) {$2$};
        \node[ftc label,anchor=south] at ([yshift=2pt]p3) {$3$};
        \node[ftc label,anchor=south east] at ([xshift=-1pt,yshift=1pt]p4) {$4$};
      \else
        \ftcOrientation{#7}%
      \fi
      \node[inner sep=0pt] at (0,-.89) {\textbf{#3}};
    \end{scope}
  }
  \newcommand{\ftcStep}[5]{%
    \pgfmathtruncatemacro{\ftcColumn}{mod(#1,8)}%
    \pgfmathsetmacro{\ftcX}{.98+2.04*\ftcColumn}%
    \pgfmathsetmacro{\ftcY}{5.82-2.32*floor(#1/8)}%
    \pgfmathsetmacro{\ftcArrowX}{\ftcX-.53}%
    \pgfmathsetmacro{\ftcArrowY}{\ftcY+.32}%
    \ifnum\ftcColumn=0\relax
      \def\ftcArrowX{16.77}%
      \pgfmathsetmacro{\ftcArrowY}{\ftcY+2.64}%
    \fi
    \draw[ftc transition] (\ftcArrowX-.97,\ftcArrowY)
      -- node[ftc operator,above=4.5pt] {$#2$} (\ftcArrowX,\ftcArrowY);
    \ifnum#1<24\relax
      \ftcPanel{\ftcX}{\ftcY}{#1}{0}{#3}{#4}{#5}%
    \fi
  }
  \ftcPanel{.98}{8.10}{}{1}{}{}{}
  \ftcPanel{.98}{5.82}{0}{0}{0/1,0/3,1/2,2/3}{}{0/1/.49,1/2/.70,2/3/.51,3/0/.28}
  \ftcStep{1}{M_{234}^{-1}}{0/1,0/3,1/2,2/4,3/4}{3}{0/1/.49}
  \ftcStep{2}{M_{124}^{-1}}{0/1,0/3,1/4,3/4}{3}{0/1/.49}
  \ftcStep{3}{M_{134}}{0/1,0/3,1/3}{1,3}{0/1/.49}
  \ftcStep{4}{M_{014}^{-1}}{0/3,0/4,1/3,1/4}{1,3}{1/3/.52}
  \ftcStep{5}{M_{034}^{-1}}{1/3,1/4,3/4}{1,3}{3/4/.52}
  \ftcStep{6}{M_{124}}{1/2,1/3,2/4,3/4}{1,3}{3/4/.52}
  \ftcStep{7}{M_{024}}{0/2,0/4,1/2,1/3,3/4}{1,3}{3/4/.52}

  \ftcStep{8}{M_{034}}{0/2,0/3,1/2,1/3}{1,3}{1/3/.52}
  \ftcStep{9}{M_{124}^{-1}}{0/2,0/3,1/3,1/4,2/4}{1,3}{1/3/.52}
  \ftcStep{10}{M_{134}^{-1}}{0/2,0/3,2/4,3/4}{1,3}{0/2/.48}
  \ftcStep{11}{M_{234}}{0/2,0/3,2/3}{1,2}{0/2/.48}
  \ftcStep{12}{M_{034}^{-1}}{0/2,0/4,2/3,3/4}{1,2}{0/2/.48}
  \ftcStep{13}{M_{024}^{-1}}{2/3,2/4,3/4}{1,2}{2/3/.63}
  \ftcStep{14}{M_{134}}{1/3,1/4,2/3,2/4}{1,2}{2/3/.63}
  \ftcStep{15}{M_{014}}{0/1,0/4,1/3,2/3,2/4}{1,2}{2/3/.63}

  \ftcStep{16}{M_{024}}{0/1,0/2,1/3,2/3}{1,2}{1/0/.51}
  \ftcStep{17}{M_{134}^{-1}}{0/1,0/2,1/4,2/3,3/4}{2}{1/0/.51}
  \ftcStep{18}{M_{234}^{-1}}{0/1,0/2,1/4,2/4}{}{1/0/.51}
  \ftcStep{19}{M_{124}}{0/1,0/2,1/2}{}{2/1/.30}
  \ftcStep{20}{M_{024}^{-1}}{0/1,0/4,1/2,2/4}{}{4/2/.49}
  \ftcStep{21}{M_{014}^{-1}}{1/2,1/4,2/4}{}{4/2/.49}
  \ftcStep{22}{M_{234}}{1/2,1/4,2/3,3/4}{}{3/2/.37}
  \ftcStep{23}{M_{034}}{0/3,0/4,1/2,1/4,2/3}{}{3/2/.37}
  \ftcStep{24}{M_{014}}{0/1,0/3,1/2,2/3}{}{}
  \draw[ftc mesh] (2.88,8.10)--(3.31,8.10);
  \node[ftc key] at (3.43,8.10) {Unoccupied edge};
  \draw[ftc outer] (6.39,8.10)--(6.82,8.10);
  \node[ftc key] at (6.94,8.10) {Loop on $\langle 0123 \rangle$};
  \draw[ftc inner] (9.67,8.10)--(10.10,8.10);
  \node[ftc key] at (10.22,8.10) {Loop on $\langle i4 \rangle$};
  \filldraw[ftc point] (13.05,8.10) circle[radius=3.2pt];
  \node[ftc key] at (13.26,8.10) {Fermion $\psi$};
  \path[use as bounding box] (0,.10) rectangle (17.05,9.56);
\end{tikzpicture}
\endgroup
\caption{The 24-step Kitaev-chain flip $\mathcal K_{24}$ in
Eq.~\eqref{eq:ftc-word}. Vertex $4$ lies inside the tetrahedron
$\langle0123\rangle$. Dashed gray edges are unoccupied; black and
blue segments carry the $\Z_2$ string, and purple dots mark
$\Z_2$ particles. The string and particle occupations obey the
3-group constraints in Eq.~\eqref{eq:ftc-parent-constraint}. A Kitaev
chain and physical fermions provide one microscopic realization.
Read from left to right, row by row. Each labeled membrane operator
toggles the three edges of its triangle and updates the particle
occupations according to Fig.~\ref{fig:eq14-framing}. Fermion parity is exchanged between the Kitaev chain and physical fermions while their combined parity is conserved.
Arrowheads follow an initial choice of loop direction through the
deformations. The final move, $M_{014}$, restores the edge and
particle occupations of panel~$0$ with the loop direction reversed.
The accumulated phase is $+i$ in the microscopic realization.
This statistical phase is quantized to a fourth root of unity:
four repetitions have phase $+1$ by the locality identities
introduced in Sec.~\ref{loc:sec}.
}
\label{fig:kitaev-process}
\end{figure*}

%% file: figures/Kitaev.tex
\begin{figure*}[t]
\centering
\begingroup
\definecolor{framefixed}{HTML}{69747E}
\definecolor{framemoved}{HTML}{2B7F9C}
\definecolor{frameface}{HTML}{70AABD}
\definecolor{framehit}{HTML}{B94E24}
\definecolor{frameink}{HTML}{172934}
\def\FrameShift{0.20}
\pgfmathsetmacro{\FrameCross}{1-\FrameShift}
\newcommand{\FrameCoordinates}{%
  \coordinate (O0) at (-1,{-1/sqrt(3)},{-1/sqrt(6)});
  \coordinate (O1) at (1,{-1/sqrt(3)},{-1/sqrt(6)});
  \coordinate (O2) at (0,{2/sqrt(3)},{-1/sqrt(6)});
  \coordinate (O3) at (0,0,{3/sqrt(6)});
  \coordinate (O4) at (0,0,0);
  \coordinate (D) at (-\FrameShift,-\FrameShift,-\FrameShift);
  \foreach \i in {0,...,4}{\coordinate (T\i) at ($(O\i)+(D)$);}
  \coordinate (P2) at ($(O0)!\FrameCross!(O2)$);
  \coordinate (P3) at ($(O0)!\FrameCross!(O3)$);
  \coordinate (P4) at ($(O0)!\FrameCross!(O4)$);
  \coordinate (Q4) at (-\FrameShift,-\FrameShift,{-\FrameShift/sqrt(2)});
  \pgfmathsetmacro{\FrameLeft}{-1+\FrameShift*(sqrt(3)-1)}
  \pgfmathsetmacro{\FrameRight}{1-\FrameShift*(sqrt(3)+1)}
  \coordinate (R0) at (\FrameLeft,{-1/sqrt(3)},{-1/sqrt(6)});
  \coordinate (R1) at (\FrameRight,{-1/sqrt(3)},{-1/sqrt(6)});
}
\newcommand{\FrameSkeleton}{%
  \foreach \i/\j in {0/1,0/2,0/3,1/2,1/3,2/3}{%
    \draw[fixed skeleton] (O\i)--(O\j);
    \draw[moved skeleton] (T\i)--(T\j);
  }
  \foreach \i in {0,...,3}{%
    \draw[fixed skeleton,densely dashed] (O\i)--(O4);
    \draw[moved skeleton,densely dashed] (T\i)--(T4);
  }
}
\newcommand{\FrameFixedLabels}{%
  \foreach \i/\where in {0/below left,1/right,2/above left,3/above}{%
    \fill[framefixed] (O\i) circle[radius=.8pt];
    \node[vertex label,text=framefixed,\where] at (O\i) {$\i$};
  }
  \fill[framefixed] (O4) circle[radius=.8pt];
  \draw[label leader] (O4)--(2.65cm,.5cm);
  \node[exterior label,text=framefixed,right] at (2.65cm,.5cm) {$4$};
}
\newcommand{\FrameMovedLabels}[1]{%
  \foreach \i/\where in {#1}{%
    \fill[framemoved] (T\i) circle[radius=.8pt];
    \ifnum\i=4
      \draw[label leader] (T4)--(2.65cm,-1.0cm);
      \node[exterior label,text=framemoved,right] at (2.65cm,-1.0cm) {$4'$};
    \else
      \node[vertex label,text=framemoved,\where] at (T\i) {$\i'$};
    \fi
  }
}
\newcommand{\FrameDot}[2]{%
  \filldraw[fill=framehit,draw=white,line width=.6pt] (#1) circle[radius=2.2pt];
  \ifcase#2
    \def\CalloutX{-2.15}\def\CalloutY{-1.6}%
  \or
    \def\CalloutX{2.65}\def\CalloutY{-1.98}%
  \or
    \def\CalloutX{-2.15}\def\CalloutY{.45}%
  \or
    \def\CalloutX{1.25}\def\CalloutY{2.0}%
  \or
    \def\CalloutX{2.65}\def\CalloutY{-.25}%
  \fi
  \draw[label leader,draw=framehit!65] (#1)--(\CalloutX cm,\CalloutY cm);
  \node[exterior label,text=framehit] at (\CalloutX cm,\CalloutY cm) {$[#2]$};
}
\resizebox{0.90\textwidth}{!}{%
\begin{tikzpicture}[
  x={(1.567926484cm,.364968053cm)},
  y={(-1.320646259cm,.433305340cm)},
  z={(0cm,1.970163649cm)},
  line cap=round,line join=round,
  fixed skeleton/.style={draw=framefixed!38,line width=.45pt},
  moved skeleton/.style={draw=framemoved!30,line width=.45pt},
  membrane/.style={line width=1.35pt},
  crossing edge/.style={draw=frameink,line width=1.55pt},
  vertex label/.style={font=\normalsize,inner sep=1pt},
  exterior label/.style={font=\normalsize,fill=white,inner sep=2pt},
  label leader/.style={draw=framefixed!55,line width=.35pt},
  every node/.style={text=frameink}
]
\begin{scope}
  \FrameCoordinates
  \FrameSkeleton
  \fill[frameface,opacity=.18] (T2)--(T3)--(T4)--cycle;
  \draw[membrane,framemoved] (T2)--(T3)--(T4)--cycle;
  \foreach \j in {2,3,4}{\draw[crossing edge] (O0)--(O\j);}
  \FrameFixedLabels
  \FrameMovedLabels{2/left,3/above left,4/below left}
  \foreach \j in {2,3,4}{\FrameDot{P\j}{\j}}
  \node[font=\large] at (-2.2cm,3.15cm) {(a)};
  \node[font=\large] at (0cm,3.15cm) {$M_{234}$};
  \node[font=\normalsize] at (0cm,-2.6cm) {$\{02,03,04\}\cap(234)'$};
  \node at (0cm,-3.08cm) {$\Delta f=\ell_{02}[2]+\ell_{03}[3]+\ell_{04}[4]$};
\end{scope}

\begin{scope}[xshift=6.7cm]
  \FrameCoordinates
  \FrameSkeleton
  \fill[frameface,opacity=.18] (T1)--(T3)--(T4)--cycle;
  \draw[membrane,framemoved] (T1)--(T3)--(T4)--cycle;
  \draw[crossing edge] (O0)--(O1);
  \FrameFixedLabels
  \FrameMovedLabels{1/below right,3/above left,4/below left}
  \FrameDot{R1}{1}
  \node[font=\large] at (-2.2cm,3.15cm) {(b)};
  \node[font=\large] at (0cm,3.15cm) {$M_{134}$};
  \node[font=\normalsize] at (0cm,-2.6cm) {$01\cap(134)'$};
  \node at (0cm,-3.08cm) {$\Delta f=\ell_{01}[1]$};
\end{scope}

\begin{scope}[xshift=13.4cm]
  \FrameCoordinates
  \FrameSkeleton
  \fill[frameface,opacity=.18] (T0)--(T3)--(T4)--cycle;
  \draw[membrane,framemoved] (T0)--(T3)--(T4)--cycle;
  \draw[crossing edge] (O0)--(O1);
  \FrameFixedLabels
  \FrameMovedLabels{0/below left,3/above left,4/below left}
  \FrameDot{R0}{0}
  \node[font=\large] at (-2.2cm,3.15cm) {(c)};
  \node[font=\large] at (0cm,3.15cm) {$M_{034}$};
  \node[font=\normalsize] at (0cm,-2.6cm) {$01\cap(034)'$};
  \node at (0cm,-3.08cm) {$\Delta f=\ell_{01}[0]$};
\end{scope}

\begin{scope}[yshift=-7.0cm]
  \FrameCoordinates
  \FrameSkeleton
  \fill[frameface,opacity=.18] (O0)--(O1)--(O4)--cycle;
  \draw[membrane,framefixed] (O0)--(O1)--(O4)--cycle;
  \draw[crossing edge] (T3)--(T4);
  \FrameFixedLabels
  \FrameMovedLabels{3/above left,4/below left}
  \FrameDot{Q4}{4}
  \node[font=\large] at (-2.2cm,3.15cm) {(d)};
  \node[font=\large] at (0cm,3.15cm) {$M_{014}$};
  \node[font=\normalsize] at (0cm,-2.6cm) {$(34)'\cap014$};
  \node at (0cm,-3.08cm) {$\Delta f=\ell_{34}[4]$};
\end{scope}

\begin{scope}[xshift=6.7cm,yshift=-7.0cm]
  \FrameCoordinates
  \FrameSkeleton
  \fill[frameface,opacity=.15] (O0)--(O2)--(O4)--cycle;
  \fill[frameface,opacity=.15] (T0)--(T2)--(T4)--cycle;
  \draw[membrane,framefixed] (O0)--(O2)--(O4)--cycle;
  \draw[membrane,framemoved] (T0)--(T2)--(T4)--cycle;
  \FrameFixedLabels
  \FrameMovedLabels{0/below left,2/left,4/below left}
  \node[font=\large] at (-2.2cm,3.15cm) {(e)};
  \node[font=\large] at (0cm,3.15cm) {$M_{024}$};
  \node[font=\normalsize] at (0cm,-2.6cm) {$E\cap(024)'=E'\cap024=\varnothing$};
  \node at (0cm,-3.08cm) {$\Delta f=0$};
\end{scope}

\begin{scope}[xshift=13.4cm,yshift=-7.0cm]
  \FrameCoordinates
  \FrameSkeleton
  \fill[frameface,opacity=.15] (O1)--(O2)--(O4)--cycle;
  \fill[frameface,opacity=.15] (T1)--(T2)--(T4)--cycle;
  \draw[membrane,framefixed] (O1)--(O2)--(O4)--cycle;
  \draw[membrane,framemoved] (T1)--(T2)--(T4)--cycle;
  \FrameFixedLabels
  \FrameMovedLabels{1/below right,2/left,4/below left}
  \node[font=\large] at (-2.2cm,3.15cm) {(f)};
  \node[font=\large] at (0cm,3.15cm) {$M_{124}$};
  \node[font=\normalsize] at (0cm,-2.6cm) {$E\cap(124)'=E'\cap124=\varnothing$};
  \node at (0cm,-3.08cm) {$\Delta f=0$};
\end{scope}
\begin{scope}[xshift=-2.15cm,yshift=-10.86cm]
  \draw[framefixed,line width=.7pt] (0cm,0cm)--(.48cm,0cm);
  \node[anchor=west,inner sep=0pt] at (.62cm,0cm) {Original complex};
  \draw[framemoved,line width=.7pt] (4.60cm,0cm)--(5.08cm,0cm);
  \node[anchor=west,inner sep=0pt] at (5.22cm,0cm) {Translated copy};
  \draw[crossing edge] (9.20cm,0cm)--(9.68cm,0cm);
  \node[anchor=west,inner sep=0pt] at (9.82cm,0cm) {Intersecting edge};
  \fill[framehit] (14.10cm,0cm) circle[radius=2.2pt];
  \node[anchor=west,inner sep=0pt] at (14.35cm,0cm) {Intersection};
\end{scope}
\end{tikzpicture}
}%
\endgroup
\caption{Fermion updates for the six Kitaev-chain membrane moves in
  Eq.~\eqref{eq:ftc-direct}. Gray and blue show the original physical
  lattice and its spatially translated copy, respectively; primed
  labels refer to the copy, and vertex $4$ lies inside tetrahedron
  $\langle0123\rangle$. The translated membrane event is placed
  slightly later within the same hop to frame the chain worldsheet.
  Each panel shows a membrane's intersections with the other copy's
  edges in that membrane's time slice. Dark segments identify the
  edges that can control a fermion update; orange dots mark their
  intersections with the membrane.
  An occupied crossing edge toggles the fermion occupation at the
  vertex approached as the displacement tends to zero. Here
  $\ell_{uv}$ records the chain occupation of edge $uv$, and $[v]$
  denotes a toggle of the fermion occupation at vertex $v$.
  In (a), an occupied edge $03$
  crosses $(234)'$ and toggles the fermion at $3$. In (e) and (f),
  neither membrane crosses the other copy's edges, so $\Delta f=0$.
  The sets $E$ and $E'$ contain all edges of the original lattice and
  translated copy, respectively.}
\label{fig:eq14-framing}
\end{figure*}

%% file: sections/microscopic_fermionic_toric.tex
A closed Kitaev chain on a line defect in a $(3+1)$D fermionic
superconductor~\cite{Barkeshli2023Codimension} can exchange fermion parity
with physical fermions as it deforms. We construct a process
$\mathcal K_{24}$ that flips a single closed chain, restores all
point occupations, and has phase $+i$.
We first describe the configurations and
membrane updates, then follow the process and calculate its phase
microscopically. Finally, we relate this phase to the 3-group anomaly.

\paragraph*{Configurations and parity.}
The physical lattice in Fig.~\ref{fig:kitaev-process} has five
vertices labeled $0,\ldots,4$. Let $\ell_{uv}=1$ when edge $uv$
carries the chain and $f_v=1$ when vertex $v$ carries a physical fermion.
An even number of occupied edges meet at each vertex, so the chain
has no ends. The allowed configurations also satisfy
\begin{equation}
 \sum_v f_v=\ell_{01}\ell_{34}\pmod2.
 \label{eq:ftc-parity}
\end{equation}
In the Majorana realization below, the chain parity relative to the
empty-edge vacuum is $(-1)^{\ell_{01}\ell_{34}}$. This constraint
therefore fixes the combined chain and point parity to be even.
As the chain deforms, its parity can change together with the total
point parity. The local membrane rule determines which point
occupations change.

To relate these physical rules to a higher group, regard the
physical lattice as dual to the boundary of the ordered auxiliary
4-simplex $\langle0'1'2'3'4'\rangle$. Physical edge $uv$ is dual to
the auxiliary triangle omitting $u'$ and $v'$, and physical vertex
$v$ is dual to the tetrahedron omitting $v'$. The edge and point
occupations thus give binary cochains $n_2$ and $n_3$, respectively,
as in the decoration constructions of
Refs.~\cite{Gu2014Symmetry,Wang2018Towards,Wang2020Construction,Chen2021Disentangling,Kapustin2017Fermionic}.
Their constraints have the 3-group form
\begin{equation}
 \delta n_2=0,\qquad \delta n_3=n_2\cup n_2.
 \label{eq:ftc-parent-constraint}
\end{equation}
Here, $\pi_2=\pi_3=\Z_2$ and $\pi_1$ is trivial. The first equation
closes the chain. Evaluating the second on the auxiliary filling
4-simplex gives Eq.~\eqref{eq:ftc-parity}. The coboundary $\delta n_3$
sums the five point occupations, and $n_2\cup n_2$ multiplies the labels
on $\langle0'1'2'\rangle$ and $\langle2'3'4'\rangle$, dual to
physical edges $34$ and $01$.

\raggedbottom
\paragraph*{Membrane updates.}
A membrane $M_{ijk}$ toggles the three edges of triangle
$\langle ijk\rangle$ and can also toggle point occupations at its
vertices.{\interfootnotelinepenalty=10000
\postdisplaypenalty=10000
\footnote{\label{fn:ftc-cochain-updates}Mathematically, membrane operators act by
\begin{equation}
 \begin{aligned}
 n_2&\mapsto n_2+\delta\lambda,\\
 n_3&\mapsto n_3+n_2\cup\lambda+\lambda\cup n_2
                 +\lambda\cup\delta\lambda,
 \end{aligned}
\end{equation}
for a 1-cochain $\lambda$.
A fermion hop sends $n_3\mapsto n_3+\delta\gamma$ for a 2-cochain $\gamma$.
}} The point changes are controlled by occupied edges
outside the triangle. These controlling edges remain unchanged
through the hop, so the inverse has the same occupation update.

For $M_{234}$, the controlling edges are $02$, $03$, and $04$.
Each occupied edge toggles the point occupation at its endpoint on
the membrane. For example, the first move of
Fig.~\ref{fig:kitaev-process}, $M_{234}^{-1}$, replaces edge $23$
with edges $24$ and $34$. Only $03$ among the controlling edges
is occupied, so this move creates a physical fermion $\psi$ at vertex
$3$. The chain parity changes at the same time, preserving the
combined parity. For the six membranes used in the process, the
point updates are
\begin{equation}
\begin{array}{c|l}
 \text{membrane} & \text{particle changes} \\ \hline
 M_{234} & \ell_{02}[2]+\ell_{03}[3]+\ell_{04}[4] \\
 M_{134} & \ell_{01}[1] \\
 M_{034} & \ell_{01}[0] \\
 M_{014} & \ell_{34}[4] \\
 M_{024},\ M_{124} & 0
\end{array}
\label{eq:ftc-direct}
\end{equation}
\flushbottom
where $[v]$ denotes a toggle of $f_v$, and the changes add modulo
two. \mbox{These updates preserve Eq.~\eqref{eq:ftc-parent-constraint}.}

Fig.~\ref{fig:eq14-framing} gives a geometric way to read the
controlling edges. In four-dimensional spacetime, the membrane
triangle joins the incoming and outgoing chain worldsheets at the
hop. To frame the worldsheet, translate a copy slightly in space
and place its membrane at a slightly later time within the same
hop interval~\cite{Tata2020Geometrically,Chen2023Higher}. The original membrane can intersect incoming edges
of the translated copy, and the translated membrane can intersect
outgoing edges of the original. Each intersection lies in its
membrane's time slice and can be read in the spatial figure.
An occupied crossing edge toggles the point occupation at the
vertex the crossing approaches as the spatial displacement tends
to zero. The primed labels in Fig.~\ref{fig:eq14-framing} belong
to the translated physical copy. The primes on the auxiliary
simplex above label a separate set of vertices.

\paragraph*{The process $\mathcal K_{24}$.}
Applying $M_{012}$ and then $M_{023}$ to the vacuum prepares a
loop on edges $01,03,12,23$ with empty point modes. Starting from
this configuration, shown in panel $0$ of
Fig.~\ref{fig:kitaev-process}, apply the following word from right
to left.
\begin{equation}
\begin{aligned}
 \mathcal K_{24}&=~M_{014}M_{034}M_{234}M_{014}^{-1}M_{024}^{-1}M_{124}M_{234}^{-1}M_{134}^{-1}\\
 &~\times M_{024}M_{014}M_{134}M_{024}^{-1}M_{034}^{-1}M_{234}M_{134}^{-1}M_{124}^{-1}\\
 &~\times M_{034}M_{024}M_{124}M_{034}^{-1}M_{014}^{-1}M_{134}M_{124}^{-1}M_{234}^{-1}.
\end{aligned}
\label{eq:ftc-word}
\end{equation}
The three blocks of eight moves are related by the permutation
$(1\,3\,2)$ of the membrane labels and loop edges. Every
intermediate chain is a single closed loop; crossings in the
drawing arise from its projection onto the page. By
Eq.~\eqref{eq:ftc-direct}, the occupied point sites change only at
the steps indicated below,
\nopagebreak[4]
\begin{equation*}
 \varnothing\xrightarrow{1}\{3\}\xrightarrow{3}\{1,3\}
 \xrightarrow{11}\{1,2\}\xrightarrow{17}\{2\}
 \xrightarrow{18}\varnothing .
\end{equation*}
At step 11, for example, both $02$ and $03$ are occupied.
The membrane $M_{234}$ therefore toggles sites $2$ and $3$,
while site $1$ remains occupied. No separate fermion hopping
operator is used.

The final move restores the occupied edges, and the point modes
are again empty. To identify the flip, orient the initial loop
along $0\to1\to2\to3\to0$ and carry this direction through each
deformation, retaining it on unchanged edges. The restored loop
then runs along $0\to3\to2\to1\to0$. These arrows track the
deformation; the configuration labels $\ell_{uv}$ remain binary.
The loop follows the trajectory of Ref.~\cite{Kobayashi2026Generalized}
with $0$ and $4$ interchanged.

At each vertex of a membrane, the local phase shift of a forward
hop cancels against that of an inverse hop with matching chain and
point occupations. The pairing can differ from vertex to vertex.
Table~\ref{tab:ftc-local-pairs} in Appendix~\ref{app:ftc-calculation} gives all the pairings.
To base the process at the vacuum, undo the preparation after step 24. The
preparation and its inverse cancel in the signed local counts.

\paragraph*{Microscopic realization.}
Moving a Kitaev chain changes its Majorana pairing pattern.
When the incoming and outgoing dimer states have different fermion
parities, projection alone has zero matrix element between them.
The Wang-Gu construction resolves this in the fermionic
$F$ moves by combining dimer projections with fermionic insertions
that preserve the total parity~\cite{Wang2018Towards}. Their $F$ moves change the
triangulation. We adapt the normalized, parity-corrected projection
to the deformation $\ell\mapsto\ell+\partial s$ on a fixed physical
lattice, with the particle changes in Eq.~\eqref{eq:ftc-direct}.

An orthogonal register $\ket\ell$ records the chain geometry.
Each edge $uv$ also carries two Majoranas, $\gamma_{u,v}$ at $u$ and
$\gamma_{v,u}$ at $v$. On an empty edge they pair across the
edge; on occupied edges they pair at their common vertices to
form the Kitaev chain. An arrow $a\to b$ specifies the dimer
condition $-i\gamma_a\gamma_b=1$, selected by
$(1-i\gamma_a\gamma_b)/2$. Appendix~\ref{app:ftc-calculation}
fixes these arrows and the Fock basis for our cylinder orientation.

For a triangle $s$, let $V_s(\ell)$ be the vertices whose point
occupations change. At each $v\in V_s(\ell)$, $u(v)$ is the
other endpoint of the occupied edge controlling that change. The point mode $c_v$
supplies $\chi_v=c_v+c_v^\dagger$. The even bilinear
$-i\chi_v\gamma_{v,u(v)}$ transfers parity between the point mode
and the chain: it flips both parities and preserves their product.
The bilinears at distinct vertices commute. With this grouping,
the membrane is
\begin{equation}
\begin{aligned}
 &M_s=\\
 &\sum_\ell\ket{\ell+\partial s}\!\bra{\ell}\otimes
 \mathcal N_s(\ell)
P_{\rm out}
 \left[\prod_{v\in V_s(\ell)}
 \bigl(-i\chi_v\gamma_{v,u(v)}\bigr)\right]P_{\rm in}.
\end{aligned}
\label{eq:ftc-membrane-main}
\end{equation}
The sum runs over closed even networks with the pairings of
Appendix~\ref{app:ftc-calculation}; $\ell+\partial s$ toggles the
three edges of $s$. The tensor product separates the geometry
register from the fermionic modes. For each $s,\ell$, the
projectors $P_{\rm in}$ and $P_{\rm out}$ select the incoming and
outgoing pairings on the Majoranas affected by the move.
On an allowed incoming state, the operator
applies the parity transfers, projects onto the outgoing pairing,
and updates the loop label. The remaining dimers are unchanged.
The dimers changed by a membrane form one alternating cycle of
$2L_s$ Majoranas in the two superimposed matchings. The
parity-corrected projection has norm $2^{-(L_s-1)/2}$, so the
positive normalization factor is $\mathcal N_s(\ell)=2^{(L_s-1)/2}$. The normalized membrane maps each
allowed configuration state to its prescribed successor with a
unit-modulus amplitude. Since the update is invertible, its
restriction to the allowed configuration sector is unitary.

Appendix~\ref{app:ftc-calculation} evaluates the product of membrane operators along the closed process, giving
\begin{equation}
 \mathcal K_{24}\ket{\ell_{\rm in},0}
      =+i\ket{\ell_{\rm in},0},
 \label{eq:ftc-phase}
\end{equation}
where $\ket{\ell_{\rm in},0}$ is the prepared loop with empty
point modes.

\paragraph*{The 3-group anomaly.}
The 3-group equations specify the allowed configurations. The
statistical phases of this realization are encoded by a
$U(1)$-valued 5-cocycle. For the 3-group $\mathbb G$ in
Eq.~\eqref{eq:ftc-parent-constraint}, the anomaly is classified by
$H^5(B\mathbb G,U(1))=\Z_4$~\cite{Kapustin2017Fermionic}.
In our convention $U(1)=\R/\Z$, we use the
representative adapted from Ref.~\cite{Barkeshli2024Higher},
\begin{equation}
\begin{aligned}
 \Omega_{\rm K}={}&\tfrac12\bigl(n_3\cup_1 n_3+n_3\cup_2\delta n_3
 +\delta n_3\cup_1 n_2+\zeta_5(n_2,n_2)\bigr)\\
 &-\tfrac18[n_2]_2\cup\delta[n_2]_2\pmod1~,
\end{aligned}
\label{eq:ftc-restricted-anomaly}
\end{equation}
where $\cup_i$ denotes a higher cup product~\cite{Steenrod1947Products,Chen2020Exact,Tata2020Geometrically,Chen2023Higher}.
The terms in parentheses
are evaluated modulo two, whereas the last term uses the $0,1$
integer lift $[n_2]_2$ and its integer coboundary.
The Cartan correction $\zeta_5$ is~\cite{MedinaMardones2020Effective}
\begin{equation}
 \zeta_5(A_2,B_2)(012345)
 =A_2(023)A_2(012)B_2(345)B_2(235).
\end{equation}
Appendix~\ref{app:ftc-calculation} verifies that
$\delta\Omega_{\rm K}=0\pmod1$.
Doubling this response cancels the terms in parentheses and gives
\begin{equation}
 2\Omega_{\rm K}=\tfrac12 n_2\cup\operatorname{Sq}^1 n_2\pmod1 .
 \label{eq:ftc-doubled-anomaly}
\end{equation}
The doubled response depends only on the loop and gives the
fermionic-loop flip sign $-1$~\cite{Fidkowski2022Gravitational, Kobayashi2026Generalized}.
The process $\mathcal K_{24}$ keeps track of the accompanying physical
fermions
and measures the phase of $\Omega_{\rm K}$ itself, namely $+i$ in
the chosen orientation.

%% file: figures/cyl_j_correspondence.tex
\begingroup
\definecolor{cjink}{HTML}{30343B}
\definecolor{cjgray}{HTML}{66717C}
\definecolor{cjblue}{HTML}{3A6EA5}
\definecolor{cjteal}{HTML}{267F83}
\definecolor{cjpurple}{HTML}{7561A8}
\definecolor{cjgold}{HTML}{C88A2A}
\begin{tikzpicture}[x=1cm,y=1cm,line cap=round,line join=round,
 font=\fontsize{9}{11}\selectfont,text=cjink,
 map/.style={line width=1.05pt,-{Stealth[length=4.8pt,width=4pt]}},
 small/.style={font=\fontsize{9}{11}\selectfont,text=cjgray},
 edge/.style={line width=.8pt,draw=cjteal},
 rear/.style={line width=.7pt,draw=cjgray!65,densely dashed}]
 \path[use as bounding box] (0,-.20) rectangle (17.1,5.8);
 \node[font=\fontsize{10}{12}\selectfont\bfseries] at (3.05,5.53)
  {One local move};
 \node[font=\fontsize{10}{12}\selectfont\bfseries] at (14.00,5.53)
  {Its labeled cylinder cells};

 \coordinate (a) at (1.55,2.78);
 \coordinate (b) at (4.65,2.78);
 \coordinate (c) at (2.50,3.42);
 \coordinate (at) at (1.55,4.18);
 \coordinate (bt) at (4.65,4.18);
 \coordinate (ct) at (2.50,4.82);
 \fill[cjblue!9] (a)--(b)--(c)--cycle;
 \fill[cjteal!5] (a)--(b)--(bt)--(at)--cycle;
 \fill[cjteal!8] (b)--(c)--(ct)--(bt)--cycle;
 \draw[rear] (a)--(c)--(b) (c)--(ct);
 \fill[cjpurple!9] (at)--(bt)--(ct)--cycle;
 \draw[draw=cjblue,line width=1pt] (a)--(b);
 \draw[edge] (a)--(at) (b)--(bt);
 \draw[draw=cjpurple,line width=1pt] (at)--(bt)--(ct)--cycle;
 \node[text=cjpurple,anchor=south] at (3.12,4.91) {$\tau_s\cfg$};
 \node[text=cjblue,anchor=north] at (3.10,2.66) {$\cfg$};
 \node[text=cjgray,fill=white,inner sep=2pt] at (3.10,3.61)
  {$T\times I$};
 \draw[map,draw=cjgray] (.77,2.90)--(.77,4.34);
 \node[small,rotate=90] at (.25,3.61) {time $I$};

 \coordinate (ra) at (12.45,2.78);
 \coordinate (rb) at (15.55,2.78);
 \coordinate (rc) at (13.40,3.42);
 \coordinate (rat) at (12.45,4.18);
 \coordinate (rbt) at (15.55,4.18);
 \coordinate (rct) at (13.40,4.82);
 \fill[cjblue!9] (ra)--(rb)--(rc)--cycle;
 \fill[cjblue!11] (ra)--(rb)--(rbt)--cycle;
 \fill[cjpurple!8] (ra)--(rbt)--(rat)--cycle;
 \fill[cjgold!10] (rb)--(rc)--(rct)--cycle;
 \fill[cjteal!9] (rb)--(rct)--(rbt)--cycle;
 \draw[rear] (ra)--(rc)--(rb) (rc)--(rct) (ra)--(rct) (rb)--(rct);
 \fill[cjpurple!9] (rat)--(rbt)--(rct)--cycle;
 \draw[draw=cjblue,line width=1pt] (ra)--(rb);
 \draw[edge] (ra)--(rat) (rb)--(rbt);
 \draw[draw=cjpurple,line width=1pt] (rat)--(rbt)--(rct)--cycle;
 \draw[draw=cjteal,line width=.85pt] (ra)--(rbt);
 \node[text=cjpurple,anchor=south] at (14.02,4.91) {$\tau_s\cfg$};
 \node[text=cjblue,anchor=north] at (14.00,2.66) {$\cfg$};

 \draw[map,draw=cjblue] (5.55,3.74)--(11.45,3.74);
 \node[text=cjblue,align=center,anchor=south] at (8.50,3.93)
  {$\operatorname{Cyl}$\\collect oriented simplices};
 \node[small] at (8.50,3.08)
  {$\sum_j(-1)^{d+1-j}[C_j;\widehat{\cfg}]$
   };
 \node[small] at (8.55,2.24)
  {Join the cylinders of all moves; closure and local cancellation give $\partial z=0$.};

 \draw[draw=cjgray!28,line width=.45pt] (.18,1.98)--(16.92,1.98);
 \begin{scope}[yshift=-.20cm]
 \node at (3.20,1.39) {Statistical class $[e]$};
 \node at (13.90,1.39) {Homology class $[z]$};
 \draw[map,draw=cjblue] (6.00,1.50)--(11.10,1.50)
  node[midway,above=2pt,inner sep=1pt,text=cjblue] {$\overline{\operatorname{Cyl}}$};
 \draw[map,draw=cjpurple] (11.10,1.08)--(6.00,1.08)
  node[midway,below=2pt,text=cjpurple] {$\bar J$};
 \node[align=center] at (4.05,.39)
  {$J(\operatorname{Cyl}(e))-e\in E_{\mathrm{id}}$\\[1pt]
   \fontsize{9}{11}\selectfont modulo locality identities};
 \node[align=center] at (12.90,.39)
  {$\operatorname{Cyl}(J(z))-z=\partial H(z)$\\[1pt]
   \fontsize{9}{11}\selectfont modulo boundaries};
 \end{scope}
\end{tikzpicture}
\endgroup

%% file: figures/anomaly_to_process.tex
\begingroup
\definecolor{flowink}{HTML}{30343B}
\definecolor{flowgray}{HTML}{66717C}
\definecolor{flowblue}{HTML}{3A6EA5}
\definecolor{flowteal}{HTML}{267F83}
\definecolor{flowpurple}{HTML}{7561A8}
\definecolor{flowgold}{HTML}{C88A2A}
\begin{tikzpicture}[x=1cm,y=1cm,
 font=\fontsize{9}{11}\selectfont,
 line cap=round,line join=round,
 major/.style={draw=flowgray,line width=.9pt,
   -{Stealth[length=5pt,width=4pt]}},
 move/.style={draw=flowpurple,line width=.9pt,
   -{Stealth[length=4pt,width=3.4pt]}},
 smalltext/.style={font=\fontsize{9}{11}\selectfont,text=flowgray},
 title/.style={font=\fontsize{10}{12}\selectfont\bfseries,text=flowink}]
 \path[use as bounding box] (0,0) rectangle (17.2,5.05);

 \foreach \x/\col in {1.75/flowblue,6.25/flowteal,10.75/flowpurple,15.25/flowgold}{
  \draw[\col,line width=1.4pt] (\x-1.55,4.43)--(\x+1.55,4.43);
 }
 \node[title] at (1.75,4.76) {Anomaly cocycle};
 \node[title] at (6.25,4.76) {Detecting cycle};
 \node[title] at (10.75,4.76) {Replace each cell by $\mathcal J$};
 \node[title] at (15.25,4.76) {Statistical process};

 \fill[flowblue!8] (.68,3.05)--(2.83,3.05)--(1.65,4.05)--cycle;
 \draw[flowblue,line width=.95pt]
  (.68,3.05)--(2.83,3.05)--(1.65,4.05)--cycle;
 \draw[flowblue!65,line width=.75pt]
  (.68,3.05)--(1.83,3.48)--(2.83,3.05)
  (1.65,4.05)--(1.83,3.48);
 \draw[flowblue!65,densely dashed,line width=.65pt]
  (.68,3.05)--(2.83,3.05);
 \node[fill=white,inner sep=2pt,text=flowblue] at (1.73,3.43) {$\Omega$};
 \node[text=flowink] at (1.75,2.28) {$\delta\Omega=0$};
 \node[smalltext,align=center] at (1.75,1.58)
  {Specify the anomaly\\to be probed};

 \fill[flowteal!8] (4.99,3.10)--(5.78,4.01)--(6.32,3.10)--cycle;
 \fill[flowteal!15] (5.78,4.01)--(6.32,3.10)--(7.02,3.94)--cycle;
 \fill[flowteal!7] (6.32,3.10)--(7.02,3.94)--(7.51,3.10)--cycle;
 \draw[flowteal,line width=.95pt]
  (4.99,3.10)--(5.78,4.01)--(7.02,3.94)--(7.51,3.10)--cycle;
 \draw[flowteal!55,line width=.65pt]
  (5.78,4.01)--(6.32,3.10)--(7.02,3.94);
 \node[text=flowteal] at (5.70,3.41) {$\sigma_1$};
 \node[text=flowteal] at (6.35,3.67) {$\sigma_2$};
 \node[text=flowteal] at (6.99,3.39) {$\cdots$};
 \node[text=flowink] at (6.25,2.12)
  {$\begin{gathered}z=\sum_a n_a\sigma_a\\\partial z=0\end{gathered}$
   };
 \node[smalltext] at (6.25,1.29) {Match complete face data};

 \coordinate (vac) at (10.75,3.97);
 \coordinate (minus) at (9.65,3.02);
 \coordinate (plus) at (11.85,3.02);
 \draw[move] ($(vac)+(-.20,-.18)$)--($(minus)+(.30,.26)$);
 \draw[move] ($(minus)+(.23,0)$)--($(plus)+(-.23,0)$);
 \draw[move] ($(plus)+(-.28,.24)$)--($(vac)+(.28,-.24)$);
 \node[fill=white,inner sep=1.5pt,text=flowpurple] at (10.75,3.97) {$\cfg_0$};
 \node[fill=white,inner sep=1.5pt,text=flowpurple] at (9.65,3.02) {$\cfg^-$};
 \node[fill=white,inner sep=1.5pt,text=flowpurple] at (11.85,3.02) {$\cfg^+$};
 \node[font=\fontsize{9}{11}\selectfont,text=flowpurple,fill=white,inner sep=1.3pt]
  at (9.86,3.77) {$P(\cfg^-)$};
 \node[font=\fontsize{9}{11}\selectfont,text=flowpurple,fill=white,inner sep=1.3pt]
  at (11.85,3.98) {$P(\cfg^+)^{-1}$};
 \node[font=\fontsize{9}{11}\selectfont,text=flowpurple,fill=white,inner sep=1.2pt]
  at (10.75,3.02) {$S(\sigma)$};
 \node[text=flowpurple] at (10.75,2.60)
  {$L(\sigma)$};
 \node[text=flowink] at (10.75,2.12)
  {$\sigma\longmapsto\mathcal J(\sigma)$ in Eq.~\eqref{dc:count-map}};
 \node[smalltext,align=center] at (10.75,1.43)
  {Prepare, sweep, return;\\add reference comparisons};

 \draw[flowgold!45,line width=.75pt] (13.83,3.44)--(16.66,3.44);
 \foreach \x in {14.35,15.05,15.75}{
  \draw[major,draw=flowgold] (\x-.24,3.44)--(\x+.22,3.44);
 }
 \foreach \x in {14.70,15.40}{
  \fill[flowgold!17] (\x-.07,3.19) rectangle (\x+.07,3.69);
  \draw[flowgold,line width=.7pt] (\x-.07,3.19) rectangle (\x+.07,3.69);
 }
 \node[fill=white,inner sep=2pt,text=flowgold] at (13.78,3.44) {$\cfg_0$};
 \node[fill=white,inner sep=2pt,text=flowgold] at (16.72,3.44) {$\cfg_0$};
 \node[smalltext] at (15.25,2.91) {closed local process};
 \node[text=flowink,font=\fontsize{9}{11}\selectfont] at (15.25,2.25)
  {$\begin{aligned}W(z)&=\mathcal J(\sigma_s)^{n_s}\cdots
   \mathcal J(\sigma_1)^{n_1}\end{aligned}$};
 \node[smalltext,align=center] at (15.25,1.43)
  {Concatenate signed blocks;\\reverse negative copies};

 \draw[major] (3.45,3.48)--(4.55,3.48);
 \node[smalltext] at (4.00,3.88) {choose};
 \draw[major] (7.98,3.48)--(9.07,3.48);
 \draw[major] (12.43,3.48)--(13.48,3.48);

 \draw[black!15,line width=.45pt] (.15,.80)--(17.05,.80);
 \node[text=flowink] at (8.6,.35)
  {Phase of the process\quad $W(z)\ket0=\exp\bigl(2\pi i\langle z,\Omega\rangle\bigr)\ket0$
   };
\end{tikzpicture}
\endgroup

%% file: sections/spt_examples.tex
\section{Detecting 2-group anomalies at SPT boundaries}
\label{sec:spt-examples}

We use local hopping operators to detect the anomalies of the
2-group boundary actions constructed in Ref.~\cite{Feng2026Defining}.
At fixed spatial dimension, the defect configurations and moves
are shared by the different actions. Composing the hops into
closed statistical processes isolates the anomaly-dependent phase.

\subsection{2-group configurations and local moves}
\label{spt:setup}

Both layers of the nonsplit 2-group are $\Z_2$, with trivial
coefficient action and Postnikov cocycle $\beta(g,h,k)=ghk$.
Write $A$ and $B$ for the background fields of the 0-form and
1-form symmetries, respectively. On the
ordered auxiliary simplex $T=[0\cdots m]$, with $m=d+1$, they obey
\begin{equation}
 \delta A=0,\qquad \delta B=A\cup A\cup A.
 \label{spt:constraints}
\end{equation}
The physical lattice is $\partial T$. A nonzero $A_{ij}$ marks a
wall on the cell complementary to $ij$. A nonzero $B_{ijk}$
marks a point in two spatial dimensions or a loop segment in three.
The second constraint allows the 1-form defects to end where the
wall labels require a junction. In one spatial dimension, $B$ on
the filling triangle instead labels a global sector of the circle;
Sec.~\ref{spt:one-dimensional} treats its local moves separately.
A configuration $\cfg=(A,B)$ is determined by the independent
bits $A_{0i}$ for $i>0$ and $B_{0ij}$ for $0<i<j$.
Evaluating Eq.~\eqref{spt:constraints} on the triangle $[0ij]$
and tetrahedron $[0ijk]$ gives
\begin{equation*}
 \begin{aligned}
 A_{ij}&=A_{0i}+A_{0j},\\
 B_{ijk}&=B_{0ij}+B_{0ik}+B_{0jk}+A_{0i}A_{ij}A_{jk}.
 \end{aligned}
\end{equation*}

\begin{figure*}[t]
 \centering
 \input{figures/spt_boundary_qubits}
 \caption{Reading boundary defects from local potentials in the
  nonsplit 2-group models. Solid lines show the primal lattice,
  dashed lines show its dual, and open circles mark the indicated
  vertex and edge qubits. (a) The parity $A_{ij}$ of the two
  vertex bits detects a wall on the face dual to $ij$.
  (b,c) The triangle label $B_{ijk}$ combines the parity of the
  three edge bits with the wall-dependent correction
  $a_i A_{ij}A_{jk}$, which reads the vertex bits of the same
  triangle. The occupied dual cell is a loop segment in three
  spatial dimensions and a point in two. All sums are modulo two.}
 \label{fig:spt-boundary-readout}
\end{figure*}
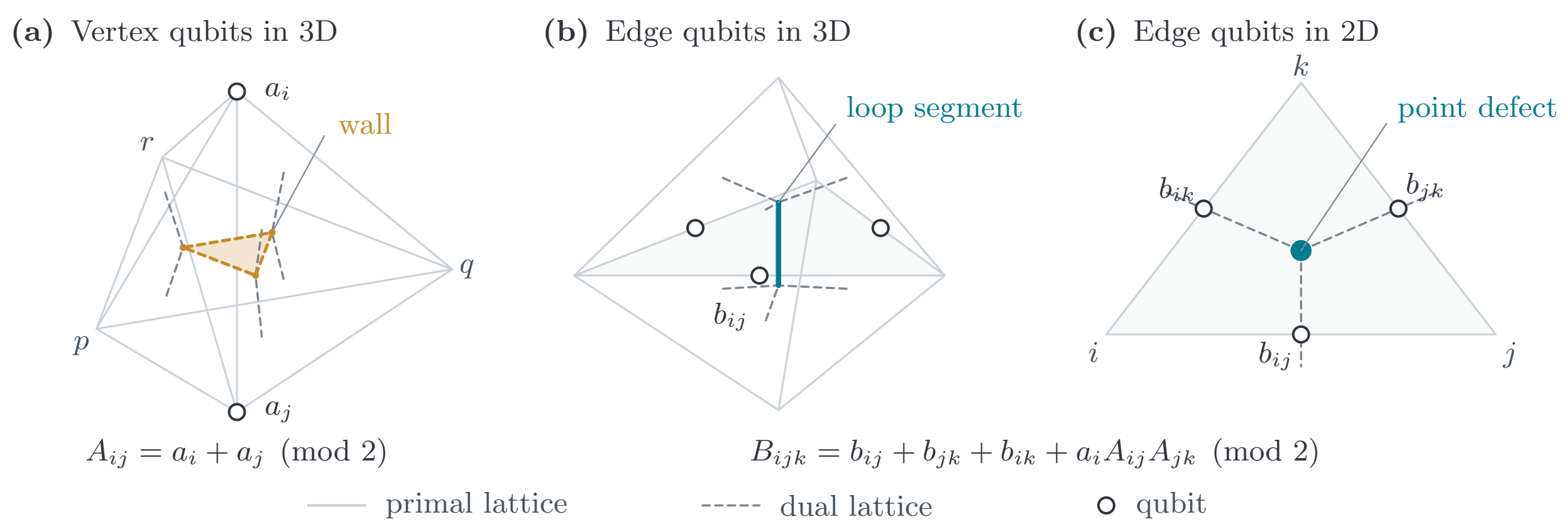

For $d\ge2$, the move $V_i$ moves a wall across the physical cell
opposite $i$, and $E_{ij}$ toggles the
boundary of the cell complementary to $ij$. The edge move fixes
$A$ and adds $\delta e_{ij}$ to $B$, where $e_{ij}$ is the unit
edge cochain. For a wall move with unit vertex cochain $h$ at $i$,
the update is
\begin{equation}
 \begin{aligned}
 A'&=A+\delta h,\\
 B'&=B+A\cup A\cup h+A\cup h\cup A'+h\cup A'\cup A'.
 \end{aligned}
 \label{spt:wall-update}
\end{equation}
The change of $B$ keeps the junction condition in
Eq.~\eqref{spt:constraints} satisfied as the wall moves. All
cochain additions are modulo two. An inverse edge move makes
the same configuration update. To undo a wall move, first
recover the incoming wall field from $A=A'+\delta h$, then
add the same $B$ increment computed from $(A,A')$.
With $\overline f$ the ordered complementary vertex set, the
physical defect chains are
\begin{equation}
 \begin{aligned}
 a^{(d-1)}(\cfg)&=\sum_{i<j}A_{ij}\langle\overline{ij}\rangle,\\
 a^{(d-2)}(\cfg)&=\sum_{i<j<k}B_{ijk}\langle\overline{ijk}\rangle.
 \end{aligned}
 \label{spt:physical-chains}
\end{equation}
\begingroup\makeatletter
\edef\@currentlabel{\p@equation\theequation}\label{spt:mixed-physical-chains}
\makeatother\endgroup
Fig.~\ref{fig:spt-boundary-readout} shows how vertex bits $a_i$ and
edge bits $b_{ij}$ determine the defects on the dual cells. The
wall-dependent correction enforces the
junction rule in Eq.~\eqref{spt:constraints}. In a wall-free
configuration it vanishes, so the triangle readout is the parity
of its three edge bits.
In two dimensions the constraint fixes the total point parity
on $T$. In three dimensions it fixes the boundary of the loop chain.

\subsection{Boundary symmetry actions and hopping operators}
\label{spt:boundary-gates}

Ref.~\cite{Feng2026Defining} constructs the anomalous boundary symmetry
actions and identifies their cocycle classes. We use the binary cochains
\begin{equation}
 \begin{aligned}
 Q_3&=A\cup B+B\cup_1B+B\cup_2\delta B,\\
 Q_4&=A\cup A\cup B+B\cup B+B\cup_1\delta B.
 \end{aligned}
 \label{spt:Q-definitions}
\end{equation}
Both are closed modulo two under Eq.~\eqref{spt:constraints}.
The anomaly generators in one, two, and three spatial dimensions are
\begin{equation}
 \begin{aligned}
 \alpha_3&=\tfrac12Q_3, & \alpha_4&=\tfrac12Q_4,\\
 \alpha_5^{\mathrm I}&=\tfrac12Q_3\cup_1Q_3,
 &\alpha_5^{\mathrm{II}}&=\tfrac12A\cup Q_4.
 \end{aligned}
 \label{spt:anomalies}
\end{equation}
Here $Q_3\cup_1Q_3=\Sq^2Q_3$ because $\delta Q_3=0$.
The factors of $1/2$ embed binary values in $\R/\Z$.
The corresponding cohomology groups are $\Z_2$, $\Z_2$, and
$\Z_2\oplus\Z_2$, respectively~\cite{Feng2026Defining,Gu2026Understanding}.
In particular, a three-dimensional boundary has response
$\Omega_5=\mu\alpha_5^{\mathrm I}+\nu\alpha_5^{\mathrm{II}}$,
with $\mu,\nu\in\Z_2$.

The microscopic potentials encode the defect fields as
$A=\delta a$ and $B=\delta b+a\cup\delta a\cup\delta a$.
The ordinary symmetry relabels these potentials by
$(a,b)\mapsto(a+1,b+a\cup\delta a)$, and a 1-form transformation
adds $\delta\lambda$ to $b$. Both preserve $(A,B)$.
We work in the common $+1$ eigenspace of the 1-form Gauss-law
operators~\cite{Feng2026Defining} and choose hopping operators
that commute with both symmetries in this sector.

For $d\ge2$, write $p=(a,b)$ for the potential bits. The edge
move $E_{ij}$ flips $b_{ij}$. The wall move lifts
Eq.~\eqref{spt:wall-update} by sending $a$ to $a+h$ and
adding $a\cup A\cup h+a\cup h\cup A'$ to $b$.
Write $gp$ for either permutation. Given $p=(a,b)$, add an
ordered cone vertex $*$ before the vertices of $T$ and set
$A_{*i}=a_i$ and $B_{*ij}=b_{ij}$. The opposite face carries
$\cfg(p)=(A,B)$. For a response $\Omega=q/2$, let $R_q(p)$
be $q$ evaluated on this cone, and let $c_g(\cfg)$ be the
cylinder simplex in Eq.~\eqref{app:conv:coupled-operators}.
The microscopic hopping operator is
\begin{equation}
 \widehat g_q\ket p=
 (-1)^{R_q(p)+R_q(gp)+q(c_g(\cfg(p)))}\ket{gp}.
 \label{spt:microscopic-cone-gate}
\end{equation}
\begingroup\makeatletter
\edef\@currentlabel{\p@equation\theequation}\label{spt:main-microscopic-gate}
\makeatother\endgroup
The two cone factors account for the phases of the anomalous
boundary symmetry action. With these factors, the gates commute
with both symmetries on the 1-form Gauss-law sector. However, they cancel out in closed processes, leaving only the cylinder signs. Cocycle closure gives the locality identities for the hopping operators.

The induced operator $g$ on configuration states acts as
\begin{equation}
 g\ket{\cfg}_\Omega=(-1)^{q(c_g(\cfg))}\ket{g\cfg}_\Omega.
 \label{spt:transport-phase}
\end{equation}
Here $\ket{\cfg}_\Omega$ is the normalized sum of the potential
states with fields $\cfg$, weighted by $(-1)^{R_q(p)}$. For a sum of cocycles, the phase exponents add while the
configuration update remains the same.

%% file: figures/spt_boundary_qubits.tex
\begingroup
\definecolor{bdyLoop}{HTML}{007C91}
\definecolor{bdyWall}{HTML}{C88A2A}
\definecolor{bdyCharge}{HTML}{30343B}
\definecolor{bdyMesh}{HTML}{CCD3D8}
\definecolor{bdyDual}{HTML}{7B8791}
\definecolor{bdyText}{HTML}{46515B}
\begin{tikzpicture}[x=1cm,y=1cm,line cap=round,line join=round,
 font=\fontsize{9}{11}\selectfont,text=bdyCharge,
 primal/.style={draw=bdyMesh,line width=0.6pt},
 dual/.style={draw=bdyDual,line width=0.65pt,
              dash pattern=on 2.2pt off 1.6pt},
 qubit/.style={circle,draw=bdyCharge,fill=white,line width=0.75pt,
               minimum size=4.7pt,inner sep=0pt},
 guide/.style={draw=bdyText!65,line width=0.45pt}]
 \path[use as bounding box] (0,0) rectangle (16.8,5.4);
 \node[anchor=west] at (0.2,5.1) {\textbf{(a)}\enspace Vertex qubits in $3$D};
 \node[anchor=west] at (5.8,5.1) {\textbf{(b)}\enspace Edge qubits in $3$D};
 \node[anchor=west] at (11.4,5.1) {\textbf{(c)}\enspace Edge qubits in $2$D};
 \begin{scope}[shift={(2.7,2.8)},
  x={(0.939693cm,0.157329cm)},
  y={(-0.342020cm,0.432276cm)},
  z={(0cm,0.887919cm)}]
  \coordinate (ai) at (0,0,1.9);
  \coordinate (aj) at (0,0,-1.9);
  \coordinate (ar0) at ({-sqrt(3)*2.3/2},-1.15,0);
  \coordinate (ar1) at ({sqrt(3)*2.3/2},-1.15,0);
  \coordinate (ar2) at (0,2.3,0);
  \foreach \r in {0,1,2}{
   \draw[primal] (ai)--(ar\r)--(aj);
  }
  \foreach \r/\s in {0/1,1/2,2/0}{
   \draw[primal] (ar\r)--(ar\s);
   \coordinate (at\r) at
    (barycentric cs:ai=1,aj=1,ar\r=1,ar\s=1);
   \coordinate (al\r) at
    (barycentric cs:ai=1,ar\r=1,ar\s=1);
   \coordinate (aq\r) at
    (barycentric cs:aj=1,ar\r=1,ar\s=1);
  }
  \draw[primal] (ai)--(aj);
  \fill[bdyWall!22]
   (at0)--(at1)--(at2)--cycle;
  \foreach \r in {0,1,2}{
   \draw[dual] (al\r)--(at\r)--(aq\r);
  }
  \draw[dual,draw=bdyWall,line width=0.95pt]
   (at0)--(at1)--(at2)--cycle;
  \foreach \r in {0,1,2}{
   \fill[bdyWall] (at\r) circle (1.05pt);
  }
  \node[qubit] at (ai) {};
  \node[qubit] at (aj) {};
  \node[anchor=west,xshift=5pt] at (ai) {$a_i$};
  \node[anchor=west,xshift=5pt] at (aj) {$a_j$};
  \node[anchor=north east,text=bdyText,inner sep=2pt] at (ar0) {$p$};
  \node[anchor=west,text=bdyText,inner sep=2pt] at (ar1) {$q$};
  \node[anchor=south east,text=bdyText,inner sep=2pt] at (ar2) {$r$};
 \end{scope}
 \node[text=bdyWall,anchor=west] at (3.65,4.15) {wall};
 \draw[guide] (3.62,4.02)--(at1);
 \node at (2.7,0.68) {$A_{ij}=a_i+a_j\pmod2$};
 \begin{scope}[xshift=5.6cm]
  \coordinate (bi) at (0.65,2.55);
  \coordinate (bj) at (4.55,2.55);
  \coordinate (bk) at (3.2,3.55);
  \coordinate (bf) at (barycentric cs:bi=1,bj=1,bk=1);
  \coordinate (bu) at ($(bf)+(0,1.75)$);
  \coordinate (bd) at ($(bf)-(0,1.75)$);
  \coordinate (bt) at (barycentric cs:bi=1,bj=1,bk=1,bu=1);
  \coordinate (bb) at (barycentric cs:bi=1,bj=1,bk=1,bd=1);
  \fill[bdyMesh!12] (bi)--(bj)--(bk)--cycle;
  \draw[primal] (bi)--(bu)--(bj) (bu)--(bk);
  \draw[primal] (bi)--(bd)--(bj) (bd)--(bk);
  \draw[primal] (bi)--(bj)--(bk)--cycle;
  \foreach \p/\q in {bi/bj,bj/bk,bk/bi}{
   \coordinate (bfu) at (barycentric cs:\p=1,\q=1,bu=1);
   \coordinate (bfd) at (barycentric cs:\p=1,\q=1,bd=1);
   \draw[dual] (bfu)--(bt);
   \draw[dual] (bfd)--(bb);
  }
  \draw[dual] (bt)--(bb);
  \draw[bdyLoop,line width=1.6pt] (bt)--(bb);
  \node[qubit] at ($(bi)!0.5!(bj)$) {};
  \node[qubit] at ($(bj)!0.5!(bk)$) {};
  \node[qubit] at ($(bk)!0.5!(bi)$) {};
  \node[anchor=north] at (2.3,2.38) {$b_{ij}$};
  \node[text=bdyLoop,anchor=west] at (3.4,4.3) {loop segment};
  \draw[guide] (3.4,4.14)--(bt);
 \end{scope}
 \begin{scope}[xshift=11.2cm]
  \coordinate (ci) at (0.65,1.93);
  \coordinate (cj) at (4.75,1.93);
  \coordinate (ck) at (2.70,4.58);
  \coordinate (ct) at (barycentric cs:ci=1,cj=1,ck=1);
  \fill[bdyMesh!12] (ci)--(cj)--(ck)--cycle;
  \draw[primal] (ci)--(cj)--(ck)--cycle;
  \foreach \p/\q in {ci/cj,cj/ck,ck/ci}{%
   \coordinate (cm) at ($(\p)!0.5!(\q)$);
   \draw[dual] (ct)--($(ct)!1.38!(cm)$);
   \node[qubit] at (cm) {};
  }
  \fill[bdyLoop] (ct) circle (3.15pt);
  \node[anchor=north east,text=bdyText,inner sep=2pt] at (ci) {$i$};
  \node[anchor=north west,text=bdyText,inner sep=2pt] at (cj) {$j$};
  \node[anchor=south,text=bdyText,inner sep=2pt] at (ck) {$k$};
  \node[anchor=north east,inner sep=2pt] at ($(ci)!0.5!(cj)$) {$b_{ij}$};
  \node[anchor=south west,inner sep=2pt] at ($(cj)!0.5!(ck)$) {$b_{jk}$};
  \node[anchor=south east,inner sep=2pt] at ($(ck)!0.5!(ci)$) {$b_{ik}$};
  \node[text=bdyLoop,anchor=west] at (3.6,4.30) {point defect};
  \draw[guide] (3.60,4.15)--(ct);
 \end{scope}
 \node at (11.1,0.68) {$B_{ijk}=b_{ij}+b_{jk}+b_{ik}+a_iA_{ij}A_{jk}\pmod2$};
 \draw[primal,line width=0.8pt] (3.45,0.13)--(4.05,0.13);
 \node[anchor=west,text=bdyText] at (4.15,0.13) {primal lattice};
 \draw[dual] (7.6,0.13)--(8.2,0.13);
 \node[anchor=west,text=bdyText] at (8.3,0.13) {dual lattice};
 \node[qubit] at (11.85,0.13) {};
 \node[anchor=west,text=bdyText] at (12.05,0.13) {qubit};
\end{tikzpicture}
\endgroup

%% file: sections/spt_d1.tex
\subsection{\texorpdfstring{$(1{+}1)$D}{(1+1)D}: Wall transport}
\label{spt:one-dimensional}

\input{figures/spt_circle_history}

On the physical circle $\partial[012]$, the wall move $V_i$ acts
on the arc $R_i$ opposite $i$. The gate $E_{ij}$ acts at the
common endpoint $p_{ij}$ of $R_i$ and $R_j$. The complete
configuration is specified by three binary labels.
\begin{equation}
 \begin{gathered}
 \cfg(x,y,\beta):\quad A_{01}=x,\quad A_{02}=y,\\
 A_{12}=x+y,\qquad B_{012}=\beta.
 \end{gathered}
 \label{spt:d1-config}
\end{equation}
The vacuum is $\cfg(0,0,0)$. The wall occupations are local,
while $\beta$ labels a global sector of the entire circle.
The gate $E_{ij}$ toggles $\beta$ at fixed wall configuration
without creating a localized point excitation. A wall move $V_i$
toggles the walls at the endpoints of $R_i$ and can also change
$\beta$. In the order $(x,y,\beta)$, the updates are
\begin{equation*}
 \begin{aligned}
 E_{ij}&:\ (x,y,\beta)\mapsto(x,y,\beta+1),\\
 V_0&:\ (x,y,\beta)\mapsto(x+1,y+1,\beta+(1+x)y),\\
 V_1&:\ (x,y,\beta)\mapsto(x+1,y,\beta+xy),\\
 V_2&:\ (x,y,\beta)\mapsto(x,y+1,\beta+x(1+y)).
 \end{aligned}
\end{equation*}
The wall rules follow by evaluating Eq.~\eqref{spt:wall-update}
on $[012]$. An inverse move reverses its displayed update.
The only local configuration label at $p_{ij}$ is $A_{ij}$, so
the gate $E_{ij}$ can change the global sector without revealing
it locally.
This lets us probe the boundary anomaly
\begin{equation}
 \kappa\alpha_3,\qquad \alpha_3=\tfrac12 Q_3,\quad
 \kappa\in\Z_2.
 \label{spt:alpha-three}
\end{equation}

The process toggles $\beta$ in the vacuum, changes the wall
background, and then reverses both updates.
\begin{equation}
 \mathcal D_6=V_2^{-1}V_1^{-1}E_{12}V_1V_2E_{12}^{-1}.
 \label{spt:d1-six}
\end{equation}
The first move flips $\beta$, and the fourth restores it. At both
uses of $E_{12}$ the walls give $A_{12}=0$ at $p_{12}$.
For each wall move, the input of the forward occurrence and
the output of its inverse differ only in $\beta$, so their
local labels agree.
The three forward--inverse pairs thus cancel their local phase
choices. Fig.~\ref{fig:spt-circle} follows the configurations.

\paragraph*{Local operators.}
Use a bit $a_i$ on arc $R_i$, copied at its two endpoints, and a
bit $b_{ij}$ at $p_{ij}$. The arc bits give $A_{ij}=a_i+a_j$, and
the global bit is $\beta=b_{01}+b_{02}+b_{12}+a_0A_{01}A_{12}$.
The gate $E_{ij}$ flips $b_{ij}$ and therefore shifts $\beta$.
In the aligned-arc space the junction operator is
\begin{equation}
 E_{ij}\ket{a,b}
 =(-1)^{\kappa(1+b_{ij}+a_i)}\ket{a,b+\mathbf1_{ij}}.
 \label{spt:d1-point-gate}
\end{equation}
For a wall move $V_v$, set $h_i=\delta_{iv}$, $a'=a+h$,
and $A'_{ij}=a'_i+a'_j$. Its action is
\begin{equation}
 \begin{aligned}
 b'_{ij}&=b_{ij}+a_i A_{ij}h_j+a_i h_i A'_{ij},\\
 V_v\ket{a,b}&=(-1)^{\kappa\nu_v(a,b)}\ket{a',b'},\\
 \nu_v(a,b)&=
 \sum_{j>v}a_v a_j(1+b_{vj})
 +\sum_{i<v}a_i b_{iv}(1+a_v).
 \end{aligned}
 \label{spt:d1-wall-gate}
\end{equation}
The gate $E_{ij}$ acts only at $p_{ij}$. The wall gate flips the
copies of $a_v$ and reads only the bits at its endpoints.
These phases follow from Eq.~\eqref{spt:microscopic-cone-gate}
with $q=\kappa Q_3$ on the circle and realize the anomaly
$\kappa\alpha_3$. On the aligned-arc space, the gates $E_{ij}$
commute with the ordinary symmetry and the closed
1-form shifts, and the wall gates commute with the closed shifts.
Interchanging a wall gate with the ordinary symmetry introduces
a closed 1-form shift, which acts trivially in the 1-form
Gauss-law sector.

\paragraph*{Phase of the process.}
On any vacuum preparation the arc bits are all equal, say to
$t$. The gate $E_{12}^{-1}$ contributes
$(-1)^{\kappa(b_{12}+t)}$ and flips $b_{12}$.
The next two wall gates change the arc bits from $(t,t,t)$ to
$(t,t+1,t+1)$ with no phase and leave the junction bits fixed.
At the forward occurrence $E_{12}$, both $a_1$ and $b_{12}$
have been toggled, so $E_{12}$ contributes
$(-1)^{\kappa(1+b_{12}+t)}$, with the bits in this exponent
read on the initial preparation. The return wall gates have
unit phase and restore the initial preparation. Hence
\begin{equation}
 \mathcal D_6\ket{\cfg(0,0,0)}_\kappa
 =(-1)^\kappa\ket{\cfg(0,0,0)}_\kappa.
 \label{spt:d1-response}
\end{equation}

The global bit places this example outside the $n\le d$ range of
Theorem~\ref{dc:correspondence-theorem}. The direct cylinder
comparison in Appendix~\ref{app:spt-examples} identifies the
process with the corresponding detector in Eq.~\eqref{spt:d1-cycle-table}.

%% file: figures/spt_circle_history.tex
\begin{figure*}[t]
\centering
\begingroup
\definecolor{hgsCircleWall}{RGB}{198,112,27}
\newcommand{\hgsCircleArcs}{%
  \foreach \start/\finish in {90/210,210/330,330/450}{%
    \draw[black!55,densely dashed,line width=.65pt]
      (\start:.93) arc[start angle=\start,end angle=\finish,radius=.93];%
  }%
}%
\newcommand{\hgsCircleVertices}{%
  \foreach \angle in {90,210,330}{%
    \fill[black!85] (\angle:.93) circle[radius=1.3pt];%
  }%
}%
\newcommand{\hgsCircleState}[6]{%
  \begin{scope}[shift={(#1,#2)}]
    \node[font=\small] at (0,1.49) {\textbf{#3}\enspace $#4$};
    \hgsCircleArcs
    \foreach \angle/\occupied in {#6}{%
      \ifnum\occupied=1\relax
        \draw[hgsCircleWall,line width=2.1pt,line cap=round]
          (\angle:.76)--(\angle:1.10);
      \fi
    }%
    \hgsCircleVertices
    \node[font=\small] at (0,-1.19) {$\beta=#5$};
  \end{scope}%
}%
\begin{tikzpicture}[x=1cm,y=1cm]
  \begin{scope}
    \node[font=\small] at (0,1.49) {Lattice key};
    \hgsCircleArcs
    \hgsCircleVertices
    \node[font=\small,anchor=south,inner sep=1pt] at (90:1.02) {$p_{12}$};
    \node[font=\small,anchor=east,inner sep=1pt] at (210:1.02) {$p_{02}$};
    \node[font=\small,anchor=west,inner sep=1pt] at (330:1.02) {$p_{01}$};
    \node[font=\small,inner sep=1pt] at (270:.63) {$R_0$};
    \node[font=\small,inner sep=1pt] at (30:.63) {$R_1$};
    \node[font=\small,inner sep=1pt] at (150:.63) {$R_2$};
  \end{scope}
  \hgsCircleState{4.20}{0}{0}{\text{vacuum}}{0}{90/0,210/0,330/0}
  \hgsCircleState{8.40}{0}{1}{E_{12}^{-1}}{1}{90/0,210/0,330/0}
  \hgsCircleState{12.60}{0}{2}{V_2}{1}{90/1,210/1,330/0}
  \hgsCircleState{0}{-3.40}{3}{V_1}{1}{90/0,210/1,330/1}
  \hgsCircleState{4.20}{-3.40}{4}{E_{12}}{0}{90/0,210/1,330/1}
  \hgsCircleState{8.40}{-3.40}{5}{V_1^{-1}}{0}{90/1,210/1,330/0}
  \hgsCircleState{12.60}{-3.40}{6}{V_2^{-1}}{0}{90/0,210/0,330/0}
  \draw[black!20,line width=.35pt] (-1.84,-1.55)--(14.44,-1.55);
\end{tikzpicture}
\endgroup
\caption{The circle process $\mathcal D_6$ of Eq.~\eqref{spt:d1-six}, read
across each row. Filled dots mark the vertices $p_{ij}$, and the
dashed arcs are the edges $R_i$. Each numbered panel shows the
configuration after the indicated move. An orange tick marks a
wall at a vertex with $A_{ij}=1$. The bit $\beta$ labels the
entire circle. The two $E_{12}$ occurrences act at $p_{12}$
and toggle the global bit $\beta$. Each $V_i$ toggles the walls
at the endpoints of $R_i$. For the reference vacuum preparation $a_i=b_{ij}=0$,
step 4 contributes the phase $(-1)^\kappa$ and all other steps
contribute unit phase.}
\label{fig:spt-circle}
\end{figure*}
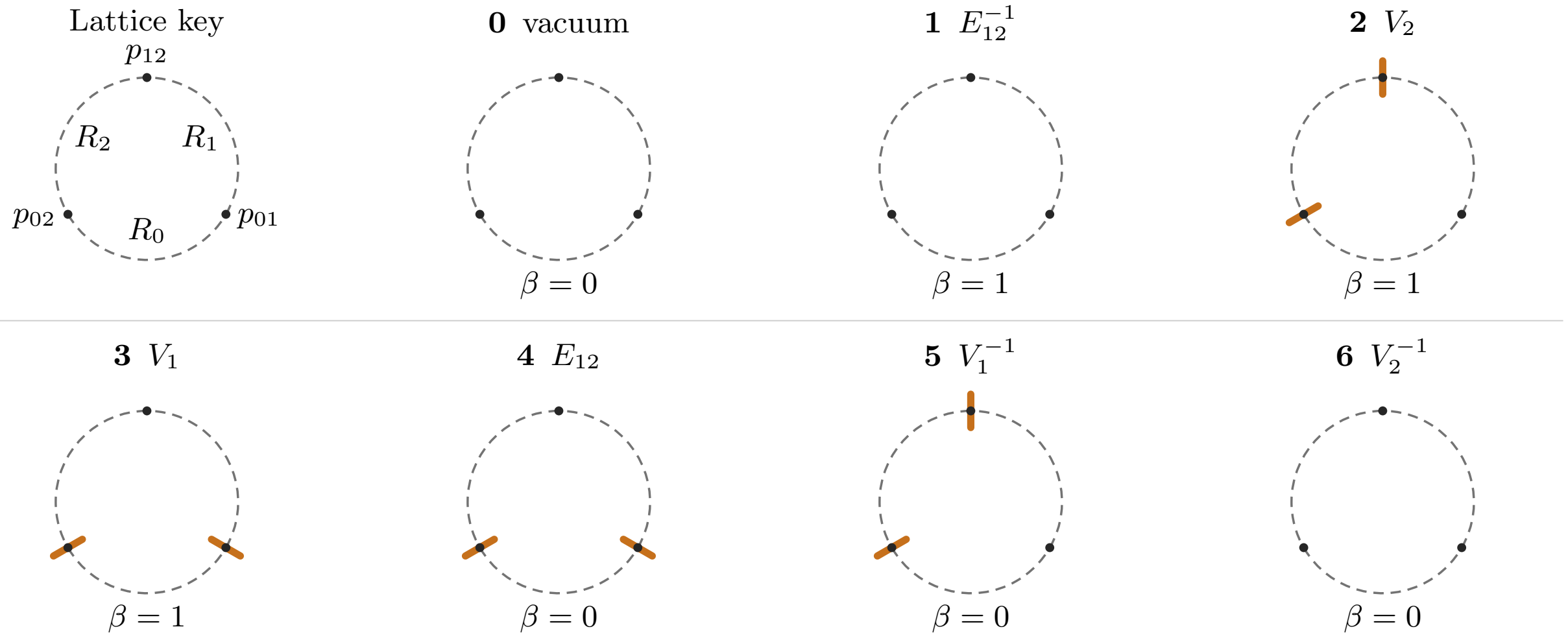

%% file: sections/spt_d2.tex
\subsection{\texorpdfstring{$(2{+}1)$D}{(2+1)D}: Particle exchange}
\label{spt:d2}

We use the standard T-junction exchange~\cite{Levin2003Fermions}
to detect the boundary 4-cocycle without producing wall excitations.
Setting $A=0$ gives
\begin{equation}
 \alpha_4=\tfrac12Q_4,\qquad
 \alpha_4\big|_{A=0}=\tfrac12 B\cup B.
 \label{spt:d2-anomaly}
\end{equation}
On $\partial[0123]$, a wall-free configuration is specified by the
four point occupations $(B_{123},B_{023},B_{013},B_{012})$ at
vertices $0,1,2,3$, respectively. Their sum is zero modulo two
by $\delta B=0$. The moves $E_{23},E_{13},E_{12}$ act on the
physical edges $01,02,03$, respectively, and toggle both
endpoint occupations. Each inverse has the same occupation
update. The exchange uses no wall moves and stays in the $A=0$
sector. Its operator word is
\begin{equation}
 \mathcal T_6=E_{12}^{-1}E_{13}E_{23}^{-1}E_{12}E_{13}^{-1}E_{23}.
 \label{spt:d2-short-word}
\end{equation}
The two pairs created on $01$ and $03$ exchange annihilation
partners. The occupations at the two endpoints of each move
match between its forward and inverse occurrences, so local
phase choices cancel.

\paragraph*{Local operators and exchange phase.}
Use the symmetry-commuting edge gates in
Eq.~\eqref{spt:microscopic-cone-gate} with $q=Q_4$.
At $A=0$, the cone phase is $R_{Q_4}=b_{01}B_{123}$.
The cylinder exponents for $E_{12},E_{13},E_{23}$ are
$0,0,B_{012}$, respectively. The last follows from
$(B\cup B)_{01234}=B_{012}B_{234}$ with $B_{234}=1$ on
the $E_{23}$ cylinder. Combining the cone and cylinder signs
gives the following actions on the wall-free basis.
\begin{equation}
 \begin{aligned}
 E_{12}\ket{a,b}&=(-1)^{b_{01}}\ket{a,b+\mathbf1_{12}},\\
 E_{13}\ket{a,b}&=(-1)^{b_{01}}\ket{a,b+\mathbf1_{13}},\\
 E_{23}\ket{a,b}&=(-1)^{b_{02}+b_{12}}\ket{a,b+\mathbf1_{23}}.
 \end{aligned}
 \label{spt:d2-matrix-elements}
\end{equation}
In this sector the displayed gates square to one since their
signs are independent of the bit they flip. The two $E_{23}$
occurrences are separated by a flip of $b_{12}$, so their signs
multiply to $-1$. The other signs cancel in pairs because
$b_{01}$ stays fixed. All edge bits return to their initial
values, and
\begin{equation}
 \mathcal T_6\ket0=-\ket0
 =e^{2\pi i\langle z_4,\alpha_4\rangle}\ket0.
 \label{spt:d2-response}
\end{equation}
The sign is the fermionic exchange phase of the point defects.
The trivial cocycle gives $+1$. The quadratic cycle $z_4$ in
Eq.~\eqref{spt:d2-cycle} is a cycle in the full 2-group and has
pairing $1/2$ with $\alpha_4$. Since $\alpha_4$ generates the
full degree-4 cohomology, Theorem~\ref{dc:correspondence-theorem}
identifies the exchange with this detecting class.

%% file: sections/spt_d3_typeI.tex
\subsection{\texorpdfstring{$(3{+}1)$D}{(3+1)D}: Loop flip}
\label{spt:typeI}

For the three-dimensional response in Eq.~\eqref{spt:anomalies},
the two processes give the following outcomes.
\begin{equation}
 \begin{array}{c|cc}
  (\mu,\nu)&\mu_{24}&\mathcal N_{50}\\ \hline
  (0,0)&+1&+1\\
  (1,0)&-1&+1\\
  (0,1)&+1&-1\\
  (1,1)&-1&-1
 \end{array}
 \label{spt:response-table}
\end{equation}

The first 5-cocycle can be detected without inserting an ordinary
symmetry wall. Set $A=0$, so that $\delta B=0$. Its restriction is
\begin{equation}
 \alpha_5^{\mathrm I}\big|_{A=0}
 =\frac12(B\cup_1B)\cup_1(B\cup_1B)
 =\frac12\Sq^2\Sq^1B.
 \label{spt:typeI-restriction}
\end{equation}
This is the fermionic-loop response for the 1-form defects~\cite{Thorngren2015Framed, Fidkowski2022Gravitational, Chen2023Exactly}.
The representative $\tfrac12 B\cup\Sq^1B$ used in
Ref.~\cite{Kobayashi2026Generalized} gives the same nonzero
wall-free anomaly class. We use the process $\mu_{24}$ from
Eq.~(26) of that reference. On the prepared loop
$1\to2\to3\to4\to1$, the published process reverses its orientation.
Here we run the same signed word directly from the vacuum, where
intermediate states can be closed loop networks.

A wall-free configuration on $\partial[01234]$ is specified by
its closed loop chain. The occupation of each physical edge is
the $B$ bit on the complementary triangle. For example, $B_{012}$
is the occupation of edge $34$. The constraint $\delta B=0$
requires even occupation at every vertex. To implement the word,
use the gate $\widehat E_{uv,q}$ for the membrane $U_{ijk}$ on
the triangle $\langle ijk\rangle$ complementary to $uv$.
Either sign of the membrane toggles the three edges of
$\partial\langle ijk\rangle$ at fixed $A=0$. Write the response as
$\Omega_5=q/2$. In the potential basis $p=(a,b)$, the gate of
Eq.~\eqref{spt:microscopic-cone-gate} acts by
\begin{equation}
 \begin{aligned}
 \widehat E_{uv,q}\ket p
 &=(-1)^{R_q(p)+R_q(p')+q(c_{E_{uv}}(\cfg(p)))}\ket{p'},\\
 p'&=(a,b+\mathbf1_{uv}).
 \end{aligned}
 \label{spt:typeI-membrane-gate}
\end{equation}
These edge gates commute with both global symmetries by the
argument in Sec.~\ref{spt:boundary-gates}.

Each membrane appears twice with each sign in $\mu_{24}$, so
the process restores the boundary potentials. The $R_q$ terms
therefore telescope. The local phase shifts also cancel when the forward inputs
and inverse outputs are grouped by support vertex as in
Eq.~\eqref{loc:cancellation-main}. For example, the forward
$U_{023}$ at step $11$ pairs with step $1$ at vertex $0$ and
with step $18$ at vertex $2$. The
type-I numerator gives a minus sign at steps $16$, $17$, and
$21$ in execution order. The type-II numerator vanishes because
$A=0$ on every move cylinder. Thus for
$\Omega_5=\mu\alpha_5^{\mathrm I}+\nu\alpha_5^{\mathrm{II}}$,
\begin{equation}
 \mu_{24}\ket0=(-1)^\mu\ket0.
 \label{spt:typeI-phase}
\end{equation}

The wall-free cycle $z_{\mathrm I}$ in Eq.~\eqref{spt:typeI-cycle}
has the same two anomaly pairings. Theorem~\ref{dc:correspondence-theorem}
identifies the class of $\mu_{24}$ with $[J(z_{\mathrm I})]$.

%% file: sections/spt_d3_typeII.tex
\subsection{\texorpdfstring{$(3{+}1)$D}{(3+1)D}: Wall--loop statistics}
\label{spt:mixed}

The factor $A$ in $\alpha_5^{\mathrm{II}}=A\cup Q_4/2$ makes its
response vanish when there are no walls. To isolate this anomaly
from $\alpha_5^{\mathrm I}$, we use a process whose two signs are
$(+1,-1)$. Appendix~\ref{spt:mixed-details} explains its detector.

We now keep both the wall faces and the occupied loop edges in
$\cfg=(A,B)$ on $\partial[01234]$. The two chains in
Eq.~\eqref{spt:physical-chains} label these cells, and the
junctions are fixed by $\delta B=A\cup A\cup A$. An edge move
$E_{ij}$ toggles the three loop edges around the triangle
complementary to $ij$ at fixed $A$. A wall move $V_i$ toggles the
four wall faces of the tetrahedron opposite $i$, and can change
the loop chain through the $B$ update in Eq.~\eqref{spt:wall-update}.
The vacuum has both chains empty. Starting there, apply
\begin{equation}
 \begin{aligned}
 \mathcal N_{50}={}&V_{0}^{-1}E_{04}E_{23}E_{04}^{-1}E_{24}E_{34}E_{23}^{-1}\\
 &\times E_{13}^{-1}E_{24}^{-1}E_{13}V_{0}V_{4}^{-1}E_{23}V_{3}^{-1}\\
 &\times E_{04}^{-1}V_{2}^{-1}E_{24}^{-1}E_{04}V_{2}V_{3}E_{02}^{-1}\\
 &\times V_{4}V_{0}^{-1}E_{34}^{-1}V_{0}E_{02}E_{13}E_{02}^{-1}\\
 &\times E_{23}V_{4}^{-1}E_{04}^{-1}V_{3}^{-1}V_{2}^{-1}E_{34}^{-1}V_{2}E_{04}\\
 &\times E_{02}V_{2}^{-1}E_{04}^{-1}E_{34}E_{24}V_{2}E_{02}^{-1}\\
 &\times V_{3}E_{23}^{-1}E_{02}E_{04}V_{4}E_{13}^{-1}E_{23}^{-1}
 \end{aligned}
 \label{spt:mixed-protocol}
\end{equation}
with the rightmost operator first. The wall and membrane moves
restore the full configuration to the vacuum. For each move and
support vertex, the forward and inverse occurrences can be paired
with matching complete local inputs. The configurations in
Table~\ref{spt:mixed-history-1} therefore satisfy the local
cancellation condition in Eq.~\eqref{loc:cancellation-main}, so
all local phase shifts cancel.

To evaluate the process in the anomalous boundary action, take the
symmetry-commuting hopping operators of
Eq.~\eqref{spt:microscopic-cone-gate} on the 1-form-invariant sector.
Track the component with $a=b=0$ of the vacuum state. Just before
step 45 the vertex potential is $a=e_0$ and the edge potential is
$b=e_{23}+e_{24}+e_{34}$, where $e_0$ is the unit vertex cochain
at $0$. For the type-II action, the cone phase is
zero on both sides of this hop. The cylinder has
$A_{01}=B_{123}=B_{345}=1$ and all $A_{ij}=0$ for $i,j\geq1$.
Thus the term $A\cup(B\cup B)$ supplies the sign in the
microscopic matrix element
\begin{equation}
 \widehat E_{34,\mathrm{II}}\ket{e_0,e_{23}+e_{24}+e_{34}}
   =-\ket{e_0,e_{23}+e_{24}} .
\end{equation}
Here the subscript $\mathrm{II}$ selects the binary exponent
$q=A\cup Q_4$.
This is the only negative matrix element for type II on this
vacuum component. For type I, the negative factors occur an even
number of times, as listed in Table~\ref{spt:nu-step-phases}.
Each vertex bit flips an even number of times. Only $V_i$, $V_j$,
and $E_{ij}$ affect $b_{ij}$. Substituting their updates into
the word gives zero net shift for each of the four initial
values of $(a_i,a_j)$. Thus the word restores every preparation,
and the cone factors telescope. In the action with
$\Omega_5=\mu\alpha_5^{\mathrm I}+\nu\alpha_5^{\mathrm{II}}$,
\begin{equation}
 \mathcal N_{50}\ket0=(-1)^\nu\ket0.
 \label{spt:mixed-response}
\end{equation}
Together with the wall-free sign $(-1)^\mu$, this distinguishes
the four boundary actions in Eq.~\eqref{spt:response-table}.

The cycle $z_\nu$ in Appendix~\ref{spt:mixed-details} has the
same pairings with the two generators, so
Theorem~\ref{dc:correspondence-theorem} identifies its class
with that of the process.

%% file: figures/inversion_square_hopping_figure.tex
\begin{figure*}[t]
 \centering
 \input{figures/inversion_square_hopping.tex}
 \caption{Local hopping operators on the square lattice.
 (a) The wall gate $U_v^{(\epsilon)}$ flips the four incident
 wall qubits, reverses outgoing electric labels, and sends
 $q_v$ to $-q_v$. Acting on the vacuum creates a closed dual
 square. The shaded star contains the labels read by the phase;
 gray leaders point to the corresponding edges. Panels (a,b)
 use the electric basis.
 (b) At $A=0$, $\widetilde Z_{ij}^{-1}$ shifts $\lambda_{ij}$
 to $\lambda_{ij}-1$ and transfers one unit from $i$ to $j$.
 Eq.~\eqref{ex:inversion-electric-endpoints-main} gives the
 wall-dependent endpoint charges.
 (c) $\widetilde X_e$ leaves $\lambda_e$ fixed and moves flux
 from the lower plaquette to the upper one at $A=0$.
 In (b,c), labels give the occupations before and after the hop.
 Acting on the vacuum creates a $(-1,+1)$ pair of the indicated
 excitation type and preserves the other type and the walls.}
 \label{fig:inversion-square-hopping}
\end{figure*}
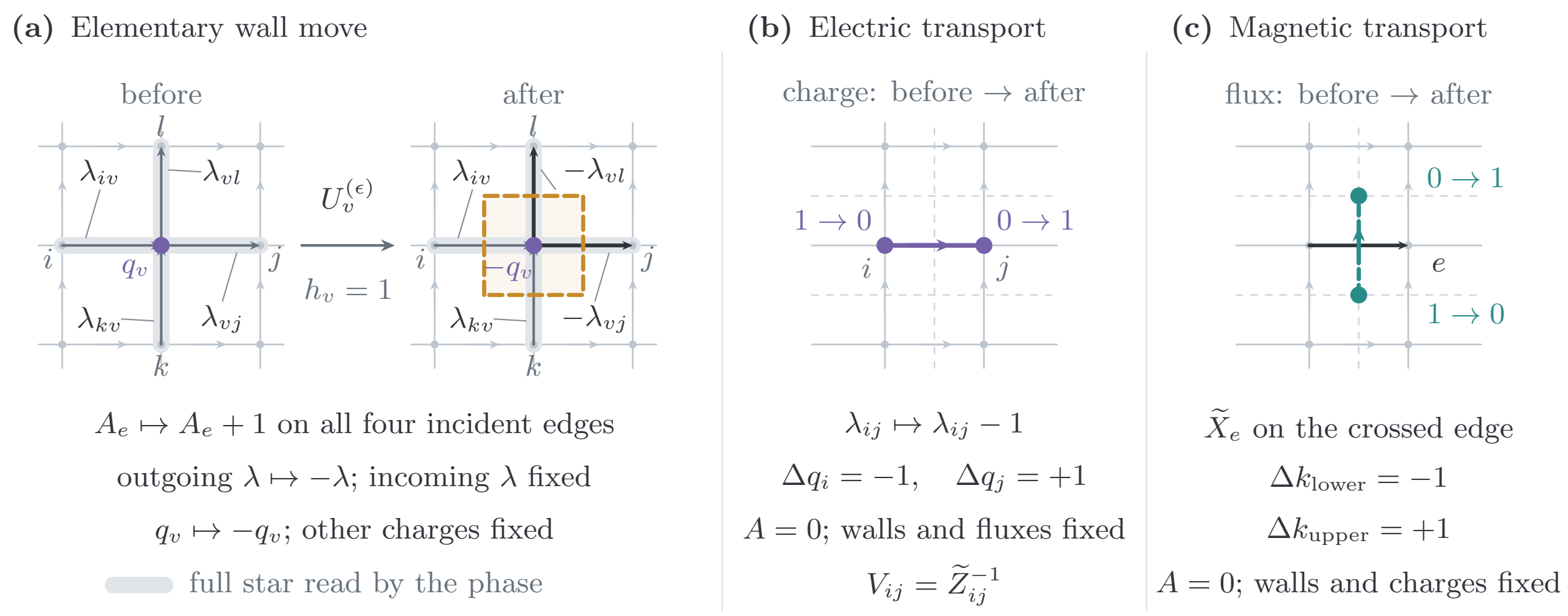

%% file: figures/inversion_square_hopping.tex
\begingroup
\definecolor{sqwall}{HTML}{C88A2A}
\definecolor{sqelectric}{HTML}{7561A8}
\definecolor{sqflux}{HTML}{2A8C89}
\definecolor{sqink}{HTML}{30343B}
\definecolor{sqgray}{HTML}{65717C}
\definecolor{sqmesh}{HTML}{BDC6CE}
\begin{tikzpicture}[x=1cm,y=1cm,>=Stealth,
 font=\fontsize{9}{11}\selectfont,line cap=round,line join=round,
 mesh/.style={draw=sqmesh,line width=.5pt},
 dual/.style={draw=sqmesh!75,line width=.45pt,dash pattern=on 2pt off 2.4pt},
 read/.style={draw=sqmesh!45,line width=5pt},
 field/.style={text=sqink,inner sep=1pt},
 callout/.style={draw=sqgray!80,line width=.4pt},
 corner/.style={text=sqgray,inner sep=1pt},
 edgearrow/.style={-{Stealth[length=3.1pt,width=2.6pt]}},
 head/.style={anchor=west,text=sqink}]
 \path[use as bounding box] (0,0) rectangle (16.8,6.6);
 \draw[sqmesh!55,line width=.4pt] (7.8,.18)--(7.8,6.1);
 \draw[sqmesh!55,line width=.4pt] (12.3,.18)--(12.3,6.1);
 \node[head] at (.14,6.34) {\textbf{(a)}\enspace Elementary wall move};
 \node[head] at (7.94,6.34) {\textbf{(b)}\enspace Electric transport};
 \node[head] at (12.44,6.34) {\textbf{(c)}\enspace Magnetic transport};

 \foreach \cx in {1.85,5.8} {
  \begin{scope}[shift={(\cx,4.05)}]
   \ifdim\cx pt>4pt
    \fill[sqwall!7] (-.525,-.525) rectangle (.525,.525);
   \fi
   \foreach \s in {-1.05,0,1.05} {
    \draw[mesh] (-1.3,\s)--(1.3,\s);
    \draw[mesh] (\s,-1.3)--(\s,1.3);
   }
   \draw[read] (-1.05,0)--(1.05,0);
   \draw[read] (0,-1.05)--(0,1.05);
   \foreach \x in {-1.05,0,1.05} {
    \foreach \y in {-1.05,0,1.05} {
     \fill[sqmesh] (\x,\y) circle[radius=1.2pt];
    }
   }
   \foreach \x in {-1.05,0} {
    \foreach \y in {-1.05,1.05} {
     \draw[mesh,edgearrow] (\x+.36,\y)--(\x+.69,\y);
    }
   }
   \foreach \x in {-1.05,1.05} {
    \foreach \y in {-1.05,0} {
     \draw[mesh,edgearrow] (\x,\y+.36)--(\x,\y+.69);
    }
   }
   \draw[sqgray,line width=.75pt,edgearrow] (-1.05,0)--(0,0);
   \draw[sqgray,line width=.75pt,edgearrow] (0,0)--(1.05,0);
   \draw[sqgray,line width=.75pt,edgearrow] (0,-1.05)--(0,0);
   \draw[sqgray,line width=.75pt,edgearrow] (0,0)--(0,1.05);
   \node[corner,anchor=east] at (-1.1,-.14) {$i$};
   \node[corner,anchor=west] at (1.1,-.14) {$j$};
   \node[corner,anchor=north] at (0,-1.12) {$k$};
   \node[corner,anchor=south] at (0,1.12) {$l$};
  \end{scope}
 }
 \node[text=sqgray] at (1.85,5.66) {before};
 \node[text=sqgray] at (5.8,5.66) {after};
 \draw[sqgray,line width=.8pt,-{Stealth[length=4.3pt,width=3.6pt]}]
  (3.34,4.05)--(4.33,4.05);
 \node[text=sqink,anchor=south] at (3.835,4.29)
  {$U_v^{(\epsilon)}$};
 \node[text=sqgray,anchor=north] at (3.835,3.82) {$h_v=1$};

 \begin{scope}[shift={(1.85,4.05)}]
  \fill[sqelectric] (0,0) circle[radius=2.55pt];
  \node[field,text=sqelectric] at (-.27,-.25) {$q_v$};
  \node[field] at (-.65,.78) {$\lambda_{iv}$};
  \node[field] at (.65,-.78) {$\lambda_{vj}$};
  \node[field] at (-.65,-.78) {$\lambda_{kv}$};
  \node[field] at (.65,.78) {$\lambda_{vl}$};
  \draw[callout] (-.65,.59)--(-.80,.08);
  \draw[callout] (.65,-.59)--(.80,-.08);
  \draw[callout] (-.42,-.78)--(-.08,-.80);
  \draw[callout] (.42,.78)--(.08,.80);
 \end{scope}
 \begin{scope}[shift={(5.8,4.05)}]
  \draw[sqink,line width=1.15pt,edgearrow] (0,0)--(1.05,0);
  \draw[sqink,line width=1.15pt,edgearrow] (0,0)--(0,1.05);
  \draw[sqwall,line width=1.25pt,dash pattern=on 4pt off 2pt]
   (-.525,-.525) rectangle (.525,.525);
  \fill[sqelectric] (0,0) circle[radius=2.55pt];
  \node[field,text=sqelectric] at (-.27,-.25) {$-q_v$};
  \node[field] at (-.65,.78) {$\lambda_{iv}$};
  \node[field] at (.65,-.78) {$-\lambda_{vj}$};
  \node[field] at (-.65,-.78) {$\lambda_{kv}$};
  \node[field,inner sep=2pt] (outgoingTop) at (.65,.80) {$-\lambda_{vl}$};
  \draw[callout] (-.65,.59)--(-.80,.08);
  \draw[callout] (.65,-.59)--(.80,-.08);
  \draw[callout] (-.42,-.78)--(-.08,-.80);
  \draw[callout] (outgoingTop.south west)--(.08,.80);
 \end{scope}
 \node[text=sqink] at (3.9,2.14)
  {$A_e\mapsto A_e+1$ on all four incident edges};
 \node[text=sqink] at (3.9,1.58)
  {outgoing $\lambda\mapsto-\lambda$; incoming $\lambda$ fixed};
 \node[text=sqink] at (3.9,1.02)
  {$q_v\mapsto-q_v$; other charges fixed};
 \draw[read] (1.34,.45)--(1.89,.45);
 \node[text=sqgray,anchor=west] at (2.03,.45)
  {full star read by the phase};

 \begin{scope}[shift={(10.05,4.05)}]
  \foreach \s in {-.525,.525} {
   \draw[mesh] (\s,-1.3)--(\s,1.3);
  }
  \foreach \s in {-1.05,0,1.05} {
   \draw[mesh] (-1.3,\s)--(1.3,\s);
  }
  \draw[dual] (0,-1.3)--(0,1.3);
  \foreach \s in {-.525,.525} {
   \draw[dual] (-1.3,\s)--(1.3,\s);
  }
  \foreach \x in {-.525,.525} {
   \foreach \y in {-1.05,0,1.05} {
    \fill[sqmesh] (\x,\y) circle[radius=1.2pt];
   }
  }
  \foreach \y in {-1.05,1.05} {
   \draw[mesh,edgearrow] (-.15,\y)--(.19,\y);
  }
  \foreach \x in {-.525,.525} {
   \foreach \y in {-1.05,0} {
    \draw[mesh,edgearrow] (\x,\y+.36)--(\x,\y+.69);
   }
  }
  \draw[sqelectric,line width=1.4pt] (-.525,0)--(.525,0);
  \draw[sqelectric,line width=1.4pt,
    -{Stealth[length=4.5pt,width=3.7pt]}] (-.3,0)--(.18,0);
  \fill[sqelectric] (-.525,0) circle[radius=2.55pt];
  \fill[sqelectric] (.525,0) circle[radius=2.55pt];
  \node[field,text=sqelectric,anchor=east] at (-.62,.27) {$1\to0$};
  \node[field,text=sqelectric,anchor=west] at (.62,.27) {$0\to1$};
  \node[corner,anchor=east] at (-.62,-.24) {$i$};
  \node[corner,anchor=west] at (.62,-.24) {$j$};
 \end{scope}
 \node[text=sqgray] at (10.05,5.66) {charge: before $\to$ after};
 \node[text=sqink] at (10.05,2.14) {$\lambda_{ij}\mapsto \lambda_{ij}-1$};
 \node[text=sqink] at (10.05,1.58) {$\Delta q_i=-1,\quad\Delta q_j=+1$};
 \node[text=sqink] at (10.05,1.02) {$A=0$; walls and fluxes fixed};
 \node[text=sqink] at (10.05,.45) {$V_{ij}=\widetilde Z_{ij}^{-1}$};

 \begin{scope}[shift={(14.55,4.05)}]
  \foreach \s in {-.525,.525} {
   \draw[mesh] (\s,-1.3)--(\s,1.3);
  }
  \foreach \s in {-1.05,0,1.05} {
   \draw[mesh] (-1.3,\s)--(1.3,\s);
  }
  \draw[dual] (0,-1.3)--(0,1.3);
  \foreach \s in {-.525,.525} {
   \draw[dual] (-1.3,\s)--(1.3,\s);
  }
  \foreach \x in {-.525,.525} {
   \foreach \y in {-1.05,0,1.05} {
    \fill[sqmesh] (\x,\y) circle[radius=1.2pt];
   }
  }
  \foreach \y in {-1.05,1.05} {
   \draw[mesh,edgearrow] (-.15,\y)--(.19,\y);
  }
  \foreach \x in {-.525,.525} {
   \foreach \y in {-1.05,0} {
    \draw[mesh,edgearrow] (\x,\y+.36)--(\x,\y+.69);
   }
  }
  \draw[sqink,line width=1.1pt,edgearrow] (-.525,0)--(.525,0);
  \draw[sqflux,line width=1.3pt,dash pattern=on 3pt off 1.8pt]
   (0,-.525)--(0,.525);
  \draw[sqflux,line width=1.3pt,
    -{Stealth[length=4.5pt,width=3.7pt]}] (0,-.17)--(0,.2);
  \fill[sqflux] (0,-.525) circle[radius=2.55pt];
  \fill[sqflux] (0,.525) circle[radius=2.55pt];
  \node[field,text=sqflux,anchor=west] at (.68,-.72) {$1\to0$};
  \node[field,text=sqflux,anchor=west] at (.68,.72) {$0\to1$};
  \node[field,anchor=north] at (.85,-.10) {$e$};
 \end{scope}
 \node[text=sqgray] at (14.55,5.66) {flux: before $\to$ after};
 \node[text=sqink] at (14.55,2.14) {$\widetilde X_e$ on the crossed edge};
 \node[text=sqink] at (14.55,1.58) {$\Delta k_{\rm lower}=-1$};
 \node[text=sqink] at (14.55,1.02) {$\Delta k_{\rm upper}=+1$};
 \node[text=sqink] at (14.55,.45) {$A=0$; walls and charges fixed};
\end{tikzpicture}
\endgroup

%% file: figures/inversion_framed_cancellation_table.tex
\begingroup
\makeatletter\def\@captype{table}\makeatother
\renewcommand{\arraystretch}{1.12}
 \centering
 \small
 \begin{tabular}{c|l|l|l}
 \hline
 Move & Neighborhood & Paired steps & Local forward input\\ \hline
 $V_{0'1'}$ & entire path & $(1,6)$ & vacuum\\ \hline
 $U_{123}$ & charge site $0'$: $q_{0'}$
   & $(2,8)$; $(4,10)$ & $-1$; $0$\\
 & black wall vertices: incident wall tuple
   & $(2,10)$; $(4,8)$ & $(0,0,0)$; $(0,1,1)$\\
 & open wall segments, including crossings: $c$
   & $(2,10)$; $(4,8)$ & $0$; $1$\\ \hline
 $V_{0'2'}$ & source $0'$: $q_{0'}$
   & $(3,7)$; $(5,9)$ & $+1$; $0$\\
 & destination $2'$: $q_{2'}$
   & $(3,9)$; $(5,7)$ & $0$; $-1$\\
 & wall crossing: $c$
   & $(3,9)$; $(5,7)$ & $1$; $0$\\ \hline
 \end{tabular}
 \caption{Local phase cancellation for the ten moves of
 $\mathcal W_{10}$ in Fig.~\ref{fig:inversion-framed-history}.
 Each pair lists a forward step followed by its inverse partner.
 The local configuration before the forward step equals that
 after the inverse step. Semicolon-separated inputs follow the
 order of the pairs. Charges are modulo four, with $3$ written
 as $-1$, and wall bits are modulo two. The incident wall tuples
 at black vertices $1,2,3$ are $(C_{01},C_{12},C_{13})$,
 $(C_{02},C_{12},C_{23})$, and $(C_{03},C_{13},C_{23})$,
 respectively; all equal $(0,c,c)$. At wall crossings, all local
 charges are zero in the compared configurations, so $c$ records
 the remaining local input. Unlisted neighborhoods have vacuum
 local labels and balanced forward and inverse counts.
 }
 \label{ex:inversion-local-pairings-main}
 \label{app:tc:inversion-primitive-pairings}
\endgroup

%% file: sections/inversion_operator_details.tex
\paragraph*{Electric and magnetic hopping.}
For the source-frame hop $V_{ij}=\widetilde Z_{ij}^{-1}$ in
Eq.~\eqref{ex:inversion-single-edge-main}, the endpoint relations are
\begin{equation}
 \begin{aligned}
 V_{ij}^\dagger\widehat{\mathcal A}_iV_{ij}
   &=i^{-1}\widehat{\mathcal A}_i,\\
 V_{ij}^\dagger\widehat{\mathcal A}_jV_{ij}
   &=i^{Z_{ij}}\widehat{\mathcal A}_j.
 \end{aligned}
 \label{ex:inversion-electric-endpoints-main}
\end{equation}
Thus the charge increments are $-1$ at the source and
$(-1)^{A_{ij}}$ at the target. Transport along a blue path in
Fig.~\ref{fig:inversion-framed-history} uses this same rule in
the successive local charge frames. Crossing a wall reverses
the transported unit, giving Eq.~\eqref{ex:inversion-updates-main}.
The magnetic hop $\widetilde X_e$ moves flux between adjacent
plaquettes and leaves the charge and wall labels fixed.
Fig.~\ref{fig:inversion-square-hopping} displays these local gates.

\paragraph*{Microscopic locality.}
We use the same local gates and their strict inverses at every
occurrence of a transfer in $\mathcal W_{10}$. Electric transport
is a prescribed sequence of single-edge shifts. The wall gate
permutes registers edge by edge and adds phases on edges and
vertex stars. Bounded valence permits a fixed number of circuit
layers, with fixed collars around their supports.

In the Fourier basis, each microscopic generator's phase is a
sum of functions on these edge and star neighborhoods. Its
register permutations preserve the registers read by each
function, so restriction to a neighborhood respects composition
and taking inverses. Include a neighborhood in a gate's prescribed
support whenever its registers change or a phase assigned there
is read. If a nested commutator has empty common support, each
neighborhood has a factor that restricts to the identity.
The restricted commutator then has trivial permutations and zero
phase. This proves Eq.~\eqref{loc:locality-identity} for the
microscopic gates at every nesting depth. The ten completed
transfers in Fig.~\ref{fig:inversion-framed-history} use these
fixed gates; the local cancellation pairs are given below.

%% file: sections/inversion_framed_details.tex
\subsection{The process on the separated lattices}
\label{app:tc:inversion-framed}

We use the configurations and ten moves of
Fig.~\ref{fig:inversion-framed-history}. The black triangle $123$
bounds the region containing $0'$, and $q_{i'}$ denotes the charge at
blue vertex $i'$. Write $c=0,1$ for the absence or presence of the
wall on this triangle. Eq.~\eqref{app:tc:inversion-framed-dictionary}
then reads $(-1)^cq_{0'}+q_{1'}+q_{2'}+q_{3'}=0$ modulo four.
All other charges and all magnetic fluxes are zero at the end of
each displayed move. These labels specify the configuration ray
in Eq.~\eqref{app:tc:inversion-reduced-state}.

The graphs admit a bounded-valence microscopic refinement on $S^2$.
Choose the charge sites, black vertices, and distinct path crossings
farther apart than the range of the allowed local modifications,
including the fixed collars of the hopping circuits. This lets a
local neighborhood detect the wall and charges at a crossing
without also reading a distant endpoint or spectator charge.

\paragraph*{The ten moves.}
The wall hop $U_{123}$ toggles $c$ and reverses $q_{0'}$. The charge
hop $V_{0'j'}$ removes one unit at $0'$ and transports it along the
fixed blue path to $j'$, reversing its sign at each wall crossing.
The process uses $V_{0'1'}$ only without a wall. The two forward occurrences
of $V_{0'2'}$ therefore change $(q_{0'},q_{2'})$ by $(-1,-1)$ and
$(-1,+1)$, respectively.

Orient the microscopic path edges from $0'$ outward. Each completed
transfer is implemented by prescribed products of
the microscopic edge hops of
Eq.~\eqref{ex:inversion-single-edge-main}. The path fixes their
order and the transported charge fixes their signs. The same edge
gates are reused throughout, with strict inverses in the reverse
transfers; their phases are not chosen independently for different
wall backgrounds. The chronological sequence is
\begin{equation}
 \begin{gathered}
 V_{0'1'},\ U_{123},\ V_{0'2'},\ U_{123},\ V_{0'2'},\\
 V_{0'1'}^{-1},\ V_{0'2'}^{-1},\ U_{123}^{-1},\
 V_{0'2'}^{-1},\ U_{123}^{-1}.
 \end{gathered}
 \label{app:tc:inversion-framed-ten-moves}
\end{equation}
\begingroup\makeatletter
\edef\@currentlabel{\p@equation\theequation}\label{app:tc:inversion-primitive-word}
\makeatother\endgroup
These are precisely the ten arrows of
Eq.~\eqref{ex:inversion-word-main}. Their updates reproduce every
configuration in Fig.~\ref{fig:inversion-framed-history} and return
to the vacuum.

\paragraph*{The statistical phase.}
In the state convention of
Eq.~\eqref{app:tc:inversion-reduced-state}, the electric transfers
used here have unit phase. Their electric-label shifts pair
trivially with the flat representative $E^0(A)$: the wall-free
parts have $A_{ij}=0$, and each outward crossing ends in the
exterior reference frame. Comparing
Eq.~\eqref{ex:inversion-wall-main} with the configuration states
gives the edge contribution
$(-1)^{\epsilon\sum_{v\text{ inside}}A(\gamma_v)Q_v}$.
Here $A(\gamma_v)=c$ inside the triangle and only $q_{0'}$ can be
nonzero. Including the charge-dependent sign in the wall gate gives
\begin{equation}
 e^{i\theta_{U_{123}}(C,q)}
 =(-1)^{\epsilon(cq_{0'}+\mathbf1_{q_{0'}=1})}.
 \label{app:tc:inversion-framed-wallphase}
\end{equation}
For an inverse, this phase is inverted and evaluated at the
outgoing configuration, the corresponding forward input.
The wall phases at steps $2,4,8,10$ are therefore
$1,1,(-1)^\epsilon,1$, respectively. Hence
\begin{equation}
 \mathcal W_{10}^{(\epsilon)}\ket{\vac}^{(\epsilon)}
 =(-1)^\epsilon\ket{\vac}^{(\epsilon)}.
 \label{app:tc:inversion-framed-response}
\end{equation}
This is a return to the complete configuration ray, including the
microscopic charge and flux labels.

\paragraph*{Local phase cancellation.}
\label{app:tc:inversion-framed-local}
Table~\ref{ex:inversion-local-pairings-main} uses the same step
numbers as Fig.~\ref{fig:inversion-framed-history}. Each pair
compares the configuration before a forward move with that after
an inverse move. At $0'$, for example, the wall hops pair as
$(2,8)$ and $(4,10)$, whereas near the wall they pair as
$(2,10)$ and $(4,8)$. For $V_{0'2'}$, the endpoint pairings also
differ: $(3,7)$ and $(5,9)$ at $0'$, but $(3,9)$ and $(5,7)$ at
$2'$. At a wall crossing the latter pairs have the same wall bit
and zero local charges.

The comparison retains the full microscopic local state. At fixed
wall registers the edge model is a generalized Pauli stabilizer
state. Its reduced density operator on a contractible disk depends
only on the flux operators and charges enclosed by that disk.
Consequently, matching the wall bits and enclosed charges, with
vacuum fluxes fixed, also matches the local electric data.
The crossing comparison includes both sides of the wall. During a
prescribed transfer, the incoming unit and the crossing bit fix
the temporary local charges; the paired inverse follows the same
local update in reverse. The microscopic edge-hop contributions
therefore cancel within each pair of completed transfers as well.
The table establishes Eq.~\eqref{loc:cancellation-main} in every
allowed neighborhood, so the sign is unaffected by local phase
changes of the hopping operators.

%% file: sections/fermionic_toric_details.tex
\section{Microscopic calculation of the Kitaev-chain process}
\label{app:ftc-calculation}

We give the microscopic calculation of the process $\mathcal K_{24}$
introduced in Sec.~\ref{sec:fermionic-toric}. We specify the
configuration states and membrane operators, show how local phase
changes cancel, and obtain the phase $+i$. We then explain how the
5-cocycle in Eq.~\eqref{eq:ftc-restricted-anomaly} gives the same
statistical phase.

\subsection{Majorana pairings and the membrane operator}

\paragraph*{Configuration states.}
The physical lattice in Fig.~\ref{fig:kitaev-process} has vertices
$0,\ldots,4$. A configuration $\cfg=(\ell,f)$ records the chain
occupation $\ell_{uv}$ of each edge and the fermion occupation
$f_v$ of each vertex. Both are binary. The chain forms a closed
even network: an even number of occupied edges meet at every
vertex. The allowed point occupations satisfy Eq.~\eqref{eq:ftc-parity}.
\begin{equation*}
 \sum_v f_v=\ell_{01}\ell_{34}\pmod2.
\end{equation*}
The label $\ell_{uv}$ specifies the chain geometry. The occupation
of the microscopic edge fermion is a separate Fock-basis label.

We put a complex fermion $c_v$ with $\chi_v=c_v+c_v^\dagger$ at each
physical vertex $v$. On each edge $uv$, with $u<v$, we put a complex
fermion $a_{uv}$ whose Majoranas are
$\gamma_{u,v}=a_{uv}+a_{uv}^\dagger$ at $u$ and
$\gamma_{v,u}=i(a_{uv}-a_{uv}^\dagger)$ at $v$.
All modes obey the canonical fermion anticommutation relations.
For a directed dimer $a\to b$, where $a,b$ label Majoranas, the
projector is
\begin{equation}
 P_{a\to b}=\frac{1-i\gamma_a\gamma_b}{2}.
 \label{eq:ftc-dimer-projector}
\end{equation}
On an empty edge $uv$ with $u<v$, the arrow points from $v$ to
$u$. With the Majoranas defined above, this projector is
$1-a_{uv}^\dagger a_{uv}$, so it selects the empty edge mode.
The Majoranas on occupied edges instead pair at their common
vertex. When two occupied edges meet, their endpoint Majoranas
form a single pair. When four meet, list the neighboring vertices
in increasing order and pair the first with the third and the second
with the fourth. For a pair at $v$ with neighbors $u<w$, the arrow
points from $\gamma_{v,w}$ to $\gamma_{v,u}$, except for the following
six pairs, whose arrows are reversed.
\begin{equation}
 \begin{gathered}
 (v;u,w)=(0;1,4),(1;2,3),(2;0,4),\\
 (3;1,2),(4;0,1),(4;0,2).
 \end{gathered}
 \label{eq:ftc-internal-arrows}
\end{equation}
These rules pair every edge Majorana exactly once and specify a
normalized chain state $\ket{D(\ell)}$ up to phase. The rule at a
degree-four vertex is needed for the locality check below. Every
configuration in the process $\mathcal K_{24}$ itself has a single
loop with degree two at its occupied vertices.
Relative to the empty-edge vacuum, the chain state has parity
$(-1)^{\ell_{01}\ell_{34}}$. One can obtain this sign by multiplying its
ten directed dimer stabilizers and comparing the resulting Majorana
ordering with the vacuum ordering. The configuration state
$\ket{\ell,f}$ consists of the orthogonal loop label, this dimer
state, and the specified point occupations. The parity constraint
therefore makes the combined chain and point parity even; it does
not determine the individual $f_v$ from $\ell$.

\paragraph*{Phase convention.}
For the matrix calculation, order the point modes as
$c_0,\ldots,c_4$, followed by the edge modes $a_{uv}$ with $u<v$
in lexicographic order of $(u,v)$.
Number these modes by $j=0,\ldots,14$. In the Fock vector
$\ket{N_0,\ldots,N_{14}}$, write the occupied creation operators
from left to right in increasing mode order, acting on the vacuum.
Order the basis vectors by the
integer $N=\sum_j2^jN_j$. For each chain and point configuration,
choose the normalized configuration state whose first nonzero
coefficient in this order is real and positive. This fixes the
individual step phases in Table~\ref{tab:ftc-complete-history}.

\paragraph*{Membrane action.}
A membrane $M_s$ on triangle $s$ toggles its three boundary edges,
$\ell\mapsto\ell+\partial s$. Its point update can be read directly
from the physical occupations. At a vertex $v\in s$, list the
other four vertices as $a<b<c<d$. Then
\begin{equation*}
 f_v\longmapsto f_v+
 \begin{cases}
  \ell_{vd},&s=\{v,a,b\},\\
  \ell_{va},&s=\{v,c,d\},\\
  0,&\text{otherwise},
 \end{cases}
 \qquad\pmod2.
\end{equation*}
Vertices outside $s$ are unchanged. This is the cochain update in
footnote~\ref{fn:ftc-cochain-updates} written in the physical variables; it
reproduces Eq.~\eqref{eq:ftc-direct} for the six membranes in the
process. In particular, each point toggle is controlled by an
occupied edge outside the flipped triangle.

As in the main text, $V_s(\ell)$ is the set of vertices whose point
occupations change, and $vu(v)$ is the occupied edge controlling
the change at $v$. For a fixed incoming $\ell$, the fermionic
block of Eq.~\eqref{eq:ftc-membrane-main} is
\begin{equation*}
 \mathcal N_s(\ell)\,P_{\rm out}
 \left[\prod_{v\in V_s(\ell)}
 (-i\chi_v\gamma_{v,u(v)})\right]P_{\rm in}.
\end{equation*}
Here $P_{\rm in}$ and $P_{\rm out}$ select the incoming and outgoing
dimers that change during the move; the remaining dimers are
unchanged. The factor $-i\chi_v\gamma_{v,u(v)}$ transfers parity
between the point mode and the chain at $v$, making the outgoing
projection compatible with the prescribed point occupations.
Each factor is Hermitian, squares to one, and has even fermion
parity. Factors at different vertices commute.

Superimpose the incoming and outgoing dimer matchings. Each
toggled edge supplies an across-edge dimer in the matching in
which it is empty. At each vertex of the membrane, the local
pairings connect the endpoint Majoranas of the two toggled edges,
possibly through other Majoranas at that vertex. Together these
links form one alternating cycle of changed dimers around the
triangle. A controlling edge is paired into this cycle because
the two toggled neighbors are the first two or the last two in
the local neighbor order.

A cycle of $2L_s$ Majoranas imposes $L_s-1$ independent outgoing
stabilizer conditions, each contributing a factor $1/2$ to the
squared overlap. The remaining condition fixes the cycle parity.
Since the other dimers are unchanged, the point--chain insertions
compensate the required parity change on the cycle. The projected
state has norm $2^{-(L_s-1)/2}$, and
$\mathcal N_s(\ell)=2^{(L_s-1)/2}$ gives a unit-norm outgoing state.
For every step of $\mathcal K_{24}$, $2L_s=8$ and
$\mathcal N_s(\ell)=\sqrt8$. The
normalized membrane maps each allowed configuration state to its
prescribed successor with a unit-modulus amplitude. Since the update
is invertible, its restriction to the allowed configuration sector
is unitary.

The controlling edges are unchanged by the flip. The reverse move
therefore has the same parity-transfer factors and normalization,
with the incoming and outgoing projectors exchanged. Its block is
the adjoint of the forward block. In this realization $M_s$ is an
involution on the allowed configuration sector. The inverse symbols
in $\mathcal K_{24}$ still record the signed occurrences needed
to test cancellation under local redefinitions of the moves;
a redefined membrane need not remain an involution.

\paragraph*{Relation to the Wang--Gu construction.}
The chain geometry also defines a binary 2-cochain $n_2$ on the
boundary of the ordered auxiliary 4-simplex
$\langle0'1'2'3'4'\rangle$: its value on the triangle omitting
$u',v'$ is $\ell_{uv}$. This is the Kitaev-chain decoration denoted
$\tilde n_2$ by Wang and Gu~\cite{Wang2018Towards}. We adapt the
parity-corrected dimer projection in their fermionic $F$
move~\cite[Eqs.~(42) and (43)]{Wang2018Towards}. In that
retriangulation move, a chain Majorana repairs the parity mismatch
between the two dimer states. The full operator also acts on the
physical fermions and contains a scalar $\nu_4$.
Eq.~\eqref{eq:ftc-membrane-main} applies the same projection and
overlap normalization to a loop deformation on the fixed physical
lattice, with an even point--chain insertion at every vertex selected
by Eq.~\eqref{eq:ftc-direct}.

For auxiliary orientation $-\partial\langle 0'1'2'3'4'\rangle$
and singular surface $\langle 0'1'3'\rangle+\langle 0'2'3'\rangle
+\langle 0'2'4'\rangle+\langle 1'2'3'\rangle+\langle 2'3'4'\rangle$,
the rules of Wang and Gu~\cite[Figs.~13 and 14]{Wang2018Towards}
give the reverse of our arrows after the endpoint sign changes
$\gamma_{v,u}\mapsto\rho_{v,u}\gamma_{v,u}$. Each column lists
$\rho_{v,u}$ in increasing neighbor order.
\begin{equation}
 \begin{array}{c|ccccc}
 v & 0 & 1 & 2 & 3 & 4\\
 \hline
 \rho_v & ++-- & +-+- & ++-- & +-+- & +-++
 \end{array}
 \label{eq:ftc-wg-signs}
\end{equation}
This comparison fixes how to translate the arrow conventions.
The membrane used here is the fixed-lattice operator of
Eq.~\eqref{eq:ftc-membrane-main}.

\subsection{Local cancellation and phase}

\begin{table}[t]
\centering\small
\setlength{\tabcolsep}{3pt}
\begin{tabular}{@{}rcllc@{}}
\toprule
Step & Operator & Chain edges & Point sites & Phase\\
\midrule
0 & $-$ & $01,03,12,23$ & $\varnothing$ & $-$ \\
1 & $M_{234}^{-1}$ & $01,03,12,24,34$ & $3$ & $-i$ \\
2 & $M_{124}^{-1}$ & $01,03,14,34$ & $3$ & $1$ \\
3 & $M_{134}$ & $01,03,13$ & $1,3$ & $-1$ \\
4 & $M_{014}^{-1}$ & $03,04,13,14$ & $1,3$ & $1$ \\
5 & $M_{034}^{-1}$ & $13,14,34$ & $1,3$ & $1$ \\
6 & $M_{124}$ & $12,13,24,34$ & $1,3$ & $1$ \\
7 & $M_{024}$ & $02,04,12,13,34$ & $1,3$ & $1$ \\
8 & $M_{034}$ & $02,03,12,13$ & $1,3$ & $1$ \\
9 & $M_{124}^{-1}$ & $02,03,13,14,24$ & $1,3$ & $1$ \\
10 & $M_{134}^{-1}$ & $02,03,24,34$ & $1,3$ & $1$ \\
11 & $M_{234}$ & $02,03,23$ & $1,2$ & $i$ \\
12 & $M_{034}^{-1}$ & $02,04,23,34$ & $1,2$ & $1$ \\
13 & $M_{024}^{-1}$ & $23,24,34$ & $1,2$ & $1$ \\
14 & $M_{134}$ & $13,14,23,24$ & $1,2$ & $1$ \\
15 & $M_{014}$ & $01,04,13,23,24$ & $1,2$ & $1$ \\
16 & $M_{024}$ & $01,02,13,23$ & $1,2$ & $1$ \\
17 & $M_{134}^{-1}$ & $01,02,14,23,34$ & $2$ & $-1$ \\
18 & $M_{234}^{-1}$ & $01,02,14,24$ & $\varnothing$ & $i$ \\
19 & $M_{124}$ & $01,02,12$ & $\varnothing$ & $1$ \\
20 & $M_{024}^{-1}$ & $01,04,12,24$ & $\varnothing$ & $1$ \\
21 & $M_{014}^{-1}$ & $12,14,24$ & $\varnothing$ & $1$ \\
22 & $M_{234}$ & $12,14,23,34$ & $\varnothing$ & $1$ \\
23 & $M_{034}$ & $03,04,12,14,23$ & $\varnothing$ & $1$ \\
24 & $M_{014}$ & $01,03,12,23$ & $\varnothing$ & $1$ \\
\bottomrule
\end{tabular}
\caption{The process $\mathcal K_{24}$ in chronological order.
Each operator takes the preceding row to the listed chain and point
configuration. The phase is its matrix element in the Fock convention
specified in the text. The initial and final configurations agree,
and the phases multiply to $+i$.}
\label{tab:ftc-complete-history}
\end{table}

\begin{table}[t]

\centering
\small
\setlength{\tabcolsep}{3pt}
\begin{tabular}{@{}clll@{}}
\toprule
Membrane & First vertex & Second vertex & Third vertex\\
\midrule
$014$ & $(15,21),(24,4)$ & $(15,4),(24,21)$ & $(15,21),(24,4)$\\
$024$ & $(7,13),(16,20)$ & $(7,20),(16,13)$ & $(7,13),(16,20)$\\
$034$ & $(8,12),(23,5)$ & $(8,5),(23,12)$ & $(8,12),(23,5)$\\
$124$ & $(6,9),(19,2)$ & $(6,2),(19,9)$ & $(6,2),(19,9)$\\
$134$ & $(3,17),(14,10)$ & $(3,10),(14,17)$ & $(3,17),(14,10)$\\
$234$ & $(11,18),(22,1)$ & $(11,1),(22,18)$ & $(11,1),(22,18)$\\
\bottomrule
\end{tabular}
\caption{Local cancellation for the process $\mathcal K_{24}$.
For a membrane $M_{ijk}$ with $i<j<k$, the three columns
correspond to vertices $i,j,k$ and cover every vertex in its
prescribed support.
For each pair $(p,q)$, the configuration before the forward move
at step $p$ agrees locally with the configuration after the inverse
move at step $q$. The compared data are $f_v$ and all four incident
chain occupations, including the controlling edges extending
outside the triangle. These controls are read at their endpoints
within the triangle. The other two vertices lie outside the
prescribed support and do not enter the local cancellation
condition for this move.
}
\label{tab:ftc-local-pairs}
\end{table}

\paragraph*{Cancellation for $\mathcal K_{24}$.}
Applying $M_{012}$ and then $M_{023}$ to the vacuum prepares the
loop on edges $01,03,12,23$ with empty point modes. Starting from
this state, Table~\ref{tab:ftc-complete-history} follows the
operators in Eq.~\eqref{eq:ftc-word} from right to left. The
preparation and its inverse are outside the listed process; adding
them bases it at the vacuum without changing its phase or local
cancellation. The point occupation can have odd parity at an
intermediate step, but Eq.~\eqref{eq:ftc-parity} ensures that the
combined chain and point parity remains even throughout.

At a membrane vertex $v$, the local input $d_v\cfg$ consists of
$f_v$ and all incident chain occupations. A local modification
adds an arbitrary function $\varphi_{s,v}(d_v\cfg)$ to the forward
hopping angle. An inverse occurrence contributes the negative of
this function evaluated after that occurrence, since this is the
input of the forward hop it undoes. Thus cancellation compares
the configuration before a forward move with the configuration
after an inverse move. Table~\ref{tab:ftc-local-pairs} pairs every
forward occurrence with an inverse having the same local input
at each support vertex. Each pair cancels the corresponding
$\varphi_{s,v}$, as required by Eq.~\eqref{loc:cancellation-main}.
The pairing may differ between vertices of the same membrane.

\paragraph*{Microscopic phase.}
Both Majoranas of an edge mode $j$ flip its occupation $N_j$.
In the endpoint order specified above, they multiply the Fock
vector by $(-1)^{\sum_{k<j}N_k}$ and
$-i(-1)^{N_j+\sum_{k<j}N_k}$, respectively.
Applying Eq.~\eqref{eq:ftc-membrane-main} with the state phases
fixed above gives Table~\ref{tab:ftc-complete-history}. Only
steps $1,3,11,17,18$ have nontrivial phases. Their product is
\begin{equation*}
 (-i)(-1)(i)(-1)(i)=+i,
\end{equation*}
which gives Eq.~\eqref{eq:ftc-phase}.

Step 11 illustrates the origin of the factor $i$. The membrane
$M_{234}$ reads the occupied controlling edges $02$ and $03$,
removing the physical fermion at $3$ and adding one at $2$.
Let $\ket{D}$ and $\ket{D'}$ be the incoming and outgoing chain
states. Edge $23$ is initially empty, so its dimer condition is
$\gamma_{2,3}\gamma_{3,2}\ket{D}=-i\ket{D}$. In the outgoing
state, the arrows are $\gamma_{2,3}\to\gamma_{2,0}$ and
$\gamma_{3,2}\to\gamma_{3,0}$. Hence the two bilinears
$\gamma_{2,0}\gamma_{2,3}$ and $\gamma_{3,0}\gamma_{3,2}$ each
have eigenvalue $-i$ on $\bra{D'}$. Using the Majorana identity
\begin{equation}
 \gamma_{2,0}\gamma_{3,0}
 =(\gamma_{2,0}\gamma_{2,3})(\gamma_{3,0}\gamma_{3,2})
   \gamma_{2,3}\gamma_{3,2},
\end{equation}
we obtain
\begin{equation*}
 \langle D'|\gamma_{2,0}\gamma_{3,0}|D\rangle
 =(-i)^3\langle D'|D\rangle
 =i\langle D'|D\rangle.
\end{equation*}
The two parity-transfer factors give
\begin{equation*}
 (-i\chi_2\gamma_{2,0})(-i\chi_3\gamma_{3,0})
 =\chi_2\chi_3\gamma_{2,0}\gamma_{3,0}.
\end{equation*}
The sign from moving $\gamma_{2,0}$ past $\chi_3$ cancels the
factor $(-i)^2$. The point part also has amplitude $+1$:
$\chi_3\ket{\{1,3\}}=-\ket{\{1\}}$ and
$\chi_2\ket{\{1\}}=-\ket{\{1,2\}}$, where the sets list occupied
point sites.
In the specified Fock convention the overlap without the
parity-transfer insertion is $\langle D'|D\rangle=1/\sqrt8$. The normalization $\sqrt8$
therefore leaves the amplitude $+i$, as in
Table~\ref{tab:ftc-complete-history}.
Individual step phases depend on the phases assigned to the
configuration states, but their product around the closed process
does not.

\paragraph*{Locality of the elementary operators.}
The locality identity requires a nested commutator to act trivially
when its operators have empty common support. We check
Eq.~\eqref{loc:locality-identity} on the allowed configuration
sector using the ten membrane moves on triangles and the ten
fermion hops $-i\chi_u\chi_v$ on edges ($u<v$).
A point hop toggles its two endpoint occupations and leaves
the chain fixed. We use the main-text
convention $[X,Y]=X^{-1}Y^{-1}XY$.
The allowed sector contains
$64$ closed even networks and, for each network, $16$ point
assignments satisfying Eq.~\eqref{eq:ftc-parity}. Each generator
acts on these $1024$ configurations as a permutation with phases
in units of $\pi/4$ and squares to the identity.

An exhaustive check forms distinct nested commutators by adding
each generator in turn and intersecting its support with the
preceding supports. Every branch becomes scalar by depth five.
Since a scalar commutes with every generator, all further nested
commutators are the identity. Every branch with empty common
support already has phase one. This verifies
Eq.~\eqref{loc:locality-identity} on the allowed configuration
sector.

The sitewise product in Eq.~\eqref{eq:ftc-membrane-main} fixes the
fermionic signs in this check. If $r$ occupations are toggled,
order the selected vertices as $v_1<\cdots<v_r$. Moving all point
Majoranas to the left requires $r(r-1)/2$ exchanges, giving
\begin{equation*}
 \begin{aligned}
 &\prod_{j=1}^{r}(-i\chi_{v_j}\gamma_{v_j,u(v_j)})\\
 &\quad=(-i)^r(-1)^{r(r-1)/2}
   \left(\prod_{j=1}^r\chi_{v_j}\right)
   \left(\prod_{j=1}^r\gamma_{v_j,u(v_j)}\right).
 \end{aligned}
\end{equation*}
Both products on the right are in increasing $j$ order.
Omitting the exchange sign $(-1)^{r(r-1)/2}$ changes the operator.
It reverses the phase of step 11 and changes
$[[M_{234},M_{012}],M_{013}]$ from $1$ to $-1$, although the three
triangles have empty common intersection. The latter two membranes
toggle the distinct controlling occupations $\ell_{02}$ and
$\ell_{03}$ of $M_{234}$, leaving a nonzero mixed difference of
this extra sign in the commutator.

\subsection{The 3-group 5-cocycle}

We now express the same statistical phase in cochain language.
On the auxiliary boundary, the chain occupations define $n_2$
as above, and $n_3$ takes value $f_v$ on the tetrahedron omitting
$v'$. Closed chains satisfy $\delta n_2=0$. On the filling
4-simplex, $\delta n_3=n_2\cup n_2$ gives the parity constraint
in Eq.~\eqref{eq:ftc-parity}.
In the cocycle realization, the hopping phase is the cylinder
evaluation of the 5-cocycle in Eq.~\eqref{eq:ftc-restricted-anomaly}.
\begin{equation*}
 \begin{aligned}
 \Omega_{\rm K}={}&\tfrac12\bigl(n_3\cup_1n_3+n_3\cup_2\delta n_3+\delta n_3\cup_1n_2+\zeta_5(n_2,n_2)\bigr)\\
 &-\tfrac18[n_2]_2\cup\delta[n_2]_2\pmod1.
 \end{aligned}
\end{equation*}
The terms in parentheses are binary, and $[n_2]_2$ is the integer
lift with values $0,1$. The higher cup products use the ordered
simplicial convention of Ref.~\cite[Appendix~C]{Barkeshli2024Higher}.
We first check closure, then evaluate this cocycle on the process
$\mathcal K_{24}$.

\paragraph*{Closure and the doubled response.}
On an ordered 5-simplex, the Cartan correction is
\begin{equation}
 \zeta_5(n_2,n_2)_{012345}
 =(n_2)_{023}(n_2)_{012}(n_2)_{345}(n_2)_{235}.
 \label{eq:ftc-cartan}
\end{equation}
For binary cochains, the higher-cup coboundary identity reads
\begin{equation*}
 \begin{aligned}
 \delta(x\cup_i y)={}&\delta x\cup_i y+x\cup_i\delta y
 +x\cup_{i-1}y+y\cup_{i-1}x,
 \end{aligned}
\end{equation*}
where $i\ge1$ and $\cup_0=\cup$. Applying it to the first two
terms in $\Omega_{\rm K}$ gives
\begin{equation*}
 \delta(n_3\cup_1n_3+n_3\cup_2\delta n_3)
 =\delta n_3\cup_2\delta n_3\pmod2.
\end{equation*}
The Cartan identity~\cite{MedinaMardones2020Effective,Barkeshli2024Higher} cancels this contribution:
\begin{equation*}
 \begin{aligned}
 \delta\zeta_5(n_2,n_2)
 ={}&(n_2\cup n_2)\cup_2(n_2\cup n_2)\\
 &+\operatorname{Sq}^1 n_2\cup\operatorname{Sq}^1 n_2\pmod2,
 \end{aligned}
\end{equation*}
where $\operatorname{Sq}^1n_2=n_2\cup_1n_2$ for the closed
binary cochain $n_2$. The remaining term
$\delta n_3\cup_1n_2$ is closed: its coboundary is
$(n_2\cup n_2)\cup n_2+n_2\cup(n_2\cup n_2)=0$ modulo two.
Thus the coboundary of the binary expression in parentheses is
$\operatorname{Sq}^1n_2\cup\operatorname{Sq}^1n_2$.

Because $n_2$ is closed modulo two, $\delta[n_2]_2$ is even.
In the same cup-one convention,
\begin{equation*}
 -\tfrac12\delta[n_2]_2
 =\operatorname{Sq}^1n_2\pmod2.
\end{equation*}
The binary contribution and the integer-lift term therefore cancel:
\begin{equation*}
 \begin{aligned}
 \delta\Omega_{\rm K}
 ={}&\tfrac12\operatorname{Sq}^1n_2\cup\operatorname{Sq}^1n_2
 -\tfrac18\delta[n_2]_2\cup\delta[n_2]_2\\
 ={}&0\pmod1.
 \end{aligned}
\end{equation*}
Doubling $\Omega_{\rm K}$ removes the half-valued binary terms
modulo one and leaves
\begin{equation*}
 \begin{aligned}
 2\Omega_{\rm K}
 &=-\tfrac14[n_2]_2\cup\delta[n_2]_2\\
 &=\tfrac12n_2\cup\operatorname{Sq}^1n_2\pmod1,
 \end{aligned}
\end{equation*}
which is Eq.~\eqref{eq:ftc-doubled-anomaly}.

\paragraph*{Cylinder evaluation.}
The auxiliary 4-simplex fills the spatial 3-sphere. Its cylinder
is therefore five-dimensional, so $\Omega_{\rm K}$ can be evaluated
on it. We orient the cylinder by the ordered auxiliary simplex
$\langle0'1'2'3'4'\rangle$ followed by the interval from the initial
to the final slice. Its five ordered simplices are
\begin{equation*}
 [(0',0),\ldots,(j',0),(j',1),\ldots,(4',1)],
 \qquad 0\le j\le4,
\end{equation*}
with orientation coefficients $(-1)^{4-j}$, as in
Eq.~\eqref{cell:eq:15}.

Let $\pi$ be spatial projection and $t$ the binary 0-cochain equal
to zero on the initial slice and one on the final slice. A membrane
on $s$ has a binary 1-cochain parameter $\lambda$ supported on the
auxiliary edge dual to $s$. Extend it to the cylinder by
$\widehat\lambda=(\pi^*\lambda)\cup t$. Applying the membrane
update in footnote~\ref{fn:ftc-cochain-updates} to the pulled-back fields gives
\begin{equation*}
 \begin{aligned}
 \widehat n_2={}&\pi^*n_2+\delta\widehat\lambda,\\
 \widehat n_3={}&\pi^*n_3+(\pi^*n_2)\cup\widehat\lambda
 +\widehat\lambda\cup(\pi^*n_2)
   +\widehat\lambda\cup\delta\widehat\lambda.
 \end{aligned}
\end{equation*}
On the initial slice $\widehat\lambda=0$; on the final slice it
equals $\lambda$. These fields therefore interpolate between the
incoming and outgoing configurations and obey
$\delta\widehat n_3=\widehat n_2\cup\widehat n_2$ throughout the
cylinder. For a point hop, the parameter $\gamma$ is a binary
2-cochain supported on the auxiliary triangle dual to the hopping
edge. Set $\widehat\lambda=0$ and add $\delta\widehat\gamma$ to
$\widehat n_3$, with $\widehat\gamma=(\pi^*\gamma)\cup t$.

This interpolation agrees with the general transport prescription
in Appendix~\ref{app:transport}. The contraction $\mathsf h$ defined
in Eq.~\eqref{dc:prism-contraction} satisfies
\begin{equation*}
 \delta\mathsf h+\mathsf h\delta=1-\pi^*i_0^*,
\end{equation*}
where $i_0$ includes the initial slice. For a membrane, the explicit
field above has $\mathsf h\widehat n_3=0$: each term contains
either a pulled-back normalized cochain on repeated spatial vertices
or a $\widehat\lambda$ factor ending on the initial slice, both of
which vanish. Using $i_0^*\widehat n_3=n_3$ and the field constraint
in the contraction identity gives
\begin{equation*}
 \widehat n_3=\pi^*n_3+\mathsf h(\widehat n_2\cup\widehat n_2),
\end{equation*}
which specializes Eq.~\eqref{dc:elementary-higher} to this
membrane move.

The hopping angle is $2\pi$ times the evaluation of
$\Omega_{\rm K}$ on these oriented simplices. An inverse contributes
the negative of the angle at the corresponding forward source.
For the positive cylinder orientation, steps $1,3,11,17,18$ of
Eq.~\eqref{eq:ftc-word} contribute $1,2,2,2,2$ in units of
$\pi/2$, and the other steps contribute zero. The total angle is
$9\pi/2=\pi/2\pmod{2\pi}$, giving the phase $+i$ in
Eq.~\eqref{eq:ftc-phase}. Thus this process evaluates
$\Omega_{\rm K}$; evaluating $2\Omega_{\rm K}$ on the same
process gives $-1$.

\paragraph*{Comparison on individual hops.}
A single move need not have the same numerical phase in the Fock
calculation and the cylinder calculation: each uses a choice of state
phases and local operator phases. To compare them, use the main-text
hopping angles $\theta_s$, with superscripts $\mathrm{mic}$ and
$\mathrm{cyl}$ for the microscopic and positively oriented cylinder
calculations. For the ten membranes and ten fermion hops,
every forward arrow in the allowed sector satisfies
\begin{equation}
 \begin{aligned}
 \theta_s^{\rm mic}(\cfg)-\theta_s^{\rm cyl}(\cfg)
 ={}&\alpha(\cfg)-\alpha(\tau_s\cfg)\\
 &+\sum_{v\in\supp(s)}\varphi_{s,v}(d_v\cfg)
 \pmod{2\pi}.
 \end{aligned}
 \label{eq:ftc-phase-matching}
\end{equation}
Here $\tau_s\cfg$ is the prescribed output, $\alpha$ specifies a
change of configuration-state phases, and $\varphi_{s,v}$ depends
only on the local input $d_v\cfg$ defined above. These are precisely
the state and operator phase changes of Sec.~\ref{loc:phase-invariance}.
Measuring all angles in units of $\pi/4$ turns the identity into
linear congruences modulo eight. We solve these congruences and
verify the solution on every forward arrow.

The state terms telescope around a closed process. For the process
$\mathcal K_{24}$, the local operator terms cancel by
Table~\ref{tab:ftc-local-pairs}. The microscopic and cylinder
calculations therefore give the same invariant phase $+i$.

%% file: sections/AppendixF_lemmas.tex
\subsection{Locality and invariance}

\begin{lemma}[Locality lemma]\label{cell:lem:word}
Let $F$ be free on letters $s$ with nonempty parameter faces
$f_s\subseteq I$, where $I$ is finite. Let $\delta_v$ delete letters
whose parameter face contains $v$. Then $\bigcap_{v\in I}\ker\delta_v$
is normally generated by right-nested commutators of signed letters
whose parameter faces cover $I$, including single letters with $f_s=I$.
\end{lemma}
\begin{proof}
For nonempty $K\subseteq I$, let $D_K$ be generated by binary
brackets whose parameter faces have union $K$, including single letters.
For such a bracket $w$ and a signed letter $s$ with $f_s\subseteq K$,
\begin{equation}
 w^s=w[w,s]\in D_K.
\label{cell:eq:60}
\end{equation}
Thus $D_K$ is normal in the group generated by these letters, and
commutator expansion gives
\begin{equation}
 [D_K,D_L]\subset D_{K\cup L}.
\label{cell:eq:61}
\end{equation}
Order subsets by increasing size. By Eq.~\eqref{cell:eq:61}, each $D_K$
normalizes the product of later factors. Their product contains all letters, so
\begin{equation}
 w=\prod_{\varnothing\ne K\subseteq I}w_K,\qquad w_K\in D_K.
\label{cell:eq:62}
\end{equation}
If every $\delta_vw=1$, deleting outside a proper $J\subsetneq I$ gives
$\prod_{\varnothing\ne K\subseteq J}w_K=1$. Induction on $|J|$ gives $w_J=1$,
hence $w\in D_I$.

To reduce to right-nested brackets, let $L_K$ be generated by those
whose parameter faces cover $K$. Since $[w,s]=[s,w]^{-1}$,
Eq.~\eqref{cell:eq:60} makes $L_K$ normal in $F$, with
\begin{equation}
 [L_K,s]\subset L_{K\cup f_s}.
\label{cell:eq:63}
\end{equation}
Write $V_u$ for the union of parameter faces in a bracket $u$.
Induct on $u=[s,v]$ to prove $[L_K,u]\subset L_{K\cup V_u}$.
Modulo $N=L_{K\cup V_v\cup f_s}$, induction gives
\begin{equation}
 [[L_K,\langle v\rangle],\langle s\rangle]=1,
 [[L_K,\langle s\rangle],\langle v\rangle]=1.
\label{cell:eq:64}
\end{equation}
The Hall--Witt identity ($[x,y]=x^{-1}y^{-1}xy$, $x^y=y^{-1}xy$)
\begin{equation}
 [[x,y^{-1}],z]^y\,
 [[y,z^{-1}],x]^z
 [[z,x^{-1}],y]^x=1,
\label{cell:eq:65}
\end{equation}
puts $[L_K,[s,v]]$ in $N$. Expanding products now gives
\begin{equation}
 [L_K,L_L]\subset L_{K\cup L}.
\label{cell:eq:66}
\end{equation}
Thus $D_I\subset L_I$, while every generator of $L_I$ vanishes
under every deletion.
\end{proof}

On $I=\{0,\ldots,m\}$, covering by parameter faces means empty
common physical support. Therefore
\begin{equation}
\begin{gathered}
 \delta_vW=1\quad\text{for all }v
\quad \Longrightarrow \quad\langle W\rangle_{\cfg}\in E_{\mathrm{id}}.
\end{gathered}
\label{cell:eq:67}
\end{equation}
Each locality word closes, so conjugating paths cancel at identical sources.

\begin{lemma}[Invariance lemma]\label{cell:invariant:word}
Restricted sources include their frames. Let $f_s$ be the parameter
face of move $s$ and restrict arrow counts by
\begin{equation} 
 \tilde{\delta}_v(s,\cfg)=
 \begin{cases}
 (s,d_v\cfg),&v\notin f_s,\\
 0,&v\in f_s.
 \end{cases}
 \label{cell:inv:restriction}
\end{equation}
and define the flow map by
\begin{equation}
 B(s,\cfg)=[\tau_s\cfg]-[\cfg].
 \label{cell:inv:flow}
\end{equation}
Note that $\tilde{\delta}_v$ and $\delta_v$ are not exactly the same, where $\tilde{\delta}_v$ acts on arrows while $\delta_v$ acts on operator words. Then
\begin{equation}
 E_{\mathrm{inv}}
   =\ker B\cap\bigcap_{v=0}^{m}\ker \tilde{\delta}_v.
 \label{cell:inv:kernel}
\end{equation}

\end{lemma}

\begin{proof}
For $v\notin f_s$, the same move on the restricted source satisfies
\begin{equation}
 d_v\tau_s \cfg=\tau_s d_v\cfg,\qquad
 d_v\tau_s^{-1}\cfg=\tau_s^{-1}d_v\cfg.
 \label{cell:inv:naturality}
\end{equation}
Thus inverse occurrences also restrict at their forward sources.

For $e=\sum_{s,\cfg}n_{s,\cfg}(s,\cfg)$,
\begin{equation}
 \begin{aligned}
    \tilde{\delta}_ve
 &=\sum_{s:\,v\notin f_s}\sum_\xi
   \left(\sum_{\cfg:\,d_v\cfg=\xi}n_{s,\cfg}\right)(s,\xi).
 \end{aligned}
 \label{cell:inv:coefficients}
\end{equation}
Restricted arrows form a free basis, so
\begin{equation}
 \tilde{\delta}_ve=0
 \quad\Longleftrightarrow\quad
 \sum_{\cfg:\,d_v\cfg=\xi}n_{s,\cfg}=0
 \quad(s:\,v\notin f_s,\ \xi).
 \label{cell:inv:counts}
\end{equation}
This is Eq.~\eqref{loc:cancellation}. The coefficient of $[\cfg]$ in $Be$ is
$\sum_s(n_{s,\tau_s^{-1}\cfg}-n_{s,\cfg})$, so $Be=0$ is
Eq.~\eqref{loc:flow}. Together these are the defining conditions for $E_{\mathrm{inv}}$.
\end{proof}

%% file: sections/AppendixFnew.tex
\section{Cycle--process correspondence}
\label{app:complete:sec}

Preparation $P$ and sweep $S$ define the closed word $L$.
Alternating sums of its counts define the map $J$. The inverse identities are
Eqs.~\eqref{cell:eq:81} and \eqref{cell:eq:110}.

For $\partial z=0$ and arbitrary $h$, we prove
\begin{equation}
 J(z)\in E_{\rm inv},\qquad J(\partial h)\in E_{\rm id}.
 \label{dc:J-quotient}
\end{equation}
On the quotients,
\begin{equation}
 \overline{\operatorname{Cyl}}\circ\bar J=\mathrm{id},\qquad
 \bar J\circ\overline{\operatorname{Cyl}}=\mathrm{id}.
 \label{loc:inverse-maps}
\end{equation}

We use $X_\ell$ to denote the set of $\ell$-cells of $B\Gg$. 

\subsection{Preparation \texorpdfstring{$P(\cfg)$}{P(c)}}
\label{cell:sec:time}

Fix $T=[0,\ldots,m]$, $m=d+1$, $D=m+1$, and let $\cfg_0$ be
the trivial configuration. For $\cfg\in X_m$, set
\begin{equation}
 P_1(\cfg)=U_{1,\{1\}}(a_{01})\cdots U_{1,\{m\}}(a_{0m}).
\label{cell:eq:26}
\end{equation}
Products act from right to left, with complete field updates.

Assume the layers below $r$ agree with those of $\cfg$. Let
\begin{equation}
 \Delta_r=B_r-B_r^{\rm cur}.
\label{cell:eq:27}
\end{equation}
For each $r$-vertex parameter face $f$, set
\begin{equation}
 \lambda_r(f)=
 \begin{cases}
 0,&0\in f,\\
 \rho_r(a_{0i}^{-1})\Delta_r(0,f),&i=\min f>0.
 \end{cases}
\label{cell:eq:28}
\end{equation}
Order the parameter faces as
$f_1<\cdots<f_N$ lexicographically and apply
\begin{equation}
 P_r(\cfg)=U_{r,f_N}(\lambda_r(f_N))\cdots U_{r,f_1}(\lambda_r(f_1)).
\label{cell:eq:29}
\end{equation}
Set $P(\cfg)=P_n(\cfg)\cdots P_1(\cfg)$, using Eq.~\eqref{cell:eq:21} at each step.

\subsection{Sweep \texorpdfstring{$S(\sigma)$}{S(sigma)}}
\label{cell:sec:sweep}

The sweep builds a simplex from a degeneracy, one layer at a time.
For $\sigma\in X_D$ and $0\le p\le m$, put
\begin{equation}
\begin{gathered}
 \cfg^-=d_{p+1}\sigma,\qquad \cfg^+=d_p\sigma,\\
 \pi:[0,\ldots,D]\longrightarrow[0,\ldots,\widehat{p+1},\ldots,D],
 \  \pi(p+1)=p.
\end{gathered}
 \label{cell:eq:41}
\end{equation}
Here $\pi$ fixes every other vertex; the target retains its original labels.
Start from $\pi^*\cfg^-$ on this auxiliary simplex and apply
complete field updates, fixing $d_{p+1}\sigma$. Begin with
\begin{equation}
 R_1^{(p)}=U_{1,\{p+1\}}(a_{p,p+1}).
 \label{cell:eq:43}
\end{equation}
For $r\ge2$, after completing the lower layers, let
$\Delta_r=B_r-B_r^{\rm cur}$, where $B_r$ is the target field from $\sigma$.
For each $r$-vertex face $f$ containing $p+1$ and avoiding $p$, set
\begin{equation}
 \begin{aligned}
 \mu_r(f)&=\begin{cases}
 \rho_r(a_{p,p+1}^{-1})\Delta_r(f\cup\{p\}),&k(f)=0,\\
 (-1)^{k(f)}\Delta_r(f\cup\{p\}),&k(f)>0,
 \end{cases}\\
 k(f)&=\#\{i\in f:i<p\}.
 \end{aligned}
 \label{cell:eq:42}
\end{equation}
List the vertices of $f\cup\{p\}$ increasingly and order $f_1<\cdots<f_N$
lexicographically. Set
\begin{equation}
 R_r^{(p)}=U_{r,f_N}(\mu_r(f_N))\cdots
       U_{r,f_1}(\mu_r(f_1))R_{r-1}^{(p)}.
 \label{cell:eq:33}
\end{equation}
The residual is closed and vanishes on $d_{p+1}$; extending
$\mu_r$ by zero gives $\delta_a\mu_r=\Delta_r$.
Thus $R^{(p)}=R_n^{(p)}$ reaches $\sigma$. Define
\begin{equation}
 S_p(\sigma)=d_pR^{(p)}:\cfg^-\longrightarrow \cfg^+,
 \qquad S(\sigma)=S_m(\sigma).
 \label{cell:eq:37}
\end{equation}
Retain zero slots in the fixed face order.

\subsection{Alternating sum \texorpdfstring{$\mathcal A_i$}{A(i)} and rotation lemma}

Use the reference cube
\begin{equation}
\begin{gathered}
 \cfg_\epsilon=\sigma(\epsilon_0,1+\epsilon_1,\ldots,
 m+\epsilon_m),\\
 \epsilon\in\{0,1\}^{m+1}.
\end{gathered}
\label{cell:eq:97}
\end{equation}
For a tuple $\epsilon$ with $\epsilon_i=0$, define $C_i(\epsilon)$ as the sweep $S_i$ on the
cell retaining both $i,i+1$ in slot $i$ and the displayed labels
in all other slots. It runs from $\cfg_\epsilon$ to
$\cfg_{\epsilon+e_i}$, where $e_i$ is the $i$th unit vector.
Define
\begin{equation}
\mathcal{A}_i(\sigma)=\sum_{\epsilon:\epsilon_i=0}
 (-1)^{\sum_{j\ne i}\epsilon_j}
 \langle C_i(\epsilon)\rangle_{\cfg_\epsilon}.
\label{cell:eq:98}
\end{equation}

\begin{lemma}[Rotation Lemma]\label{cell:lem:rotation}
For all $i,j$, $\mathcal{A}_i(\sigma)-\mathcal{A}_j(\sigma)\in E_{\mathrm{id}}$.
\end{lemma}
\begin{proof}
For $i\ne j$, start at $\cfg_{\mathbf0}$ with the square
$C_j^{-1}C_i^{-1}C_jC_i$. In each remaining direction $k$, replace
the current loop $N$ by $C_k^{-1}N^{-1}C_kN$.
Since $N$ is closed, connecting paths cancel at
identical sources. The resulting word $W_{ij}$ has counts
\begin{equation}
 \langle W_{ij}\rangle_{\cfg_{\mathbf0}}=\mathcal{A}_i(\sigma)-\mathcal{A}_j(\sigma).
\label{cell:eq:99}
\end{equation}
On facet $v$, direction-$v$ edges disappear and the remaining
restricted data are independent of bit $v$. If $v=i$ or $j$, the
initial square cancels; otherwise the $v$-difference cancels.
Every facet word therefore freely cancels, so Lemma~\ref{cell:lem:word} gives the claim.
\end{proof}

\subsection{Closed comparison \texorpdfstring{$L$}{L} and the map \texorpdfstring{$J$}{J}}
\label{cell:sec:references}

Close the direction-$m$ sweep at $\cfg_0$:
\[
 L(\sigma)=P(d_m\sigma)^{-1}S(\sigma)P(d_D\sigma).
\]
For the alternating sum, put
\begin{equation}
    \begin{aligned}
 \tau_\epsilon
 =\sigma(\epsilon_0,1+\epsilon_1,\ldots,m-1+\epsilon_{m-1},m,D),\\
 \epsilon\in\{0,1\}^m, \quad c_\epsilon=(-1)^{|\epsilon|},
 \quad |\epsilon|=\sum_i\epsilon_i .
\label{cell:eq:45}
\end{aligned}
\end{equation}
Here $\sigma(j_0,\ldots,j_k)$ is the pullback along the displayed
nondecreasing list. Concatenate $L(\tau_\epsilon)^{c_\epsilon}$
lexicographically to define $\mathcal J(\sigma)$, based at $\cfg_0$.
Retain the words of degenerate reference simplices.

All chains are understood modulo degenerate simplices. Define
\begin{equation}
 J(\sigma)=\langle \mathcal J(\sigma)\rangle_{\cfg_0}.
 \label{cell:eq:46}
\end{equation}
Extend linearly. Eq.~\eqref{cell:eq:56} verifies that degenerate
inputs vanish.
Realize integer chains by concatenation, using strict inverses for negative coefficients.

\subsection{Induced map on homology}
\label{cell:sec:step1}

Write $Z_D$ and $B_D$ for cycles and boundaries.
It suffices to show
\begin{equation}
\begin{aligned}
 \operatorname{Cyl}(E_{\mathrm{inv}})&\subset Z_D,\,
 \operatorname{Cyl}(E_{\mathrm{id}})\subset B_D,\\
 J(Z_D)&\subset E_{\mathrm{inv}},
 J(B_D)\subset E_{\mathrm{id}}.
\end{aligned}
\label{cell:eq:47}
\end{equation}

\begin{proposition}\label{cell:prop:firstthree}
\label{cell:cor:locality-boundaries}
The first three inclusions in Eq.~\eqref{cell:eq:47} hold, and $E_{\mathrm{id}}\subset E_{\mathrm{inv}}$.
\end{proposition}
\begin{proof}
For the first inclusion, the boundary of the cylinder is
\begin{equation}
\begin{aligned}
 \partial\operatorname{Cyl}(s,\cfg)={}&(-1)^m
 \bigl([\tau_s\cfg]-[\cfg]\bigr)\\
 &+\sum_{v=0}^m(-1)^v
 \operatorname{Cyl}_{d_vT}(s,d_v\cfg).
\end{aligned}
\label{cell:eq:58}
\end{equation}
Closure cancels the first term and
local cancellation cancels the second, proving
$\operatorname{Cyl}(E_{\mathrm{inv}})\subset Z_D$.

For the second inclusion, adjoin $*$ before $T=[0,1,\dots,m]$, duplicate vertex
$0$ in the source, and execute the same letters on their original
parameter faces. The initial configuration is symmetric under interchange of $*$ and $0$, but only $0$ is involved in the moves. Deleting any original vertex gives a freely cancelling word by Lemma~\ref{cell:lem:word}. Thus all side-face contributions of $[*,0,\dots,m]\times I$ except that from $T\times I$ cancel. The bottom and top contributions also cancel by closure. Only the contribution from $T\times I$ remains. Therefore,
\begin{equation}
 \partial  B=\operatorname{Cyl}(\langle W\rangle_{\cfg}),
\label{cell:eq:59}
\end{equation}
where $B$ is the cylinder sum on $[*,0,\dots,m]\times I$. Hence $\operatorname{Cyl}(E_{\mathrm{id}})\subset B_D$.

For the third inclusion, consider a cycle $z$. Eq.~\eqref{cell:eq:57} removes
preparations. Sweep factors contain slot $m$; for each $v<m$,
the two references differing in $\epsilon_v$ have equal restricted
words and sources with opposite signs. Thus their local counts
cancel and $Jz\in E_{\mathrm{inv}}$ according to Lemma~\ref{cell:invariant:word}.

Finally, deleting any vertex from a locality word removes a commutator leaf,
so every restricted word freely cancels; thus its restricted counts vanish, giving
$E_{\mathrm{id}}\subset E_{\mathrm{inv}}$.
\end{proof}

\subsection{Proof of \texorpdfstring{$J(B_D)\subset E_{\mathrm{id}}$}{J(BD) is contained in E(id)}}
\label{cell:sec:boundary-comparison}

Define the Moore projections by
\begin{equation}
 \Pi_k=(1-s_{k-1}d_{k-1})\cdots(1-s_0d_0).
\label{cell:eq:51}
\end{equation}

The standard normalization maps $\Pi_k$ kill degenerate $k$-chains,
act as the identity modulo them, and satisfy
\begin{equation}
\begin{gathered}
 d_i\Pi_k=0\ (i<k),\quad 
 \partial\Pi_k=\Pi_{k-1}\partial.
\end{gathered}
\label{cell:eq:52}
\end{equation}
Furthermore,
\begin{equation}
 \Pi_m[\sigma]=\sum_\epsilon c_\epsilon[\tau_\epsilon],\ 
 (d_D-d_m)\Pi_m=(-1)^D\Pi_m\partial.
\label{cell:eq:53}
\end{equation}

Let $p[\cfg]=\langle P(\cfg)\rangle_{\cfg_0}$ record preparations and set
\begin{equation}
\mathcal A_m[\sigma]=\sum_\epsilon c_\epsilon
 \langle S(\tau_\epsilon)\rangle_{d_D\tau_\epsilon}.
\label{cell:eq:55}
\end{equation}
Since $\Pi_m$ kills degenerate chains, $\bar p(x)=p(\Pi_mx)$ is well defined.
Preparation counts give
\begin{equation}
 Jz=\mathcal A_m z+(-1)^D\bar p(\partial z).
\label{cell:eq:56}
\end{equation}
For a degenerate input, coinciding slots $j,j+1$ with $j<m$
pair equal sweeps with opposite signs; a coinciding last pair makes
every sweep stationary. Thus $\mathcal A_m$ kills degeneracies.
Eq.~\eqref{cell:eq:56} follows from
$p(d_D-d_m)\Pi_m=(-1)^Dp\Pi_m\partial$. It also shows that $J$ kills
degenerate inputs and gives Eq.~\eqref{cell:eq:57} on cycles.

On cycles, preparation counts cancel:
\begin{equation}
 Jz=\mathcal A_m z\ (z\in Z_D),\quad
 J(\partial\tau)=\mathcal A_m(\partial\tau).
\label{cell:eq:57}
\end{equation}

\begin{lemma}\label{cell:lem:boundary}
For every $\tau\in X_{D+1}$, $J(\partial[\tau])\in E_{\mathrm{id}}$.
\end{lemma}
\begin{proof}
At $\cfg=\tau(0,\ldots,m)$, start with the triangle loop and iterate:
\begin{equation}
\begin{gathered}
 N_0=C_m(m,m+2)^{-1}C_m(m+1,m+2)C_m(m,m+1),\\
 N_{i+1}=C_i^{-1}N_i^{-1}C_iN_i\quad(0\le i<m),\ W=N_m.
\end{gathered}
\label{cell:eq:69}
\end{equation}
Here $C_m(x,y)$ uses the same reference-sweep construction with
last-slot labels $x,y$; all other slots retain their current
reference values. The $C_i$ connecting paths lie on the last-slot
$m$ layer.

Each $N_i$ is closed, so the connecting paths cancel in its counts.
Thus $\langle W\rangle_{\cfg}$ is the alternating sum of the
triangle counts over the first $m$ reference bits. In
$J(\partial[\tau])=\mathcal A_m(\partial[\tau])$, the last two
faces give $(-1)^m$ times the $(m,m+1)$ and $-(m,m+2)$ sums.
Splitting each skipped difference in the other faces as
$[j-1]-[j+1]=([j-1]-[j])+([j]-[j+1])$ cancels adjacent face terms,
leaving $(-1)^m$ times the $(m+1,m+2)$ sum. Hence
\begin{equation}
 \langle W\rangle_{\cfg}=(-1)^mJ(\partial[\tau]).
\label{cell:eq:72}
\end{equation}
On facet $i<m$, $C_i$ disappears and the restricted loops at its
two ends agree, so the $i$th comparison and all later stages
freely cancel. On facet $m$, $N_0$ disappears. The locality lemma
therefore gives $\langle W\rangle_{\cfg}\in E_{\mathrm{id}}$.
\end{proof}

Hence
\begin{equation}
\begin{aligned}
 \bar J&:H_D(X;\mathbb{Z})\longrightarrow T,\\
 \overline{\operatorname{Cyl}}&:T
 \longrightarrow H_D(X;\mathbb{Z}).
\end{aligned}
\label{cell:eq:73}
\end{equation}

\subsection{Recovering the input cycle}
\label{cell:sec:cycle-recovery}

Fix $\sigma\in X_D$. Its direct history is the cylinder
$\Gamma=\phi_m^*\sigma$, where $\phi_m$ sends $i^0,i^1$ to $i$
for $i<m$, and sends $m^0,m^1$ to $m,D$.
Its bottom is $d_D\sigma$ and its top is $d_m\sigma$.

Take the update word $R^{(m)}$ from Sec.~\ref{cell:sec:sweep}.
After each move, pull back the current labeled simplex along $\phi_m$.
This gives a sequence of cylinders, starting with the constant
history of $d_D\sigma$ and ending with $\Gamma$.
The interpolation for each move connects two successive cylinders;
joining these interpolations gives fields on $T\times I_t\times I_u$,
where $t$ is the cylinder time and $u$ follows the updates in $R^{(m)}$.
The bottom $t=0$ stays fixed, while the top $t=1$ follows
$S(\sigma)=d_mR^{(m)}$.

Let $\Sigma=I_t\times I_u$, with the triangulation supplied by
these interpolations. Its four sides carry $S(\sigma)$, $\Gamma$,
and two constant histories. The constant sides form the stationary
connector. Orient $\Sigma$ so that its boundary follows $S(\sigma)$
forward and $\Gamma$ backward.
Write $[T\times\Sigma]$ and $[\partial T\times\Sigma]$ for the
resulting integral chains of labeled product simplices, and set
\begin{equation}
 F(\sigma)=(-1)^m[T\times\Sigma],\ 
 V(\sigma)=(-1)^m[\partial T\times\Sigma].
\label{cell:eq:75}
\end{equation}
The connector contributes only degenerate simplices. In $\Gamma$,
all but the last-pivot cylinder cell are degenerate, giving
$\operatorname{Cyl}(\Gamma)=[\sigma]$. The product boundary formula
therefore gives
\begin{equation}
 \partial F(\sigma)=\operatorname{Cyl}\bigl(\langle S(\sigma)\rangle_{d_D\sigma}\bigr)
 -[\sigma]+V(\sigma).
\label{cell:eq:76}
\end{equation}
Take the alternating sum over the references in
Eq.~\eqref{cell:eq:45}, defining on nondegenerate generators
\begin{equation}
 H([\sigma])=\sum_\epsilon c_\epsilon F(\tau_\epsilon)
\label{cell:eq:77}
\end{equation}
and extending linearly. On the facet opposite $v<m$, the
$\epsilon_v$-paired disks have identical restricted fields;
opposite $m$, the construction is stationary. Hence
\begin{equation}
 \sum_\epsilon c_\epsilon V(\tau_\epsilon)=0.
\label{cell:eq:78}
\end{equation}
Summing Eq.~\eqref{cell:eq:76} and discarding degenerate terms gives
\begin{equation}
 \operatorname{Cyl}(\mathcal A_m z)=z+\partial H(z).
\label{cell:eq:79}
\end{equation}
Set
\begin{equation}
 K=(-1)^D\operatorname{Cyl}\,\bar p.
\label{cell:eq:80}
\end{equation}
Eq.~\eqref{cell:eq:56} now yields
\begin{equation}
\boxed{\begin{aligned}
 \operatorname{Cyl} J(z)={}&z+\partial H(z)
 +K(\partial z).
\end{aligned}}
\label{cell:eq:81}
\end{equation}
For cycles this proves $\overline{\operatorname{Cyl}}\,\bar J=1$.
Pairing with a cocycle and multiplying its transport phases gives
\begin{equation}
\begin{aligned}
 \langle \operatorname{Cyl} J(z),\Omega\rangle
 &=\langle z,\Omega\rangle,\ 
 W(z)|\cfg_0\rangle
 =e^{2\pi i\langle z,\Omega\rangle}|\cfg_0\rangle.
\end{aligned}
\label{cell:eq:82}
\end{equation}

\subsection{Recovering the input process}
\label{cell:sec:process-recovery}

For $e\in E_{\rm inv}$ we prove
\begin{equation}
 \boxed{J\operatorname{Cyl}(e)-e\in E_{\mathrm{id}}.}
 \label{cell:eq:83}
\end{equation}

Consider a move $s=(r,f,u)$ at source $\cfg$, with cylinder
$\Gamma$. Put $p=\max f$. Split $p$ into consecutive vertices
$p^-,p^+$ to obtain an ordered $D$-simplex $Y$, and let
$\pi:Y\to T$ identify them. Starting from $\pi^*\cfg$, apply
the updates in Eqs.~\eqref{cell:eq:19} and \eqref{cell:eq:21}
with parameter $(r,f^+,u)$, where $f^+$ replaces $p$ in $f$ by
$p^+$. Denote the resulting labeled simplex by $\sigma$.
Let $\phi_p:T\times I\to Y$ send $p^0,p^1$ to $p^-,p^+$
and identify both copies of every other vertex. Since $p=\max f$,
the active cochain pulls back to the extension in
Eq.~\eqref{app:conv:time-extension}. The complete updates commute
with this pullback, giving

\begin{equation}
 \Gamma=\phi_p^*\sigma.
\label{cell:eq:95}
\end{equation}
All cylinder cells $C_j$ with $j\ne p$ become degenerate, while
$C_p$ maps increasingly onto $Y$ with coefficient $(-1)^{m-p}$.
Thus

\begin{equation}
 \operatorname{Cyl}(s,\cfg)=(-1)^{m-p}[\sigma].
\label{cell:eq:94}
\end{equation}

\begin{lemma}[Alternating sum decomposition]\label{cell:lem:remainder}
Let $e=\sum_{s,\cfg}n_{s,\cfg}(s,\cfg)\in E_{\mathrm{inv}}$.
In each summand, take $p=\max f$ and the simplex $\sigma$
constructed above. Then
\begin{equation}
\begin{aligned}
 &\sum_{s,\cfg}n_{s,\cfg}(-1)^{m-p}\mathcal A_p(\sigma)=\sum_{s,\cfg}n_{s,\cfg}(s,\cfg).
\end{aligned}
\label{cell:eq:105}
\end{equation}
\end{lemma}
\begin{proof}

For a fixed $(s,\cfg)$, the direction-$p$ reference edge with bits
\begin{equation}
 \epsilon_i^*=0\quad(i<p),\qquad
 \epsilon_i^*=1\quad(i>p).
\label{cell:eq:100}
\end{equation}
is $\Gamma$, with coefficient $+1$ in
$(-1)^{m-p}\mathcal A_p(\sigma)$. The defining update has no mixed
residual below layer $r$; at layer $r$, Eq.~\eqref{cell:eq:42} gives
only the parameter $u$ on $f^+$ (Eq.~\eqref{cell:eq:43} if $r=1$).
Applying it reaches $\sigma$, so all later parameters vanish. Hence the sweep needs only one move:
\begin{equation}
 \langle S_p(\sigma)\rangle_{\cfg}=(s,\cfg).
\label{cell:eq:101}
\end{equation}
Fix $s$ and any other reference. Its spatial projection is
nondecreasing and not the identity, so it omits a vertex of $T$.
If it omits a vertex of $f$, its sweep is stationary. Otherwise,
choose an omitted vertex $v\notin f$. The sweep counts depend
only on $d_v\cfg$, so grouping sources with the same restriction
makes this reference's total contribution vanish by
Eq.~\eqref{loc:cancellation}:
\begin{equation}
\begin{gathered}
 \sum_{\cfg:d_v\cfg=\xi}n_{s,\cfg}=0 \quad 
 \text{for every fixed }s,\ v\notin f,\ \xi.
\end{gathered}
\label{cell:eq:104}
\end{equation}
Only the reference in Eq.~\eqref{cell:eq:100} remains, giving
Eq.~\eqref{cell:eq:105}.
\end{proof}

\paragraph*{Recovery.}
Since $\operatorname{Cyl}(e)\in Z_D$, Eq.~\eqref{cell:eq:57}
removes the preparation counts. Eq.~\eqref{cell:eq:94} gives
\begin{equation}
\begin{aligned}
 J\operatorname{Cyl}(e)={}&\sum_{s,\cfg}n_{s,\cfg}(-1)^{m-p}\mathcal{A}_m(\sigma).
\end{aligned}
\label{cell:eq:109}
\end{equation}
Lemma~\ref{cell:lem:rotation} and
Eq.~\eqref{cell:eq:105} give
\begin{equation}
\begin{aligned}
 J\operatorname{Cyl}(e)-e&=\sum_{s,\cfg}n_{s,\cfg}(-1)^{m-p}
 (\mathcal{A}_m-\mathcal A_p)(\sigma)\\
 &\quad\in E_{\mathrm{id}}.
\end{aligned}
\label{cell:eq:110}
\end{equation}

%% file: sections/Appendix_finite_order_and_localization.tex
\section{Finite order and localization property of statistics}
\label{app:disk:section}

Let $X=B\mathbb G$ have finite $G_r=\pi_r(X)$ for $1\le r\le n\le d$
and no higher homotopy groups.

\subsection{Finite order}
\label{app:finiteorder}

\begin{proposition}\label{app:complete:finite}
The group $T=E_{\mathrm{inv}}/E_{\mathrm{id}}$ is finite and has
the same exponent as $H_D(X;\mathbb Z)$.
\end{proposition}
\begin{proof}
Transfer makes $BF$ rationally acyclic for finite $F$.
The rational Serre spectral sequence gives~\cite{HatcherNDSpectral}
\begin{equation}
 H_k(Y;\mathbb Q)\cong H_k(B;\mathbb Q)
 \label{finite:serre}
\end{equation}
for a fibration $F\to Y\to B$ with connected rationally acyclic fiber.
Starting with $K(A,1)=BA$, the path fibrations
$K(A,r-1)\to PK(A,r)\to K(A,r)$ therefore make every
$K(A,r)$ for finite Abelian $A$ rationally acyclic.
Apply Eq.~\eqref{finite:serre} along the Postnikov tower
\begin{equation}
 K(G_r,r)\longrightarrow X^{(r)}\longrightarrow X^{(r-1)},
 \qquad 2\le r\le n,
\end{equation}
with $X^{(1)}=BG_1$ and $X^{(n)}\simeq X$. Hence $H_{k>0}(X;\mathbb Q)=0$.
Finite labels give finite rank in each chain degree, so
$H_{k>0}(X;\mathbb Z)$ is finite. The correspondence
Eq.~\eqref{classification} proves the claim.

More explicitly, set $M=\exp H_D(X;\mathbb Z)$, with $M=1$
for the trivial group. Completeness gives
\begin{gather}
 \partial\operatorname{Cyl}(e)=0,\qquad
 e-J\operatorname{Cyl}(e)\in E_{\mathrm{id}}
 \quad(e\in E_{\mathrm{inv}}),\label{finite:recovery}\\
 J(\partial b)\in E_{\mathrm{id}}
 \quad\text{for every $(D+1)$-chain $b$}.\label{finite:boundary}
\end{gather}
For $z=\operatorname{Cyl}(e)$, choose $b$ with $\partial b=Mz$. Then
\begin{equation}
 Me=M(e-Jz)+J(\partial b)\in E_{\mathrm{id}}.
 \label{finite:certificate}
\end{equation}
\end{proof}

\begin{remark}
Every realization evaluates a statistical class to an $M$th root of unity.
\end{remark}

\subsection{Localization in a spatial ball}
\label{app:disk:localization}

Prepare at anchor $m$ and sweep in direction $0$.
For $T=[0,\ldots,m]$ and $F=[0,\ldots,m-1]$, let
\begin{equation}
 B=m*\partial F,\qquad \partial T=B\cup F,\qquad
 B\cap F=\partial F.
 \label{app:disk:domain}
\end{equation}
The cone $B$ is a closed $d$-ball. For the complementary face $f^\vee$,
\begin{equation}
 \operatorname{supp}U_{r,f}(u)=f^\vee,\qquad \{m\}^\vee=F.
 \label{app:disk:support}
\end{equation}
Enlarge $B$ to include fixed microscopic thickenings.
For each cycle $z$, we construct a closed word $W$ based at $\cfg_0$ with
\begin{equation}
 \langle W\rangle_{\cfg_0}-Jz\in E_{\mathrm{id}},\qquad
 \operatorname{supp}U\subset B\quad(U\text{ a factor of }W).
 \label{app:disk:equivalence}
\end{equation}

Use the preparation of Sec.~\ref{cell:sec:time} with anchor $m$.
Set $p_i=a_{mi}$, $p_m=1$, and apply vertex moves in order
$m-1,\ldots,0$. At layer $r$, put $\Delta_r=B_r-B_r^{\rm cur}$ and
write it in the $m$-frame as
$\bar\Delta_r(i_0,\ldots,i_r)=\rho_r(p_{i_0})\Delta_r(i_0,\ldots,i_r)$.
For each increasing parameter face $f=(i_0,\ldots,i_{r-1})$ avoiding $m$, set
\begin{equation}
 \lambda_r(f)=\rho_r(p_{i_0}^{-1})\bar\Delta_r(m,i_0,\ldots,i_{r-1}).
 \label{app:disk:prepare}
\end{equation}
Extend $\bar\Delta_r$ alternately, with repeated vertices giving zero,
and execute the faces lexicographically. Cone contraction gives
$\delta_a\lambda_r=\Delta_r$, so the resulting word $P_m(\cfg)$ prepares $\cfg$.
Its parameter faces avoid $m$, hence all its moves lie in $B$.

Direction-$0$ sweeps lie in $[1,\ldots,m]\subset B$.
For cycles, completeness and rotation give
\begin{gather}
 Jz=\mathcal A_m(z)\in E_{\mathrm{inv}},\label{app:disk:cycle}\\
 \mathcal A_0(z)-\mathcal A_m(z)\in E_{\mathrm{id}}.\label{app:disk:rotation}
\end{gather}
Close each signed sweep in $\mathcal A_0(z)$ using $P_m$ at its
source and inverse preparation at its target; concatenate these loops to form $W$.
If $n_{s,\cfg}$ are the coefficients of $\mathcal A_0(z)$, flow closure gives
\begin{equation}
\begin{aligned}
 \langle W\rangle_{\cfg_0}
 &=\mathcal A_0(z)+\sum_{\cfg}\Bigl[\sum_s
 (n_{s,\cfg}-n_{s,\tau_s^{-1}\cfg})\Bigr]\\
 &\qquad\qquad\times\langle P_m(\cfg)\rangle_{\cfg_0}
 =\mathcal A_0(z).
\end{aligned}
 \label{app:disk:flow}
\end{equation}
Thus
\begin{equation}
 \langle W\rangle_{\cfg_0}-Jz
 =\mathcal A_0(z)-\mathcal A_m(z)\in E_{\mathrm{id}}.
 \label{app:disk:final}
\end{equation}
Every factor lies in $B$, so this representative embeds in any lattice
preserving the updates and locality relations on the ball.

%% file: sections/spt_examples_details.tex
\section{Detecting cycles for the 2-group boundary processes}
\label{app:spt-examples}

\begin{table}[!t]
\centering
\small
\setlength{\tabcolsep}{3pt}
\begin{tabular}{cll}
\hline\hline
Anomaly & Cylinder signs & Gate signs\\
\hline
$\alpha_5^{\rm I}$ & $11,12,27,28,45,47$ & \begin{tabular}[t]{@{}l@{}}$6,8,11,12,17,$\\$22,33,36,44,47$\end{tabular}\\
$\alpha_5^{\rm II}$ & $27,28,45$ & $45$\\
\hline\hline
\end{tabular}
\caption{Steps with sign $-1$ in $\mathcal N_{50}$ for the
  vacuum component $a=b=0$. Cylinder signs use the defect fields.
  The microscopic signs include the two cone factors in
  Eq.~\eqref{spt:microscopic-cone-gate}.}
\label{spt:nu-step-phases}
\end{table}

\begin{table*}[!tp]
\centering
\footnotesize
\setlength{\tabcolsep}{4pt}
\begin{tabular}{c c l l}
\hline\hline
$j$ & Operation & $a_j^{(1)}$ & $a_j^{(2)}$\\
\hline
0 & $1$ & $0$ & $0$\\
1 & $E_{23}^{-1}$ & $\langle01\rangle+\langle04\rangle+\langle14\rangle$ & $0$\\
2 & $E_{13}^{-1}$ & $\langle01\rangle+\langle02\rangle+\langle14\rangle+\langle24\rangle$ & $0$\\
3 & $V_{4}$ & $\langle01\rangle+\langle02\rangle+\langle14\rangle+\langle24\rangle$ & $\langle012\rangle+\langle013\rangle+\langle023\rangle+\langle123\rangle$\\
4 & $E_{04}$ & $\langle01\rangle+\langle02\rangle+\langle12\rangle+\langle13\rangle+\langle14\rangle+\langle23\rangle+\langle24\rangle$ & $\langle012\rangle+\langle013\rangle+\langle023\rangle+\langle123\rangle$\\
5 & $E_{02}$ & $\langle01\rangle+\langle02\rangle+\langle12\rangle+\langle23\rangle+\langle24\rangle+\langle34\rangle$ & $\langle012\rangle+\langle013\rangle+\langle023\rangle+\langle123\rangle$\\
6 & $E_{23}^{-1}$ & $\langle02\rangle+\langle04\rangle+\langle12\rangle+\langle14\rangle+\langle23\rangle+\langle24\rangle+\langle34\rangle$ & $\langle012\rangle+\langle013\rangle+\langle023\rangle+\langle123\rangle$\\
7 & $V_{3}$ & $\langle02\rangle+\langle04\rangle+\langle12\rangle+\langle14\rangle+\langle23\rangle+\langle24\rangle+\langle34\rangle$ & $\langle013\rangle+\langle014\rangle+\langle023\rangle+\langle024\rangle+\langle123\rangle+\langle124\rangle$\\
8 & $E_{02}^{-1}$ & $\langle02\rangle+\langle04\rangle+\langle12\rangle+\langle13\rangle+\langle23\rangle+\langle24\rangle$ & $\langle013\rangle+\langle014\rangle+\langle023\rangle+\langle024\rangle+\langle123\rangle+\langle124\rangle$\\
9 & $V_{2}$ & $\langle02\rangle+\langle04\rangle+\langle12\rangle+\langle13\rangle+\langle23\rangle+\langle24\rangle$ & $\langle023\rangle+\langle024\rangle+\langle034\rangle+\langle123\rangle+\langle124\rangle+\langle134\rangle$\\
10 & $E_{24}$ & $\langle01\rangle+\langle02\rangle+\langle03\rangle+\langle04\rangle+\langle12\rangle+\langle23\rangle+\langle24\rangle$ & $\langle023\rangle+\langle024\rangle+\langle034\rangle+\langle123\rangle+\langle124\rangle+\langle134\rangle$\\
11 & $E_{34}$ & $\langle03\rangle+\langle04\rangle+\langle23\rangle+\langle24\rangle$ & $\langle023\rangle+\langle024\rangle+\langle034\rangle+\langle123\rangle+\langle124\rangle+\langle134\rangle$\\
12 & $E_{04}^{-1}$ & $\langle03\rangle+\langle04\rangle+\langle12\rangle+\langle13\rangle+\langle24\rangle$ & $\langle023\rangle+\langle024\rangle+\langle034\rangle+\langle123\rangle+\langle124\rangle+\langle134\rangle$\\
13 & $V_{2}^{-1}$ & $\langle03\rangle+\langle04\rangle+\langle12\rangle+\langle13\rangle+\langle24\rangle$ & $\langle013\rangle+\langle014\rangle+\langle023\rangle+\langle024\rangle+\langle123\rangle+\langle124\rangle$\\
14 & $E_{02}$ & $\langle03\rangle+\langle04\rangle+\langle12\rangle+\langle14\rangle+\langle24\rangle+\langle34\rangle$ & $\langle013\rangle+\langle014\rangle+\langle023\rangle+\langle024\rangle+\langle123\rangle+\langle124\rangle$\\
15 & $E_{04}$ & $\langle03\rangle+\langle04\rangle+\langle13\rangle+\langle14\rangle+\langle23\rangle+\langle24\rangle+\langle34\rangle$ & $\langle013\rangle+\langle014\rangle+\langle023\rangle+\langle024\rangle+\langle123\rangle+\langle124\rangle$\\
16 & $V_{2}$ & $\langle03\rangle+\langle04\rangle+\langle13\rangle+\langle14\rangle+\langle23\rangle+\langle24\rangle+\langle34\rangle$ & $\langle023\rangle+\langle024\rangle+\langle034\rangle+\langle123\rangle+\langle124\rangle+\langle134\rangle$\\
17 & $E_{34}^{-1}$ & $\langle01\rangle+\langle02\rangle+\langle03\rangle+\langle04\rangle+\langle12\rangle+\langle13\rangle+\langle14\rangle+\langle23\rangle+\langle24\rangle+\langle34\rangle$ & $\langle023\rangle+\langle024\rangle+\langle034\rangle+\langle123\rangle+\langle124\rangle+\langle134\rangle$\\
18 & $V_{2}^{-1}$ & $\langle01\rangle+\langle02\rangle+\langle03\rangle+\langle04\rangle+\langle12\rangle+\langle13\rangle+\langle14\rangle+\langle23\rangle+\langle24\rangle+\langle34\rangle$ & $\langle013\rangle+\langle014\rangle+\langle023\rangle+\langle024\rangle+\langle123\rangle+\langle124\rangle$\\
19 & $V_{3}^{-1}$ & $\langle01\rangle+\langle02\rangle+\langle03\rangle+\langle04\rangle+\langle12\rangle+\langle13\rangle+\langle14\rangle+\langle23\rangle+\langle24\rangle+\langle34\rangle$ & $\langle012\rangle+\langle013\rangle+\langle023\rangle+\langle123\rangle$\\
20 & $E_{04}^{-1}$ & $\langle01\rangle+\langle02\rangle+\langle03\rangle+\langle04\rangle+\langle14\rangle+\langle24\rangle+\langle34\rangle$ & $\langle012\rangle+\langle013\rangle+\langle023\rangle+\langle123\rangle$\\
21 & $V_{4}^{-1}$ & $\langle01\rangle+\langle02\rangle+\langle03\rangle+\langle04\rangle+\langle14\rangle+\langle24\rangle+\langle34\rangle$ & $0$\\
22 & $E_{23}$ & $\langle02\rangle+\langle03\rangle+\langle24\rangle+\langle34\rangle$ & $0$\\
23 & $E_{02}^{-1}$ & $\langle02\rangle+\langle03\rangle+\langle13\rangle+\langle14\rangle+\langle24\rangle$ & $0$\\
24 & $E_{13}$ & $\langle03\rangle+\langle04\rangle+\langle13\rangle+\langle14\rangle$ & $0$\\
25 & $E_{02}$ & $\langle03\rangle+\langle04\rangle+\langle34\rangle$ & $0$\\
26 & $V_{0}$ & $\langle03\rangle+\langle04\rangle+\langle34\rangle$ & $\langle123\rangle+\langle124\rangle+\langle134\rangle+\langle234\rangle$\\
27 & $E_{34}^{-1}$ & $\langle01\rangle+\langle02\rangle+\langle03\rangle+\langle04\rangle+\langle12\rangle+\langle34\rangle$ & $\langle123\rangle+\langle124\rangle+\langle134\rangle+\langle234\rangle$\\
28 & $V_{0}^{-1}$ & $\langle01\rangle+\langle02\rangle+\langle03\rangle+\langle04\rangle+\langle12\rangle+\langle34\rangle$ & $0$\\
29 & $V_{4}$ & $\langle01\rangle+\langle02\rangle+\langle03\rangle+\langle04\rangle+\langle12\rangle+\langle34\rangle$ & $\langle012\rangle+\langle013\rangle+\langle023\rangle+\langle123\rangle$\\
30 & $E_{02}^{-1}$ & $\langle01\rangle+\langle02\rangle+\langle03\rangle+\langle04\rangle+\langle12\rangle+\langle13\rangle+\langle14\rangle$ & $\langle012\rangle+\langle013\rangle+\langle023\rangle+\langle123\rangle$\\
31 & $V_{3}$ & $\langle01\rangle+\langle02\rangle+\langle03\rangle+\langle04\rangle+\langle12\rangle+\langle13\rangle+\langle14\rangle$ & $\langle013\rangle+\langle014\rangle+\langle023\rangle+\langle024\rangle+\langle123\rangle+\langle124\rangle$\\
32 & $V_{2}$ & $\langle01\rangle+\langle02\rangle+\langle03\rangle+\langle04\rangle+\langle12\rangle+\langle13\rangle+\langle14\rangle$ & $\langle023\rangle+\langle024\rangle+\langle034\rangle+\langle123\rangle+\langle124\rangle+\langle134\rangle$\\
33 & $E_{04}$ & $\langle01\rangle+\langle02\rangle+\langle03\rangle+\langle04\rangle+\langle14\rangle+\langle23\rangle$ & $\langle023\rangle+\langle024\rangle+\langle034\rangle+\langle123\rangle+\langle124\rangle+\langle134\rangle$\\
34 & $E_{24}^{-1}$ & $\langle02\rangle+\langle04\rangle+\langle13\rangle+\langle14\rangle+\langle23\rangle$ & $\langle023\rangle+\langle024\rangle+\langle034\rangle+\langle123\rangle+\langle124\rangle+\langle134\rangle$\\
35 & $V_{2}^{-1}$ & $\langle02\rangle+\langle04\rangle+\langle13\rangle+\langle14\rangle+\langle23\rangle$ & $\langle013\rangle+\langle014\rangle+\langle023\rangle+\langle024\rangle+\langle123\rangle+\langle124\rangle$\\
36 & $E_{04}^{-1}$ & $\langle02\rangle+\langle04\rangle+\langle12\rangle+\langle14\rangle$ & $\langle013\rangle+\langle014\rangle+\langle023\rangle+\langle024\rangle+\langle123\rangle+\langle124\rangle$\\
37 & $V_{3}^{-1}$ & $\langle02\rangle+\langle04\rangle+\langle12\rangle+\langle14\rangle$ & $\langle012\rangle+\langle013\rangle+\langle023\rangle+\langle123\rangle$\\
38 & $E_{23}$ & $\langle01\rangle+\langle02\rangle+\langle12\rangle$ & $\langle012\rangle+\langle013\rangle+\langle023\rangle+\langle123\rangle$\\
39 & $V_{4}^{-1}$ & $\langle01\rangle+\langle02\rangle+\langle12\rangle$ & $0$\\
40 & $V_{0}$ & $\langle01\rangle+\langle02\rangle+\langle12\rangle$ & $\langle123\rangle+\langle124\rangle+\langle134\rangle+\langle234\rangle$\\
41 & $E_{13}$ & $\langle01\rangle+\langle04\rangle+\langle12\rangle+\langle24\rangle$ & $\langle123\rangle+\langle124\rangle+\langle134\rangle+\langle234\rangle$\\
42 & $E_{24}^{-1}$ & $\langle03\rangle+\langle04\rangle+\langle12\rangle+\langle13\rangle+\langle24\rangle$ & $\langle123\rangle+\langle124\rangle+\langle134\rangle+\langle234\rangle$\\
43 & $E_{13}^{-1}$ & $\langle02\rangle+\langle03\rangle+\langle12\rangle+\langle13\rangle$ & $\langle123\rangle+\langle124\rangle+\langle134\rangle+\langle234\rangle$\\
44 & $E_{23}^{-1}$ & $\langle01\rangle+\langle02\rangle+\langle03\rangle+\langle04\rangle+\langle12\rangle+\langle13\rangle+\langle14\rangle$ & $\langle123\rangle+\langle124\rangle+\langle134\rangle+\langle234\rangle$\\
45 & $E_{34}$ & $\langle03\rangle+\langle04\rangle+\langle13\rangle+\langle14\rangle$ & $\langle123\rangle+\langle124\rangle+\langle134\rangle+\langle234\rangle$\\
46 & $E_{24}$ & $\langle01\rangle+\langle04\rangle+\langle14\rangle$ & $\langle123\rangle+\langle124\rangle+\langle134\rangle+\langle234\rangle$\\
47 & $E_{04}^{-1}$ & $\langle01\rangle+\langle04\rangle+\langle12\rangle+\langle13\rangle+\langle14\rangle+\langle23\rangle$ & $\langle123\rangle+\langle124\rangle+\langle134\rangle+\langle234\rangle$\\
48 & $E_{23}$ & $\langle12\rangle+\langle13\rangle+\langle23\rangle$ & $\langle123\rangle+\langle124\rangle+\langle134\rangle+\langle234\rangle$\\
49 & $E_{04}$ & $0$ & $\langle123\rangle+\langle124\rangle+\langle134\rangle+\langle234\rangle$\\
50 & $V_{0}^{-1}$ & $0$ & $0$\\
\hline\hline
\end{tabular}
\caption{Defect configurations along $\mathcal N_{50}$, from
  the initial vacuum at $j=0$ through step $50$. The loop chain
  $a_j^{(1)}$ and wall chain $a_j^{(2)}$ are recorded after the
  displayed operation. All chain coefficients are modulo two.
  The junction zero-chain $a_j^{(0)}=\partial a_j^{(1)}$ vanishes
  in every row.}
\label{spt:mixed-history-1}
\label{spt:mixed-history-2}
\end{table*}

We give the labeled simplices of the detecting cycles used in
Sec.~\ref{sec:spt-examples}. A labeled $\ell$-simplex $\sigma$ has
ordered vertices $0,\ldots,\ell$, with an $A$ bit on each edge and a
$B$ bit on each triangle. The bit strings below record only the
independent labels $A_{0i}$ and $B_{0ij}$. The constraints in
Eq.~\eqref{spt:constraints} determine all other labels by
\begin{equation*}
 \begin{aligned}
 A_{ij}&=A_{0i}+A_{0j},\\
 B_{ijk}&=B_{0ij}+B_{0ik}+B_{0jk}+A_{0i}A_{ij}A_{jk}
 \end{aligned}
\end{equation*}
for $0<i<j$ and $0<i<j<k$, respectively, with all sums modulo two.

Brackets $[\sigma]$ denote the integral chain generator with this
vertex order. The coefficients outside the brackets are signed
integers. To take a boundary, delete vertex $i$, restrict the
fields to the remaining vertices in their inherited order, and
include the face with sign $(-1)^i$. Degenerate faces are zero
in the normalized complex. To pair a cycle with a cocycle,
evaluate on each labeled simplex and sum with the displayed
integer coefficients modulo one.

\subsection{The circle and point-exchange detectors}

Encode a 3-simplex by the bits
$(A_{01},A_{02},A_{03};B_{012},B_{013},B_{023})$.
For example, $(111;001)$ means that all three edges from vertex
$0$ have $A=1$ and only $B_{023}=1$ among the listed triangles.
The constraints then give $A_{12}=A_{13}=A_{23}=0$ and $B_{123}=1$.
The cycle
\begin{equation}
 z_3=[(111;001)]-[(011;011)]+[(001;110)]
 \label{spt:d1-cycle-table}
\end{equation}
is the signed shuffle of the circle edge with $A_{01}=1$ and
the sphere triangle with $A=0$ and $B_{012}=1$. Its three $Q_3$
values in the displayed order are $(1,0,0)$, so
$\langle z_3,\alpha_3\rangle=1/2$.
Let $\tau$ be the 4-simplex with labels
\begin{equation}
 \begin{aligned}
 (A_{01},A_{02},A_{03},A_{04})&=(1,1,1,1),\\
 (B_{012},B_{013},B_{014},B_{023},B_{024},B_{034})
 &= (0,0,0,1,0,0).
 \end{aligned}
 \label{spt:d1-chain-filler}
\end{equation}
Summing the signed cylinders of the 6 steps of $\mathcal D_6$ gives
\begin{equation}
 \operatorname{Cyl}(\langle\mathcal D_6\rangle_{\cfg_0})
 =z_3-\partial[\tau].
 \label{spt:d1-chain-certificate}
\end{equation}

For the point-exchange detector, all $A$ labels vanish. Each bracket
below gives a 4-simplex by listing the bits
$(B_{012},B_{013},B_{014},B_{023},B_{024},B_{034})$ in that order.
The chain
\begin{equation}
 z_4=[001111]-[010110]+[100001]
 \label{spt:d2-cycle}
\end{equation}
has zero normalized boundary. The $Q_4$ values are $(0,0,1)$,
giving $\langle z_4,\alpha_4\rangle=1/2$.

\subsection{The wall-free loop detector}
\label{spt:typeI-certificate}

With $A=0$, encode a 5-simplex by the ten bits $B_{0ij}$ in the order
\begin{equation}
 (ij)=(12,13,14,15,23,24,25,34,35,45).
 \label{spt:typeI-cell-order}
\end{equation}
Each bracket lists these bits, as for $z_4$; the remaining labels
follow from $\delta B=0$. The integral chain
\begin{equation}
 \begin{aligned}
 z_{\rm I}
 &= [0100000010]+[1100001110]-[0010110001]
 \end{aligned}
 \label{spt:typeI-cycle}
\end{equation}
has zero normalized boundary. Each cell evaluates to $1/2$ under
$\alpha_5^{\rm I}$ and zero under $\alpha_5^{\rm II}$, yielding
\begin{equation}
 \langle z_{\rm I},\alpha_5^{\rm I}\rangle=\frac12,
 \qquad
 \langle z_{\rm I},\alpha_5^{\rm II}\rangle=0 .
 \label{spt:typeI-pairings}
\end{equation}

\subsection{The mixed wall--loop detector}
\label{spt:mixed-details}

Take the three 4-simplices in Eq.~\eqref{spt:d2-cycle}, with
signs $(n_1,n_2,n_3)=(1,-1,1)$, and denote their full $B$ fields
by $x^{(r)}$ in the displayed order. For $r=1,2,3$ and
$t=0,\ldots,4$, define $\sigma_{rt}$ as the ordered 5-simplex
on vertices $0,\ldots,5$ obtained by shuffling simplex $r$ with
the circle edge carrying $A=1$. The circle step occupies
$(t,t+1)$, and projection back to simplex $r$ collapses this edge:
$p_t(i)=i$ for $i\leq t$ and $p_t(i)=i-1$ for $i>t$.
Its edge and triangle labels are
\begin{equation}
 \begin{aligned}
 A_{ij}&=\mathbf 1_{i\leq t<j},\\
 B_{ijk}&=x^{(r)}_{p_t(i)p_t(j)p_t(k)},
 \end{aligned}
 \label{spt:mixed-cells}
\end{equation}
where $B_{ijk}=0$ whenever projected vertices repeat.
Thus $A$ records the circle step and $B$ is the pullback
of $x^{(r)}$.
A single jump implies $A\cup A=0$, so the pulled-back $B$ satisfies
the 2-group constraint. The shuffle sign is $(-1)^{4-t}$ because
the circle step passes the $4-t$ remaining steps of the 4-simplex.
This gives the integral cycle
\begin{equation}
 z_{\rm mix}=\sum_{r=1}^3\sum_{t=0}^4
 n_r(-1)^{4-t}[\sigma_{rt}].
 \label{spt:mixed-cycle}
\end{equation}
Evaluating the two 5-cocycles on these cells gives
\begin{equation}
 \bigl(\langle z_{\rm mix},\alpha_5^{\rm I}\rangle,
       \langle z_{\rm mix},\alpha_5^{\rm II}\rangle\bigr)
 =\left(\frac12,\frac12\right).
 \label{spt:mixed-pairing}
\end{equation}
Thus the cycle
\begin{equation}
 z_\nu=z_{\rm mix}+z_{\rm I}
 \label{spt:nu-cycle}
\end{equation}
isolates the type-II pairing,
\begin{equation}
 \bigl(\langle z_\nu,\alpha_5^{\rm I}\rangle,
       \langle z_\nu,\alpha_5^{\rm II}\rangle\bigr)
 =\left(0,\frac12\right).
 \label{spt:nu-pairing}
\end{equation}
Tables~\ref{spt:nu-step-phases} and \ref{spt:mixed-history-1}
give the phases and configurations for the word in
Eq.~\eqref{spt:mixed-protocol}. In the configuration table,
angle brackets list occupied physical cells of $\partial[01234]$
via Eq.~\eqref{spt:physical-chains}. For example, $\langle01\rangle$
in the loop chain corresponds to $B_{234}=1$, and
$\langle012\rangle$ in the wall chain corresponds to $A_{34}=1$.
Cells absent from a chain have coefficient zero.

%% file: bibliography.bib
@CONTROL{PRXControl,
  author = {48},
  editor = {1},
  pages = {0},
  title = {0},
  year = {1},
  eprint = {1}
}

@article{Li2025HowMuch,
  author = {Li, Zhi and Lee, Dongjin and Yoshida, Beni},
  title = {How Much Entanglement Is Needed for Topological Codes and Mixed States with Anomalous Symmetry?},
  journal = {Phys. Rev. X},
  volume = {15},
  pages = {021090},
  year = {2025},
  doi = {10.1103/pw12-kdjx},
  eprint = {2405.07970},
  archivePrefix = {arXiv},
  primaryClass = {quant-ph}
}

@article{Bravyi2025HowMuch,
  title = {How Much Entanglement Is Needed for Quantum Error Correction?},
  author = {Bravyi, Sergey and Lee, Dongjin and Li, Zhi and Yoshida, Beni},
  journal = {Phys. Rev. Lett.},
  volume = {134},
  issue = {21},
  pages = {210602},
  numpages = {6},
  year = {2025},
  month = {May},
  publisher = {American Physical Society},
  doi = {10.1103/PhysRevLett.134.210602},
  url = {https://link.aps.org/doi/10.1103/PhysRevLett.134.210602}
}

@inproceedings{Bravyi2026Identity,
  author = {Bravyi, Sergey and Parham, Natalie and Tran, Minh},
  title = {Identity check problem for shallow quantum circuits},
  booktitle = {17th Innovations in Theoretical Computer Science Conference (ITCS 2026)},
  editor = {Saraf, Shubhangi},
  series = {Leibniz International Proceedings in Informatics (LIPIcs)},
  volume = {362},
  pages = {27:1--27:17},
  publisher = {Schloss Dagstuhl -- Leibniz-Zentrum f{\"u}r Informatik},
  address = {Dagstuhl, Germany},
  year = {2026},
  doi = {10.4230/LIPIcs.ITCS.2026.27}
}

@article{Lan2019Fermion,
  title = {Fermion decoration construction of symmetry-protected trivial order for fermion systems with any symmetry and in any dimension},
  author = {Lan, Tian and Zhu, Chenchang and Wen, Xiao-Gang},
  journal = {Phys. Rev. B},
  volume = {100},
  issue = {23},
  pages = {235141},
  numpages = {46},
  year = {2019},
  month = {Dec},
  publisher = {American Physical Society},
  doi = {10.1103/PhysRevB.100.235141},
  url = {https://link.aps.org/doi/10.1103/PhysRevB.100.235141}
}

@article{Kobayashi2026Generalized,
  author = {Kobayashi, Ryohei and Li, Yuyang and Xue, Hanyu and Hsin, Po-Shen and Chen, Yu-An},
  title = {Generalized statistics on lattices},
  journal = {Phys. Rev. X},
  volume = {16},
  pages = {011010},
  year = {2026},
  doi = {10.1103/6k88-w52n},
  eprint = {2412.01886},
  archivePrefix = {arXiv},
  primaryClass = {quant-ph}
}

@misc{Feng2026Defining,
  author = {Feng, Yitao and Chen, Yu-An},
  title = {Defining 2-group symmetries on lattices and extracting their anomalies},
  year = {2026},
  eprint = {2610.37745},
  archivePrefix = {arXiv}
}

@article{Barkeshli2024Higher,
  author = {Barkeshli, Maissam and Chen, Yu-An and Hsin, Po-Shen and Kobayashi, Ryohei},
  title = {Higher-group symmetry in finite gauge theory and stabilizer codes},
  journal = {SciPost Phys.},
  volume = {16},
  pages = {089},
  year = {2024},
  doi = {10.21468/SciPostPhys.16.4.089},
  eprint = {2211.11764},
  archivePrefix = {arXiv},
  primaryClass = {cond-mat.str-el}
}

@article{Benini2019TwoGroup,
  author = {Benini, Francesco and C\'ordova, Clay and Hsin, Po-Shen},
  title = {On 2-group global symmetries and their anomalies},
  journal = {J. High Energy Phys.},
  volume = {2019},
  number = {03},
  pages = {118},
  year = {2019},
  doi = {10.1007/JHEP03(2019)118},
  eprint = {1803.09336},
  archivePrefix = {arXiv},
  primaryClass = {hep-th}
}

@incollection{Kapustin2017Higher,
  author = {Kapustin, Anton and Thorngren, Ryan},
  title = {Higher symmetry and gapped phases of gauge theories},
  booktitle = {Algebra, Geometry, and Physics in the 21st Century},
  editor = {Auroux, Denis and Katzarkov, Ludmil and Pantev, Tony and Soibelman, Yan and Tschinkel, Yuri},
  series = {Progress in Mathematics},
  volume = {324},
  publisher = {Birkh{\"a}user},
  address = {Cham},
  pages = {177--202},
  year = {2017},
  doi = {10.1007/978-3-319-59939-7_5}
}

@article{Gaiotto2015Generalized,
  author = {Gaiotto, Davide and Kapustin, Anton and Seiberg, Nathan and Willett, Brian},
  title = {Generalized global symmetries},
  journal = {J. High Energy Phys.},
  volume = {2015},
  number = {02},
  pages = {172},
  year = {2015},
  doi = {10.1007/JHEP02(2015)172},
  eprint = {1412.5148},
  archivePrefix = {arXiv},
  primaryClass = {hep-th}
}

@article{Kitaev2003FaultTolerant,
title = {Fault-tolerant quantum computation by anyons},
journal = {Annals of Physics},
volume = {303},
number = {1},
pages = {2-30},
year = {2003},
issn = {0003-4916},
doi = {10.1016/S0003-4916(02)00018-0},
url = {https://www.sciencedirect.com/science/article/pii/S0003491602000180},
author = {A.Yu. Kitaev}
}

@article{Kitaev2001Unpaired,
  author = {Kitaev, A. {\relax Yu}},
  title = {Unpaired {Majorana} fermions in quantum wires},
  journal = {Phys. Usp.},
  volume = {44},
  number = {10S},
  pages = {131--136},
  year = {2001},
  doi = {10.1070/1063-7869/44/10S/S29},
  eprint = {cond-mat/0010440},
  archivePrefix = {arXiv}
}

@article{Arovas1984Fractional,
  author = {Arovas, Daniel and Schrieffer, J. R. and Wilczek, Frank},
  title = {Fractional statistics and the quantum {Hall} effect},
  journal = {Phys. Rev. Lett.},
  volume = {53},
  pages = {722--723},
  year = {1984},
  doi = {10.1103/PhysRevLett.53.722}
}

@article{Kesselring2024Anyon,
  author = {Kesselring, Markus S. and Magdalena de la Fuente, Julio C. and Thomsen, Felix and Eisert, Jens and Bartlett, Stephen D. and Brown, Benjamin J.},
  title = {Anyon condensation and the color code},
  journal = {PRX Quantum},
  volume = {5},
  pages = {010342},
  year = {2024},
  doi = {10.1103/PRXQuantum.5.010342},
  eprint = {2212.00042},
  archivePrefix = {arXiv},
  primaryClass = {quant-ph}
}

@article{Levin2003Fermions,
  author = {Levin, Michael and Wen, Xiao-Gang},
  title = {Fermions, strings, and gauge fields in lattice spin models},
  journal = {Phys. Rev. B},
  volume = {67},
  pages = {245316},
  year = {2003},
  doi = {10.1103/PhysRevB.67.245316},
  eprint = {cond-mat/0302460},
  archivePrefix = {arXiv}
}

@article{Kawagoe2020Microscopic,
  author = {Kawagoe, Kyle and Levin, Michael},
  title = {Microscopic definitions of anyon data},
  journal = {Phys. Rev. B},
  volume = {101},
  pages = {115113},
  year = {2020},
  doi = {10.1103/PhysRevB.101.115113},
  eprint = {1910.11353},
  archivePrefix = {arXiv},
  primaryClass = {cond-mat.str-el}
}

@book{Hatcher2002Algebraic,
  author = {Hatcher, Allen},
  title = {Algebraic Topology},
  publisher = {Cambridge University Press},
  address = {Cambridge},
  year = {2002},
  url = {https://pi.math.cornell.edu/~hatcher/AT/AT.pdf}
}

@article{Carolyn2023entanglers,
  title = {Topological invariants for symmetry-protected topological phase entanglers},
  author = {Zhang, Carolyn},
  journal = {Phys. Rev. B},
  volume = {107},
  issue = {23},
  pages = {235104},
  numpages = {23},
  year = {2023},
  month = {Jun},
  publisher = {American Physical Society},
  doi = {10.1103/PhysRevB.107.235104},
  url = {https://link.aps.org/doi/10.1103/PhysRevB.107.235104}
}

@article{Dijkgraaf1990Topological,
  author = {Dijkgraaf, Robbert and Witten, Edward},
  title = {Topological gauge theories and group cohomology},
  journal = {Commun. Math. Phys.},
  volume = {129},
  pages = {393--429},
  year = {1990},
  doi = {10.1007/BF02096988}
}

@article{Chen2013Symmetry,
  author = {Chen, Xie and Gu, Zheng-Cheng and Liu, Zheng-Xin and Wen, Xiao-Gang},
  title = {Symmetry protected topological orders and the group cohomology of their symmetry group},
  journal = {Phys. Rev. B},
  volume = {87},
  pages = {155114},
  year = {2013},
  doi = {10.1103/PhysRevB.87.155114},
  eprint = {1106.4772},
  archivePrefix = {arXiv},
  primaryClass = {cond-mat.str-el}
}

@article{Else2014Classifying,
  author = {Else, Dominic V. and Nayak, Chetan},
  title = {Classifying symmetry-protected topological phases through the anomalous action of the symmetry on the edge},
  journal = {Phys. Rev. B},
  volume = {90},
  pages = {235137},
  year = {2014},
  doi = {10.1103/PhysRevB.90.235137},
  eprint = {1409.5436},
  archivePrefix = {arXiv},
  primaryClass = {cond-mat.str-el}
}

@article{Levin2012Braiding,
  author = {Levin, Michael and Gu, Zheng-Cheng},
  title = {Braiding statistics approach to symmetry-protected topological phases},
  journal = {Phys. Rev. B},
  volume = {86},
  pages = {115109},
  year = {2012},
  doi = {10.1103/PhysRevB.86.115109},
  eprint = {1202.3120},
  archivePrefix = {arXiv},
  primaryClass = {cond-mat.str-el}
}

@article{Wang2014Braiding,
  author = {Wang, Chenjie and Levin, Michael},
  title = {Braiding statistics of loop excitations in three dimensions},
  journal = {Phys. Rev. Lett.},
  volume = {113},
  pages = {080403},
  year = {2014},
  doi = {10.1103/PhysRevLett.113.080403},
  eprint = {1403.7437},
  archivePrefix = {arXiv},
  primaryClass = {cond-mat.str-el}
}

@article{Fidkowski2022Gravitational,
  author = {Fidkowski, Lukasz and Haah, Jeongwan and Hastings, Matthew B.},
  title = {Gravitational anomaly of {$(3+1)$}-dimensional {$\mathbb{Z}_2$} toric code with fermionic charges and fermionic loop self-statistics},
  journal = {Phys. Rev. B},
  volume = {106},
  pages = {165135},
  year = {2022},
  doi = {10.1103/PhysRevB.106.165135},
  eprint = {2110.14654},
  archivePrefix = {arXiv},
  primaryClass = {cond-mat.str-el}
}

@article{Feng2026Higher,
  author = {Feng, Yitao and Kobayashi, Ryohei and Chen, Yu-An and Ryu, Shinsei},
  title = {Higher-form anomalies on lattices},
  journal = {Phys. Rev. Lett.},
  volume = {136},
  pages = {046504},
  year = {2026},
  doi = {10.1103/2jz1-m1lb},
  eprint = {2509.12304},
  archivePrefix = {arXiv},
  primaryClass = {cond-mat.str-el}
}

@article{Feng2026Anyonic,
  author = {Feng, Yitao and Xue, Hanyu and Li, Yuyang and Cheng, Meng and Kobayashi, Ryohei and Hsin, Po-Shen and Chen, Yu-An},
  title = {Anyonic membranes and {Pontryagin} statistics},
  journal = {Phys. Rev. Lett.},
  volume = {136},
  pages = {086601},
  year = {2026},
  doi = {10.1103/4jww-6b6t},
  eprint = {2509.14314},
  archivePrefix = {arXiv},
  primaryClass = {quant-ph}
}

@article{Steenrod1947Products,
  author = {Steenrod, N. E.},
  title = {Products of cocycles and extensions of mappings},
  journal = {Ann. Math.},
  volume = {48},
  pages = {290},
  year = {1947},
  doi = {10.2307/1969172}
}

@article{Chen2020Exact,
  author = {Chen, Yu-An},
  title = {Exact bosonization in arbitrary dimensions},
  journal = {Phys. Rev. Research},
  volume = {2},
  pages = {033527},
  year = {2020},
  doi = {10.1103/PhysRevResearch.2.033527},
  eprint = {1911.00017},
  archivePrefix = {arXiv},
  primaryClass = {cond-mat.str-el}
}

@article{Chen2019Bosonization,
  author = {Chen, Yu-An and Kapustin, Anton},
  title = {Bosonization in three spatial dimensions and a 2-form gauge theory},
  journal = {Phys. Rev. B},
  volume = {100},
  pages = {245127},
  year = {2019},
  doi = {10.1103/PhysRevB.100.245127},
  eprint = {1807.07081},
  archivePrefix = {arXiv},
  primaryClass = {cond-mat.str-el}
}

@article{Levin2006Quantum,
  author = {Levin, Michael and Wen, Xiao-Gang},
  title = {Quantum ether: Photons and electrons from a rotor model},
  journal = {Phys. Rev. B},
  volume = {73},
  pages = {035122},
  year = {2006},
  doi = {10.1103/PhysRevB.73.035122},
  eprint = {hep-th/0507118},
  archivePrefix = {arXiv}
}

@article{Barkeshli2023Codimension,
  author = {Barkeshli, Maissam and Chen, Yu-An and Huang, Sheng-Jie and Kobayashi, Ryohei and Tantivasadakarn, Nathanan and Zhu, Guanyu},
  title = {Codimension-2 defects and higher symmetries in {(3+1)D} topological phases},
  journal = {SciPost Phys.},
  volume = {14},
  pages = {065},
  year = {2023},
  doi = {10.21468/SciPostPhys.14.4.065},
  eprint = {2208.07367},
  archivePrefix = {arXiv},
  primaryClass = {cond-mat.str-el}
}

@misc{Wen2026Holographic,
  author = {Xue, Hanyu and Wen, Xiao-Gang},
  title = {Holographic theory of mixed-dimensional statistics and conservation-encoding hopping-operator algebras},
  year = {2026},
  eprint = {2607.08583},
  archivePrefix = {arXiv}
}

@misc{Hsin2026Bockstein,
  author = {Hsin, Po-Shen and Chen, Yu-An},
  title = {{Bockstein} braiding statistics},
  year = {2026},
  eprint = {2607.02280},
  archivePrefix = {arXiv}
}

@article{Wang2015Topological,
  author = {Wang, Chenjie and Levin, Michael},
  title = {Topological invariants for gauge theories and symmetry-protected topological phases},
  journal = {Phys. Rev. B},
  volume = {91},
  pages = {165119},
  year = {2015},
  doi = {10.1103/PhysRevB.91.165119}
}

@misc{Walker2021Universal,
  author = {Walker, Kevin},
  title = {A universal state sum},
  year = {2021},
  eprint = {2104.02101},
  archivePrefix = {arXiv}
}

@article{Tarantino2016Symmetry,
  author = {Tarantino, Nicolas and Lindner, Netanel H. and Fidkowski, Lukasz},
  title = {Symmetry fractionalization and twist defects},
  journal = {New J. Phys.},
  volume = {18},
  pages = {035006},
  year = {2016},
  doi = {10.1088/1367-2630/18/3/035006},
  eprint = {1506.06754},
  archivePrefix = {arXiv},
  primaryClass = {cond-mat.str-el}
}

@article{Kitaev2006Anyons,
title = {Anyons in an exactly solved model and beyond},
journal = {Annals of Physics},
volume = {321},
number = {1},
pages = {2-111},
year = {2006},

issn = {0003-4916},
doi = {10.1016/j.aop.2005.10.005},
url = {https://www.sciencedirect.com/science/article/pii/S0003491605002381},
author = {Alexei Kitaev}
}

@article{Liujun2024Symmetries,
	title = {Symmetries and anomalies of {Kitaev} spin-{$S$} models: Identifying symmetry-enforced exotic quantum matter},
	pages = {100},
	author = {Liu, Ruizhi and Lam, Ho Tat and Ma, Han and Zou, Liujun},
	journal = {SciPost Phys.},
	volume = {16},
	year = {2024},
	publisher = {SciPost},
	doi = {10.21468/SciPostPhys.16.4.100},
	url = {https://scipost.org/10.21468/SciPostPhys.16.4.100}
}

@article{Bombin2010Topological,
  author = {Bombin, H.},
  title = {Topological order with a twist: {Ising} anyons from an {Abelian} model},
  journal = {Phys. Rev. Lett.},
  volume = {105},
  pages = {030403},
  year = {2010},
  doi = {10.1103/PhysRevLett.105.030403},
  eprint = {1004.1838},
  archivePrefix = {arXiv},
  primaryClass = {cond-mat.str-el}
}

@article{Barkeshli2019Symmetry,
  author = {Barkeshli, Maissam and Bonderson, Parsa and Cheng, Meng and Wang, Zhenghan},
  title = {Symmetry fractionalization, defects, and gauging of topological phases},
  journal = {Phys. Rev. B},
  volume = {100},
  pages = {115147},
  year = {2019},
  doi = {10.1103/PhysRevB.100.115147},
  eprint = {1410.4540},
  archivePrefix = {arXiv},
  primaryClass = {cond-mat.str-el}
}

@article{Tsui2020Lattice,
  author = {Tsui, Lokman and Wen, Xiao-Gang},
  title = {Lattice models that realize {$\mathbb{Z}_n$}-1 symmetry-protected topological states for even {$n$}},
  journal = {Phys. Rev. B},
  volume = {101},
  pages = {035101},
  year = {2020},
  doi = {10.1103/PhysRevB.101.035101},
  eprint = {1908.02613},
  archivePrefix = {arXiv},
  primaryClass = {cond-mat.str-el}
}

@misc{Gu2026Understanding,
  author = {Gu, Zhenbang and Luo, Ran and Wang, Yi-Nan and Zhang, Yi},
  title = {Understanding non-split {2-group} symmetry: {(3+1)D} {SymTFT}, anomaly and bordism},
  year = {2026},
  eprint = {2608.08766},
  archivePrefix = {arXiv}
}

@misc{HatcherNDSpectral,
  author = {Hatcher, Allen},
  title = {Spectral Sequences},
  note = {Additional Chapter 5 of Algebraic Topology},
  url = {https://pi.math.cornell.edu/~hatcher/AT/ATch5.pdf}
}

@article{Gu2014Symmetry,
  author = {Gu, Zheng-Cheng and Wen, Xiao-Gang},
  title = {Symmetry-protected topological orders for interacting fermions: Fermionic topological nonlinear $\sigma$ models and a special group supercohomology theory},
  journal = {Phys. Rev. B},
  volume = {90},
  pages = {115141},
  year = {2014},
  doi = {10.1103/PhysRevB.90.115141},
  eprint = {1201.2648},
  archivePrefix = {arXiv},
  primaryClass = {cond-mat.str-el}
}

@article{Tarantino2016Discrete,
  author = {Tarantino, Nicolas and Fidkowski, Lukasz},
  title = {Discrete spin structures and commuting projector models for two-dimensional fermionic symmetry-protected topological phases},
  journal = {Phys. Rev. B},
  volume = {94},
  pages = {115115},
  year = {2016},
  doi = {10.1103/PhysRevB.94.115115},
  eprint = {1604.02145},
  archivePrefix = {arXiv},
  primaryClass = {cond-mat.str-el}
}

@article{Wang2018Towards,
  author = {Wang, Qing-Rui and Gu, Zheng-Cheng},
  title = {Towards a complete classification of symmetry-protected topological phases for interacting fermions in three dimensions and a general group supercohomology theory},
  journal = {Phys. Rev. X},
  volume = {8},
  pages = {011055},
  year = {2018},
  doi = {10.1103/PhysRevX.8.011055}
}

@article{Wang2020Construction,
  author = {Wang, Qing-Rui and Gu, Zheng-Cheng},
  title = {Construction and classification of symmetry-protected topological phases in interacting fermion systems},
  journal = {Phys. Rev. X},
  volume = {10},
  pages = {031055},
  year = {2020},
  doi = {10.1103/PhysRevX.10.031055},
  eprint = {1811.00536},
  archivePrefix = {arXiv},
  primaryClass = {cond-mat.str-el}
}

@article{Feng2026Pauli,
  author = {Feng, Yitao and Xue, Hanyu and Kobayashi, Ryohei and Hsin, Po-Shen and Chen, Yu-An},
  title = {{Pauli} stabilizer formalism for topological quantum field theories and generalized statistics},
  journal = {J. High Energy Phys.},
  volume = {2026},
  number = {08},
  pages = {142},
  year = {2026},
  doi = {10.1007/JHEP08(2026)142},
  eprint = {2601.00064},
  archivePrefix = {arXiv},
  primaryClass = {quant-ph}
}

@article{Chen2023Exactly,
  author = {Chen, Yu-An and Hsin, Po-Shen},
  title = {Exactly solvable lattice {Hamiltonians} and gravitational anomalies},
  journal = {SciPost Phys.},
  volume = {14},
  pages = {089},
  year = {2023},
  doi = {10.21468/SciPostPhys.14.5.089},
  eprint = {2110.14644},
  archivePrefix = {arXiv},
  primaryClass = {cond-mat.str-el}
}

@article{Liang2024Extracting,
  author = {Liang, Zijian and Xu, Yijia and Iosue, Joseph T. and Chen, Yu-An},
  title = {Extracting topological orders of generalized {Pauli} stabilizer codes in two dimensions},
  journal = {PRX Quantum},
  volume = {5},
  pages = {030328},
  year = {2024},
  doi = {10.1103/PRXQuantum.5.030328},
  eprint = {2312.11170},
  archivePrefix = {arXiv},
  primaryClass = {quant-ph}
}

@misc{Liang2024Operator,
  author = {Liang, Zijian and Yang, Bowen and Iosue, Joseph T. and Chen, Yu-An},
  title = {Operator algebra and algorithmic construction of boundaries and defects in {(2+1)D} topological {Pauli} stabilizer codes},
  year = {2024},
  eprint = {2410.11942},
  archivePrefix = {arXiv},
  primaryClass = {quant-ph}
}

@article{Ellison2022Pauli,
  author = {Ellison, Tyler D. and Chen, Yu-An and Dua, Arpit and Shirley, Wilbur and Tantivasadakarn, Nathanan and Williamson, Dominic J.},
  title = {{Pauli} stabilizer models of twisted quantum doubles},
  journal = {PRX Quantum},
  volume = {3},
  pages = {010353},
  year = {2022},
  doi = {10.1103/PRXQuantum.3.010353},
  eprint = {2112.11394},
  archivePrefix = {arXiv},
  primaryClass = {quant-ph}
}

@misc{Sun2026Majorana,
  author = {Sun, Meng and Wang, Zongyuan and Tantivasadakarn, Nathanan and Chen, Yu-An},
  title = {{Majorana--Pauli} stabilizer codes and duality webs of fermionic topological phases},
  year = {2026},
  eprint = {2606.25048},
  archivePrefix = {arXiv},
  primaryClass = {quant-ph}
}

@article{Chen2021Disentangling,
  author = {Chen, Yu-An and Ellison, Tyler D. and Tantivasadakarn, Nathanan},
  title = {Disentangling supercohomology symmetry-protected topological phases in three spatial dimensions},
  journal = {Phys. Rev. Research},
  volume = {3},
  pages = {013056},
  year = {2021},
  doi = {10.1103/PhysRevResearch.3.013056},
  eprint = {2008.05652},
  archivePrefix = {arXiv},
  primaryClass = {cond-mat.str-el}
}

@article{Burnell2018Anyon,
  author = {Burnell, F. J.},
  title = {Anyon condensation and its applications},
  journal = {Annu. Rev. Condens. Matter Phys.},
  volume = {9},
  pages = {307},
  year = {2018},
  doi = {10.1146/annurev-conmatphys-033117-054154},
  eprint = {1706.04940},
  archivePrefix = {arXiv},
  primaryClass = {cond-mat.str-el}
}

@article{Kong2014Anyon,
  author = {Kong, Liang},
  title = {Anyon condensation and tensor categories},
  journal = {Nucl. Phys. B},
  volume = {886},
  pages = {436},
  year = {2014},
  doi = {10.1016/j.nuclphysb.2014.07.003},
  eprint = {1307.8244},
  archivePrefix = {arXiv},
  primaryClass = {cond-mat.str-el}
}

@article{Jiang2014Generalized,
  author = {Jiang, Shenghan and Mesaros, Andrej and Ran, Ying},
  title = {Generalized modular transformations in {$(3+1)$D} topologically ordered phases and triple linking invariant of loop braiding},
  journal = {Phys. Rev. X},
  volume = {4},
  pages = {031048},
  year = {2014},
  doi = {10.1103/PhysRevX.4.031048},
  eprint = {1404.1062},
  archivePrefix = {arXiv},
  primaryClass = {cond-mat.str-el}
}

@article{Jian2014Layer,
  author = {Jian, Chao-Ming and Qi, Xiao-Liang},
  title = {Layer construction of {3D} topological states and string braiding statistics},
  journal = {Phys. Rev. X},
  volume = {4},
  pages = {041043},
  year = {2014},
  doi = {10.1103/PhysRevX.4.041043},
  eprint = {1405.6688},
  archivePrefix = {arXiv},
  primaryClass = {cond-mat.str-el}
}

@article{Putrov2017Braiding,
title = {Braiding statistics and link invariants of bosonic/fermionic topological quantum matter in 2+1 and 3+1 dimensions},
journal = {Annals of Physics},
volume = {384},
pages = {254-287},
year = {2017},
issn = {0003-4916},
doi = {10.1016/j.aop.2017.06.019},
url = {https://www.sciencedirect.com/science/article/pii/S0003491617301859},
author = {Pavel Putrov and Juven Wang and Shing-Tung Yau}
}

@article{Thorngren2015Framed,
  author = {Thorngren, Ryan},
  title = {Framed {Wilson} operators, fermionic strings, and gravitational anomaly in {4d}},
  journal = {J. High Energy Phys.},
  volume = {2015},
  number = {02},
  pages = {152},
  year = {2015},
  doi = {10.1007/JHEP02(2015)152},
  eprint = {1404.4385},
  archivePrefix = {arXiv},
  primaryClass = {hep-th}
}

@article{Hsin2019Comments,
  author = {Hsin, Po-Shen and Lam, Ho Tat and Seiberg, Nathan},
  title = {Comments on one-form global symmetries and their gauging in {3d} and {4d}},
  journal = {SciPost Phys.},
  volume = {6},
  pages = {039},
  year = {2019},
  doi = {10.21468/SciPostPhys.6.3.039},
  eprint = {1812.04716},
  archivePrefix = {arXiv},
  primaryClass = {hep-th}
}

@article{Wang2016Bulk,
  author = {Wang, Chenjie and Lin, Chien-Hung and Levin, Michael},
  title = {Bulk-boundary correspondence for three-dimensional symmetry-protected topological phases},
  journal = {Phys. Rev. X},
  volume = {6},
  pages = {021015},
  year = {2016},
  doi = {10.1103/PhysRevX.6.021015},
  eprint = {1512.09111},
  archivePrefix = {arXiv},
  primaryClass = {cond-mat.str-el}
}

@article{Lin2015Loop,
  author = {Lin, Chien-Hung and Levin, Michael},
  title = {Loop braiding statistics in exactly soluble three-dimensional lattice models},
  journal = {Phys. Rev. B},
  volume = {92},
  pages = {035115},
  year = {2015},
  doi = {10.1103/PhysRevB.92.035115},
  eprint = {1503.00142},
  archivePrefix = {arXiv},
  primaryClass = {cond-mat.str-el}
}

@article{Cordova2019Exploring,
  author = {C\'ordova, Clay and Dumitrescu, Thomas T. and Intriligator, Kenneth},
  title = {Exploring 2-group global symmetries},
  journal = {J. High Energy Phys.},
  volume = {2019},
  number = {02},
  pages = {184},
  year = {2019},
  doi = {10.1007/JHEP02(2019)184},
  eprint = {1802.04790},
  archivePrefix = {arXiv},
  primaryClass = {hep-th}
}

@article{Wen2019Emergent,
  author = {Wen, Xiao-Gang},
  title = {Emergent anomalous higher symmetries from topological order and from dynamical electromagnetic field in condensed matter systems},
  journal = {Phys. Rev. B},
  volume = {99},
  pages = {205139},
  year = {2019},
  doi = {10.1103/PhysRevB.99.205139},
  eprint = {1812.02517},
  archivePrefix = {arXiv},
  primaryClass = {cond-mat.str-el}
}

@article{McGreevy2023Generalized,
  author = {McGreevy, John},
  title = {Generalized symmetries in condensed matter},
  journal = {Annu. Rev. Condens. Matter Phys.},
  volume = {14},
  pages = {57},
  year = {2023},
  doi = {10.1146/annurev-conmatphys-040721-021029},
  eprint = {2204.03045},
  archivePrefix = {arXiv},
  primaryClass = {cond-mat.str-el}
}

@article{Yoshida2015Topological,
  author = {Yoshida, Beni},
  title = {Topological color code and symmetry-protected topological phases},
  journal = {Phys. Rev. B},
  volume = {91},
  pages = {245131},
  year = {2015},
  doi = {10.1103/PhysRevB.91.245131},
  eprint = {1503.07208},
  archivePrefix = {arXiv},
  primaryClass = {quant-ph}
}

@article{Yoshida2016Topological,
  author = {Yoshida, Beni},
  title = {Topological phases with generalized global symmetries},
  journal = {Phys. Rev. B},
  volume = {93},
  pages = {155131},
  year = {2016},
  doi = {10.1103/PhysRevB.93.155131},
  eprint = {1508.03468},
  archivePrefix = {arXiv},
  primaryClass = {cond-mat.str-el}
}

@article{Kapustin2017Fermionic,
  author = {Kapustin, Anton and Thorngren, Ryan},
  title = {Fermionic {SPT} phases in higher dimensions and bosonization},
  journal = {J. High Energy Phys.},
  volume = {2017},
  number = {10},
  pages = {080},
  year = {2017},
  doi = {10.1007/JHEP10(2017)080},
  eprint = {1701.08264},
  archivePrefix = {arXiv},
  primaryClass = {cond-mat.str-el}
}

@misc{Tata2020Geometrically,
  author = {Tata, Sri},
  title = {Geometrically interpreting higher cup products, and application to combinatorial {Pin} structures},
  year = {2020},
  eprint = {2008.10170},
  archivePrefix = {arXiv},
  primaryClass = {hep-th}
}

@article{MedinaMardones2020Effective,
  author = {Medina-Mardones, Anibal M.},
  title = {An effective proof of the {Cartan} formula: The even prime},
  journal = {J. Pure Appl. Algebra},
  volume = {224},
  pages = {106444},
  year = {2020},
  doi = {10.1016/j.jpaa.2020.106444},
  eprint = {1907.12113},
  archivePrefix = {arXiv},
  primaryClass = {math.AT}
}

@article{Barkeshli2013Twist,
  author = {Barkeshli, Maissam and Jian, Chao-Ming and Qi, Xiao-Liang},
  title = {Twist defects and projective non-{Abelian} braiding statistics},
  journal = {Phys. Rev. B},
  volume = {87},
  pages = {045130},
  year = {2013},
  doi = {10.1103/PhysRevB.87.045130},
  eprint = {1208.4834},
  archivePrefix = {arXiv},
  primaryClass = {cond-mat.str-el}
}

@article{Barkeshli2013Theory,
  author = {Barkeshli, Maissam and Jian, Chao-Ming and Qi, Xiao-Liang},
  title = {Theory of defects in {Abelian} topological states},
  journal = {Phys. Rev. B},
  volume = {88},
  pages = {235103},
  year = {2013},
  doi = {10.1103/PhysRevB.88.235103},
  eprint = {1305.7203},
  archivePrefix = {arXiv},
  primaryClass = {cond-mat.str-el}
}

@article{Baez2004Higher,
  author = {Baez, John C. and Lauda, Aaron D.},
  title = {Higher-dimensional algebra {V}: 2-groups},
  journal = {Theory Appl. Categ.},
  volume = {12},
  pages = {423},
  year = {2004},
  eprint = {math/0307200},
  archivePrefix = {arXiv},
  url = {http://www.tac.mta.ca/tac/volumes/12/14/12-14abs.html}
}

@article{Chen2023Higher,
  author = {Chen, Yu-An and Tata, Sri},
  title = {Higher cup products on hypercubic lattices: Application to lattice models of topological phases},
  journal = {J. Math. Phys.},
  volume = {64},
  pages = {091902},
  year = {2023},
  doi = {10.1063/5.0095189},
  eprint = {2106.05274},
  archivePrefix = {arXiv},
  primaryClass = {cond-mat.str-el}
}

@misc{Kapustin2025Higher,
  author = {Kapustin, Anton and Xu, Shixiong},
  title = {Higher symmetries and anomalies in quantum lattice systems},
  year = {2025},
  eprint = {2505.04719},
  archivePrefix = {arXiv},
  primaryClass = {math-ph}
}

@misc{Kapustin2026Anomalies,
  author = {Kapustin, Anton and Xu, Shixiong},
  title = {Anomalies, higher symmetries, and homotopy theory {I}: crossed precosheaves and a nonabelian {{\v{C}}ech} construction},
  year = {2026},
  eprint = {2610.19004},
  archivePrefix = {arXiv},
  primaryClass = {math-ph}
}

@article{Shirley2026AnomalyFree,
  author = {Shirley, Wilbur and Zhang, Carolyn and Ji, Wenjie and Levin, Michael},
  title = {Anomaly-free symmetries with obstructions to gauging and onsiteability},
  journal = {Phys. Rev. Lett.},
  volume = {136},
  pages = {216602},
  year = {2026},
  doi = {10.1103/zrs1-j2xd},
  eprint = {2507.21267},
  archivePrefix = {arXiv},
  primaryClass = {cond-mat.str-el}
}

@article{Tu2026Anomalies,
  author = {Tu, Yi-Ting and Long, David M. and Else, Dominic V.},
  title = {Anomalies of global symmetries on the lattice},
  journal = {Phys. Rev. X},
  volume = {16},
  pages = {011027},
  year = {2026},
  doi = {10.1103/m188-w1ct},
  eprint = {2507.21209},
  archivePrefix = {arXiv},
  primaryClass = {cond-mat.str-el}
}
